\documentclass[preprint]{aastex}
\usepackage{lscape, longtable, hyperref}
\usepackage{graphicx}
\usepackage[dvips]{color}

\clearpage
\usepackage{lscape}

\begin{document}

\newcommand{\adam}[1] { \textcolor{red} {\ensuremath{\clubsuit} {\bf Adam:}  {#1}\ensuremath{\clubsuit} }}
\newcommand{\lynne}[1] { \textcolor{blue} {\ensuremath{\diamondsuit} {\bf Lynne:}  {#1}\ensuremath{\diamondsuit} }}
\newcommand{\kevin}[1] { \textcolor{green} {\ensuremath{\heartsuit} {\bf Kevin:}  {#1}\ensuremath{\heartsuit} }}
\newcommand{\will}[1] { \textcolor{magenta} {\ensuremath{\spadesuit} {\bf Will:}  {#1}\ensuremath{\spadesuit} }}
\newcommand{\john}[1] { \textcolor{cyan} {\ensuremath{\spadesuit} {\bf John:}  {#1}\ensuremath{\spadesuit} }}

\newcommand{\lapprox }{{\lower0.8ex\hbox{$\buildrel <\over\sim$}}}
\newcommand{\gapprox }{{\lower0.8ex\hbox{$\buildrel >\over\sim$}}}
\newcommand\iontoo[2]{#1$\;${\scshape{#2}}}
\newcommand\rptf{$R_{\rm PTF}$}
\newcommand\msun{$M_{\sun}$}
\newcommand\pyr{yr$^{-1}$}

\slugcomment{Submitted to AJ on 20120731}
\slugcomment{Revised on 20121107}
\shorttitle{PTF10nvg2}
\shortauthors{Hillenbrand et al.}

\title{
Highly Variable Extinction and Accretion in the Jet-driving Class I Type Young Star
PTF~10nvg (V2492 Cyg, IRAS 20496+4354) 
}
\author{
Lynne A. Hillenbrand\altaffilmark{1}, 
Adam A. Miller\altaffilmark{2}, 
Kevin R. Covey\altaffilmark{3,4}, 
John M. Carpenter\altaffilmark{1},
S. Bradley Cenko\altaffilmark{2},
Jeffrey M. Silverman\altaffilmark{2}
Philip S. Muirhead\altaffilmark{1}, 
William J. Fischer\altaffilmark{5}, 
Justin R. Crepp\altaffilmark{1,6}, 
Joshua S. Bloom\altaffilmark{2},
Alexei V. Filippenko\altaffilmark{2} 
}
\altaffiltext{1}{Astronomy Department, California Institute of Technology, Pasadena, CA 91125, USA.}
\altaffiltext{2}{Department of Astronomy, University of California, Berkeley, CA 94720--3411, USA.}
\altaffiltext{3}{Department of Astronomy, Cornell University, 226 Space Sciences Building, Ithaca, NY 14853, USA.}
\altaffiltext{4}{Lowell Observatory, 1400 West Mars Hill Road, Flagstaff, AZ 86001, USA}
\altaffiltext{5}{Department of Physics and Astronomy, University of Toledo, 2801 West Bancroft Street, Toledo, OH 43606, USA.}
\altaffiltext{6}{current address: Department of Physics, University of Notre Dame, 225 Nieuwland Science Hall, Notre Dame, IN 46556}

\begin{abstract}
We report extensive new photometry and spectroscopy of the highly variable 
young stellar object PTF 10nvg (also known as IRAS 20496+4354 and V2492 Cyg), 
including optical and near-infrared time series data as well as mid-infrared 
and millimeter data. Following the previously reported 2010 rise to
\rptf $\lapprox 13.5^m$ and subsequent fade, during 2011 and 2012 
the source underwent additional episodes of brightening,  
followed by several magnitude dimming events including prolonged faint states 
at \rptf $\gapprox 20^m$.  The observed high-amplitude variations are 
largely consistent with extinction changes ($\Delta A_V$ up to 30 mag) 
having a $\sim 220$ day quasi-periodic signal.  However, 
photometry measured when the source was near maximum brightness 
in mid-2010 as well as in late-2012 does not phase well to this period.
Spectral evolution includes not only changes in the spectral slope but
correlated variation in the prominence of TiO/VO/CO bands 
and atomic line emission, as well as anticorrelated variation 
in forbidden line emission which, along with H$_2$, dominates 
optical and infrared spectra at faint epochs. 
Notably, night-to-night variations in several forbidden 
doublet strengths and ratios are observed. 
High-dispersion spectra were obtained in a variety of photometric states
and reveal time-variable line profiles. 
Neutral and singly-ionized atomic species 
are likely formed in an accretion flow and/or impact while
the origin of zero-velocity atomic \ion{Li}{1} $\lambda$ 6707 
in {\it emission} is unknown.  
Forbidden lines, 
including several rare species, exhibit
blueshifted emission profiles and likely arise from an outflow/jet.
Several of these lines are also seen spatially offset 
from the continuum source position, presumably in a shocked region 
of an extended jet.  Blueshifted absorption components
of the \ion{Na}{1} D doublet, \ion{K}{1} $\lambda$ 7665,7669 doublet,
and the \ion{O}{1} 7774 triplet, as well as blueshifted absorption components 
seen against the broad H$\alpha$ and \ion{Ca}{2} triplet emission lines,  
similarly are formed in the outflow.  CARMA maps resolve on 
larger scales a spatially extended outflow in mm-wavelength CO.
We attribute the recently observed photometric and spectroscopic behavior 
to rotating circumstellar disk material located at separation 
$a \approx 0.7 ({M_\ast}/{M_\odot})^{1/3}$ AU
from the continuum source, causing the semi-periodic dimming. 
Occultation of the central star as well as the bright inner disk and 
the accretion/outflow zones renders shocked gas in the inner part 
of the jet amenable to observation at the faint epochs.
We discuss PTF~10nvg as a source exhibiting both accretion-driven
(perhaps analogous to V1647 Ori) and extinction-driven 
(perhaps analogous to UX Ori or GM Cep) high-amplitude variability phenomena.
\end{abstract}


\section{Introduction}

PTF~10nvg, in the North America Nebula region of recent star formation,
was identified by the Palomar Transient Factory \citep[PTF;][]{Law09,Rau09} 
on 2010 July 8, as an optical transient based on
automatic discovery and classification codes 
\citep{Bloom11}.  
The source was rapidly followed up by the PTF collaboration.
We announced our findings concerning the 2010 outburst 
of this Class I type young star in \citet{Covey11} 
where we presented optical and infrared light curves and multi-epoch
optical and infrared spectroscopy.  Therein, we made the analogy of
PTF~10nvg (now called V2492 Cyg) to the behavior of V1647 Ori,
an embedded Class I-type young star that was observed over the
past decade to undergo
several large amplitude photometric events on few year time scales.

V1647 Ori and PTF~10nvg both displayed $\sim$4--6 mag photometric rises 
from their faint states that were similar in the early stages 
to the outbursts of FU Ori stars.  However, these particular objects 
and others like them do not have the spectroscopic characteristics 
of FU Ori stars; rather than being absorption line dominated
(especially at high--dispersion),
they are emission-line dominated, with absorption seen in only  
a handful of blueshifted features arising in strong winds.
Further, the V1647 Ori-type objects do not remain in the 
elevated photometric state for the long time scale
associated with FU Ori outbursts (estimated at roughly a century); 
instead, their brightening episodes last only a few months to a few years 
and are characterized by large amplitude fluctuations on month to few month 
time scales.  Members of this category possibly undergo 
repeated episodes of their outbursting (and/or extinction dominated)
behavior at several year intervals. 
In this regard they are similar to the lower amplitude ($\sim$2-4 mag) 
but repeating outbursts of EX Lup-type systems, which last months to 
$>$1 year each and repeat on few year to decade intervals. 

Whether small-scale, low-amplitude events such as EX Lup-type outbursts,
larger scale, high--amplitude events such as the V1647 Ori-type events, 
or similarly large amplitude but also longer duration FU Ori-type events,
the star/disk system in such outbursts is interpreted as undergoing 
an episode of enhanced mass accretion and associated mass outflow.  
The accretion mechanisms for the different
categories of objects are likely related, and are attributed to instabilities 
in the inner accretion disks, possibly associated with cyclic
magnetosphere-disk interactions.  For PTF~10nvg in particular,
\citet{Covey11} estimated a mass accretion rate of $2.5\times 10^{-7}$\msun\pyr\
in the elevated state of 2010\footnote{The accretion rates quoted in
\citet{Aspin11} are a factor of 2.5--10 higher.}, 
similar to the value estimated for EX Lup
during its 2008 outburst (e.g. \citep{Juhasz12}).

Time-variable accretion, however, may not be the entire explanation
for many large-amplitude young star variables. 
As demonstrated herein, time-variable extinction is clearly an
important part of the PTF~10nvg interpretation, and may also play 
a significant role in the observed photometric behavior of many of
the so-called ``outburst" light curves highlighted to date 
in the literature.  For example, GM Cep was discussed by 
\citet{SiciliaAguilar08} as an EX Lup-type object,
but later assessed by \citet{Xiao10}, \citet{SemkovPeneva12},
 and \citet{Chen12} as having
extinction-dominated rather than accretion-dominated time series
behavior.  Both phenomena are perhaps simultaneously relevant, as we 
argue here for PTF~10nvg. 
Other well-known low-amplitude young star variables such 
as UX Ori, RR Tau, VV Ser, AA Tau as well as the ``dippers" discussed by
\citet{Morales11} and \citet{CodyHillenbrand10,CodyHillenbrand11}
also appear to be undergoing short time scale extinction events.   
Larger amplitude and -- importantly -- long period examples include 
KH15D \citep{Hamilton12}, WL4 \citep{Plavchan08}, 
and YLW16A \citep{Plavchan10} which have found
explanation in binary interactions with circumstellar disk material.

The 2010 brightening of PTF~10nvg 
was independently detected by  K. Itagaki and reported in 2010, August in the
subscription service of the Central Bureau for Astronomical Telegrams as 
CBET 2426.  Additional papers to date discussing the post-outburst source 
include those by \citet{Kospal11} and \citet{Aspin11} who also
present multi-color photometric, spectroscopic, and SED analysis of the source,
often using the \citet{Covey11} data.

In this paper we present new data gathered by us since the
publication of \citet{Covey11}. Intensive time series photometry
shows that the source has continued its large amplitude and color photometric
fluctuations.  Time series spectroscopy demonstrates 
1) continuum changes;
2) variation in the broad TiO / VO optical band emission that was detected for the 
first time during the 2010 outburst of PTF~10nvg, along with 
variation in other molecules such as CO and H$_2$; and 
3) drastic changes in the atomic emission lines indicative of accretion and 
outflow that are correlated (permitted lines) and anti-correlated 
(forbidden lines) with the photometric brightness.  
We also present high--dispersion spectral data; the line profiles are used 
to quantify the velocities relevant to the inflowing and outflowing material.
An updated post-outburst spectral energy distribution is discussed,
demonstrating that the source brightened relative to its historical
spectral energy distribution in the mid-infrared, as well as in
the near-infrared and optical, during 2010.
A spatially extended, low velocity molecular outflow is detected at
millimeter wavelengths.  New high spatial resolution near-infrared imaging 
rules out the presence of stellar companions within several hundred AU. 

In $\S$\ref{sec:obs} we describe our observations from 2009-2012. 
We then present our analysis of the multi-wavelength light curves 
($\S$\ref{sec:phot}), 
changes in the overall spectral energy distribution ($\S$\ref{sec:sed}), 
and
analysis of the continuum, absorption line, and emission line spectroscopy
including discussion of spatially offset forbidden line emission
($\S$\ref{sec:spec}). 
In $\S$\ref{sec:mult} we present constraints on the 
source multiplicity from high angular resolution direct imaging. 
In $\S$\ref{sec:disc} we interpret our results and discuss 
the broader implications of our findings for accretion and extinction
evolution of young stars. 
Finally in $\S$\ref{sec:summ} we summarize and conclude.

\section{Observations, Data Reduction, and Results}\label{sec:obs}

Our photometric and spectroscopic monitoring of PTF~10nvg (V2492 Cyg)
between 2009 and 2012 includes continued monitoring
in the R-band as well as J-, H-, and K$_s$-band time series photometry. 
Almost thirty epochs of low dispersion spectroscopy 
at optical or infrared wavelengths were obtained.   
High dispersion spectroscopy in the red optical was acquired at six epochs
and in the 1 $\mu$m atmospheric window at three epochs.  
In this section we describe the details of these various observations
and resulting data sets.

\subsection{New Photometry}

\subsubsection{Optical data from the Palomar Transient Factory}\label{sec:P48-data}

Continued optical monitoring of PTF~10nvg was conducted as part of the regular observations of the greater North America
Nebula field with the main PTF Survey Camera on the 48 inch Samuel Oschin Telescope at Palomar Observatory (hereafter P48). 
All observations were taken with 60 sec exposures in the $R_{\rm PTF}$ filter, a Mould $R$ filter which 
is similar to the SDSS-$r$ band (see \citealt{Law09}). 
P48 observations from 2009 August -- 2010 December are presented in 
\citeauthor{Covey11}; the new observations 
reported here started in 2011 February and continued through 2012 October. 
The typical cadence during the 2011 observing season was two observations 
separated by roughly one hour obtained every night during astronomical
bright time, though there was a period of high cadence observations 
during 2011 July with as many as eight observations taken in a single night.  
From 2012, April 18 onward, a typical cadence of 3 observations per night 
was adopted.

\begin{figure}
\includegraphics[angle=0,width=0.90\hsize]{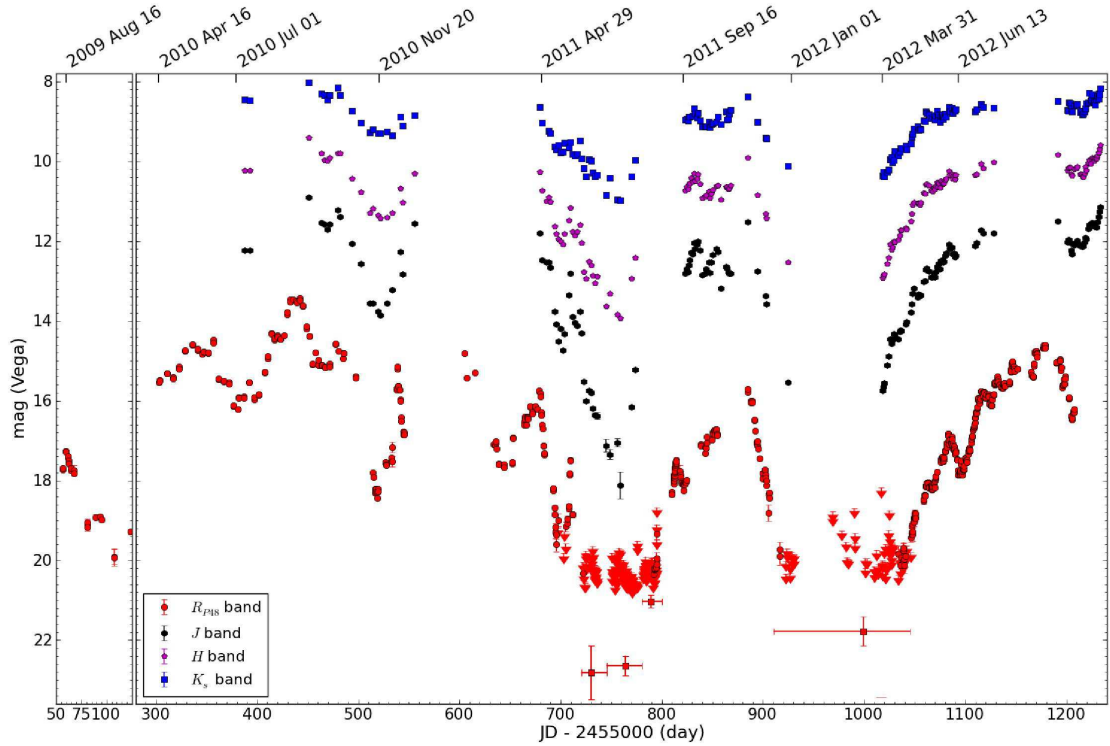}
\caption{Multi-wavelength lightcurve of PTF~10nvg 
with UT dates indicated above the figure.  From bottom to top the data streams
represent variability in the R-band (red; data from PTF) and in the 
J-, H-, and K-bands (black, purple, and blue respectively; data from PAIRITEL).
Error bars are shown, but the uncertainty in magnitude is typically smaller than 
the size of the symbols. 
During faint states when the source was not detected in individual frames,
photometry was measured from stacked PTF images (red squares, in the 21-23 mag range);
horizontal errorbars indicate the time range of measurements included in each stack. 
}
\label{fig:lightcurve}
\end{figure}

Brightness variations of PTF~10nvg are measured via a point-spread function (PSF) fitting routine as described in \citet{Sullivan2011}. 
Briefly, in each image frame the PSF is determined from several field stars and the average 
PSF is then fit to the position of PTF~10nvg with each pixel weighted according to photon statistics. 
Saturated pixels are masked from the fit, enabling photometry on even moderately saturated point sources (relevant near the absolute peak in 2010). 
The flux calibration is performed relative to SDSS, based on nightly observations 
of SDSS calibration fields (see \citealt{Ofek12} for further details), and all reported magnitudes are given for the native 
$R_{\rm PTF}$ filter in the Vega system.\footnote{We convert from the AB mag system to the Vega system using the 
following transformation: $m_R ({\rm Vega}) = m_R ({\rm AB}) - 0.19$.} Finally, we note that while the photometric 
measurements presented here are consistent with those we previously published in \citeauthor{Covey11}, the improved calibration 
and PSF model used here supersede our previous photometry.

The optical light curve for PTF~10nvg 
is shown in Figure~\ref{fig:lightcurve} and photometric measurements are reported in Table~\ref{tab:P48}. 
From 2011 June\ 08 through 2011 August\ 21, and 
2011 December\ 26 to 2012 April\ 09, PTF~10nvg was below the 5-$\sigma$ detection 
limit of the P48 images.  In order to constrain the brightness of PTF~10nvg 
during these prolonged faint periods, we took the average flux
from the PTF measurements over $\sim$1 month intervals 
and measured time-averaged photometry.  Uncertainties were determined 
by adding in quadrature the flux uncertainties from the individual 
non-detections and dividing by the total number of images 
included in the average flux measurement. 
The averages and the dates over which they were measured are shown in 
Figure~\ref{fig:lightcurve} and summarized in Table~\ref{tab:P48-stack}. 
During the summer of 2011 PTF~10nvg was on average $R \sim 23$ mag,
which can be compared to its summer of 2010 peak brightness $R \sim 13.5$ mag.

\subsubsection{Near-Infrared Data from PAIRITEL}

Continued near-infrared monitoring of PTF~10nvg was conducted with the 1.3~m Peters Automated Infrared Imaging 
Telescope \citep[PAIRITEL;][]{Bloom06} on Mt.~Hopkins, AZ. PAIRITEL is a roboticized system using the former 
2MASS southern hemisphere survey camera that employs two dichroics to observe simultaneously in
the $J$, $H$, and $K_s$ bands. Observations were scheduled and executed via a robotic system. PAIRITEL is 
operated in a fixed observing mode in which 7.8 s double-correlated images are created
from the difference of a 7.851 s and a 51 ms integration taken in rapid succession (see \citealt{Blake08}). 
The standard observing procedure involves taking three image pairs prior to dithering the telescope.

The raw data from these images are reduced using standard IR reduction methods via \hbox{PAIRITEL} PIPELINE III and 
the flux for all sources is measured via aperture photometry using 
{\tt SExtractor} \citep{Bertin96}. 
The final absolute calibration is determined using high signal-to-noise 
detections from 2MASS. When PTF~10nvg is very 
bright, corresponding to roughly $m < 10$ mag, it saturates the 7.851 s frames. This occurs in most of the reported 
$K_s$ band imaging and at some of the brightest epochs also in $H$ band. PIPELINE III produces ``short-frame''
mosaics consisting of reduced, stacked 51 ms images that are used for bright stage photometry
(see \citealt{Bloom09}). 
The ``short-frame'' mosaics contain a few dozen bright 2MASS stars which serve as calibration sources.

PAIRITEL has a known systematic uncertainty of a few percent on all flux measurements in each of the $J$, $H$, and $K_s$ bands (see \citealt{Blake08}, \citealt{Perley10}), which is larger than the typical statistical uncertainties. Following the method in \citeauthor{Perley10} we estimate the systematic uncertainties by measuring the scatter in the calibrated flux measurements of several (11 in this case) bright stars, arriving at  0.025, 0.03, and 0.06 mag in the $J$, $H$, and $K_s$ bands, respectively. These systematic uncertainties are added in quadrature with the statistical uncertainties to provide the final uncertainties on each measurement. We note that occasionally large thermal backgrounds prevented the detection of PTF~10nvg in the $K_s$ band images, which has affected 
some of the epochs included in our analysis and results 
in only $J-$ and $H-$band measurements being reported.  

The near-infrared light curve of PTF~10nvg is shown in 
Figure~\ref{fig:lightcurve} and the photometric measurements are summarized in 
Table~\ref{tab:PTEL}, which marks measurements made on short-frame mosaics as
well as those affected by thermal background.

\subsubsection{Mid-Infrared data from WISE}

The Wide-field Infrared Survey Explorer (WISE) observed PTF~10nvg 
as part of routine survey operations, obtaining several tens of scans
in late May and early June of 2010. 
This was around the time of the first peak but before the absolute peak
brightness in the optical light curve shown in Figure ~\ref{fig:lightcurve}.
The reported WISE photometry is partially saturated in bands W1 (3.35 $\mu$m), 
W2 (4.60 $\mu$m), and W3  (11.56 $\mu$m) with 12-15\% of the pixels
in the point-spread function in the saturated regime.  
The profile-fitted magnitudes account for the saturation
while the aperture magnitudes are lower limits to the fluxes.  
The W4 (22.24 $\mu$m) band photometry is not affected by saturation.
Additional mid-infrard observations with WISE 
\footnote{
The post-cryogenic phase of the WISE mission covered the field
again in late November of 2010.  These would consist of only W1 and W2 data,
but as the observations took place 
near the source's photometric minimum for the 2010 season
they could be unsaturated.  These data are not yet
available in the WISE archive.}.  
as well as Spitzer and Herschel
\footnote{
Spitzer has observed this source at several epochs during
2011 and 2012 (program PI Abraham).  
Herschel has also observed PTF~10nvg (program PI A. K{\'o}sp{\'a}l)  
obtaining imaging photometry at 70, 100, and 160 $\mu$m at several epochs  
as well as single epoch
250, 350, and 500 $\mu$m data.
}.
have also been obtained.

\subsubsection{Millimeter-Wave data from CARMA}

PTF 10nvg was observed in the 2.7~mm continuum and several molecular lines
with the Combined Array for Research in Millimeter-wave
Astronomy (CARMA), located at an altitude of 7200 feet in the Inyo
mountains of eastern California. The observations used six 10.4 m
diameter antennas and nine 6.1 m diameter antennas. 
C-configuration, which provides baseline lengths between 26 and 370 m, data 
were acquired on 2010 November 1 and 2 UT, near the 2010 photometric minimum.
Additional observations in the D-configuration, 
with baseline lengths between 11 and
148 m, were obtained on 2012, June 22 and 23 UT.

For the C-configuration observations, the local oscillator was set to a
frequency of 107.7 GHz. The CARMA correlator contains 8 bands. Six bands were
configured to have 487~MHz bandwidth to observe the continuum. The
remaining two bands were set to observe the J=1-0 rotational transitions of
$^{12}$CO (115.271204 GHz) and $^{13}$CO (110.201353GHz) with 0.98~kHz 
resolution ($\sim$ 0.25 km s$^{-1}$). Time-variable atmospheric and
instrumental gains were calibrated by observing the source J2038+513 every 20
min. Between gain calibrator observations, PTF 10nvg was observed for 7 min,
followed by a 7 minute observation of a second embedded star in the North America Nebula.
The passband amplitude and phases were calibrated using observations of 3C454.3
on each night. The absolute flux was calibrated by observing Uranus on November
1 and Neptune on November 2; the estimated absolute flux uncertainty is 10\%.
The visibility data were calibrated using the MIRIAD reduction package.

The D-configuration observations were similar to the C-configuration
observations, with the following differences. Two of the correlator bands were
centered on CN, N=1-0 J=3/2-1/2 (113.490982 GHz) and C$^{18}$O, J=1-0 (109.78216
GHz), but these data are not presented here. The passband was calibrated with
observations of 3C84. Flux calibration was performed by observing MWC~349 on
the first night and Neptune on the second night. Finally, on the second night
in D-configuration, a seven point mosaic was obtained to image the larger scale
molecular cloud.

\begin{figure}
\includegraphics[angle=0,width=0.50\hsize]{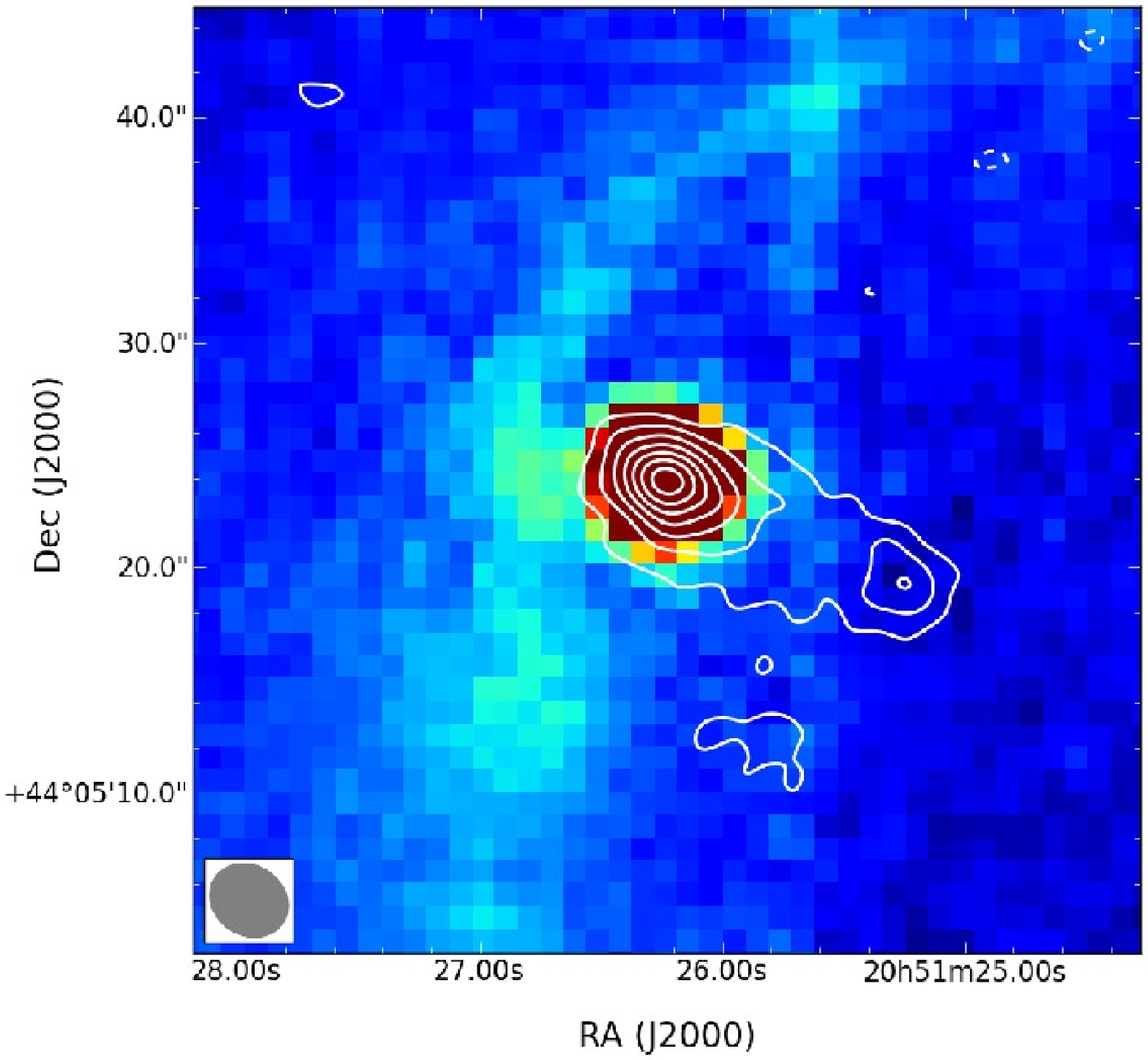}
\includegraphics[angle=0,width=0.50\hsize]{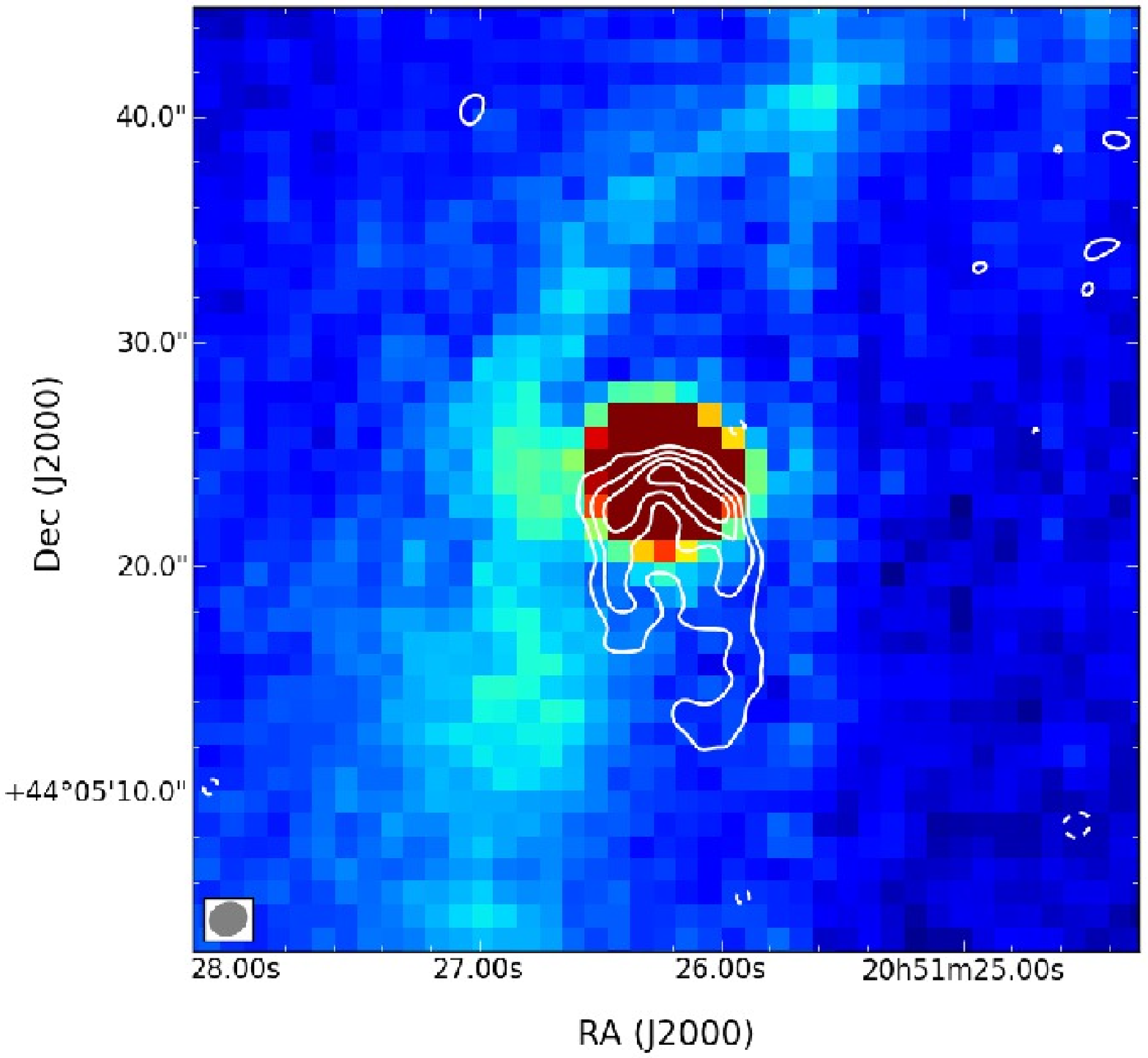}
\caption{{\it Left panel:}
CARMA 2.7 mm continuum image (contours) overlaid on a PTF R-band image 
(color scale) from 2010.  
Solid contours begin at 3$\sigma$ with 3$\sigma$ increments, where $\sigma$ =
0.18 mJy/beam and the beam size is $3.6''\times3.1''$.
Dashed contours start at $-3\sigma$. The strongest 2.7 mm continuum
source is coincident with PTF 10nvg. A second continuum source is located
11$''$ to the southwest, but has no counterpart in optical or infrared 
images out to 70 $\mu$m.
{\it Right panel:}
Contour map of the $^{12}$CO J=1-0 integrated 
intensity between 6.5 and 11\,km~s$^{-1}$ overlaid on the PTF optical image.
Contours begin at 3$\sigma$ with 3$\sigma$
increments, where $\sigma$ = 96~mJy~beam$^{-1}$~km~s$^{-1}$ 
and the beam size is $1.7''\times1.4''$.
The $^{12}$CO emission at these velocities traces a red-shifted lobe of an
outflow with PTF 10nvg at its apex. The corresponding blue shifted lobe of
the outflow was not detected. 
}
\label{fig:contoutflow}
\end{figure}

\begin{figure}
\includegraphics[angle=0,width=1.00\hsize]{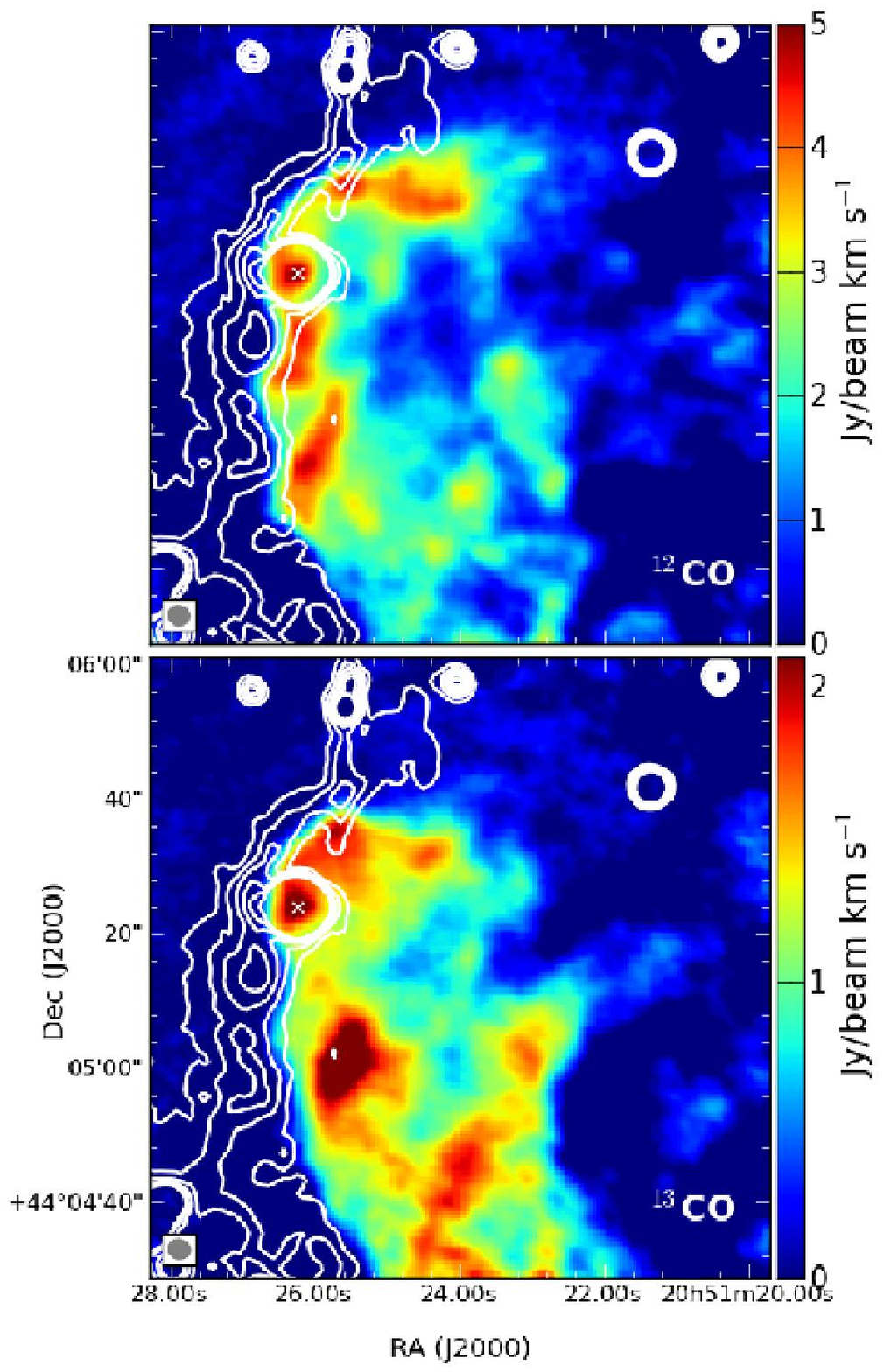}
\caption{Wide-field view showing  velocity-integrated (3-6.25 km~s$^{-1}$) 
$^{12}$CO (top panel) and $^{13}$CO (bottom panel) images 
in the color scale compared with PTF R-band intensity in the contours.
Note the reversal in color vs contour representation
compared to Figure~\ref{fig:contoutflow}.
The location of PTF~10nvg is marked by the ``{\rm x}" towards the upper left.
A remarkable alignment is found between the arc of H$\alpha$ nebulosity
(see also Figure~\ref{fig:optimages})
and the eastern edge of the CO emission.
}
\label{fig:co}
\end{figure}

Continuum images were generated by inverting the complex visibilities using
``natural" weighting for optimal point-source sensitivity. The images were then
deconvolved using the MIRIAD task mossdi. The RMS noise in the image measured
in emission-free regions near the vicinity of PTF 10nvg is 0.18~mJy/beam, with
a full-width-at-half-maximum (FWHM) beam size of $3.6''\times3.1''$.
Figure~\ref{fig:contoutflow} shows the 2.7~mm continuum map overlaid on a PTF
optical image. A compact source with a signal-to-noise of 21 is detected in the
field at a right ascension and declination of 20:51:26.2, $+$44:05:23.8 J2000,
corresponding to the optical/near-infrared position of PTF 10nvg. A secondary
compact peak in the continuum image is located 11$''$ to the southwest; this
peak has no counterpart in the optical image shown, or in archival 
ground-based or Spitzer images between 1.2 and 70 $\mu$m.

The integrated 2.7~mm continuum flux toward PTF 10nvg is 5.60 $\pm$ 0.26 mJy
(random error), obtained by fitting an elliptical gaussian to the image.
Formally the source is extended relative to the synthesized beam with a
deconvolved size of $2.7''\times 1.5''$ and a position angle of 66$^\circ$. 
However, when imaged with the C-configuration data only, 
the source is unresolved with a synthesized beam of
$1.8''\times1.5''$ and an integrated flux of $3.2\pm0.30$~mJy. Similar
combinations of extended and point source components have been observed in
protostellar sources (e.g., J{\"o}rgensen et al.~2004), and have been interpreted
to represent a resolved envelope and a compact disk. The presence of these
structures in PTF~10nvg are consistent with the Class~I classification from the
shape of the spectral energy distribution. 

The $^{12}$CO and $^{13}$CO spectral line images were generated using a Brigg's
robust factor of zero to yield a beam size of $3.2''\times2.8''$. Since the
emission is extended over several arcminutes, we used the maximum entropy
program in MIRIAD (mosmem) to deconvolve the images. However, we caution that
since no zero spacing data are available, the extended features are not 
reliably recovered in these images. Figure~\ref{fig:co} shows images of the
integrated $^{12}$CO and $^{13}$CO intensity between velocities of 3 and
6.25\,km~s$^{-1}$, which covers the main part of the cloud. 

PTF 10nvg, denoted by the cross in the figure, is coincident with a local
maximum in the $^{12}$CO and $^{13}$CO emission. Emission from the larger scale
extends to the west and southwest over a spatial extent of at least 0.18~pc.
The full extent of the local cloud has not been encompassed with the
CARMA mosaic, however. The detailed spatial structure of the extended cloud differs in
the $^{12}$CO and $^{13}$CO images, which we attribute both to differences in
optical depth and the lack of sensitivity to extended emission with an
interferometer. 

To the east, both the $^{12}$CO and $^{13}$CO emissions falls off
sharply, suggesting that this structural feature has been robustly
imaged with CARMA; PTF~10nvg is located along this ridge. The white contours in
Figure~\ref{fig:co} show the PTF optical image, which encompasses the H$\alpha$ line. The
H$\alpha$ emission peaks to the east of the $^{12}$CO and $^{13}$CO emission,
and closely follows the edge of the molecular cloud. Thus the eastern edge of
the cloud appears to be illuminated by stars that are photo-dissociating the
H$_2$ on the cloud surface.

\begin{figure}
\epsscale{0.5}
\includegraphics[angle=0,width=1.00\hsize]{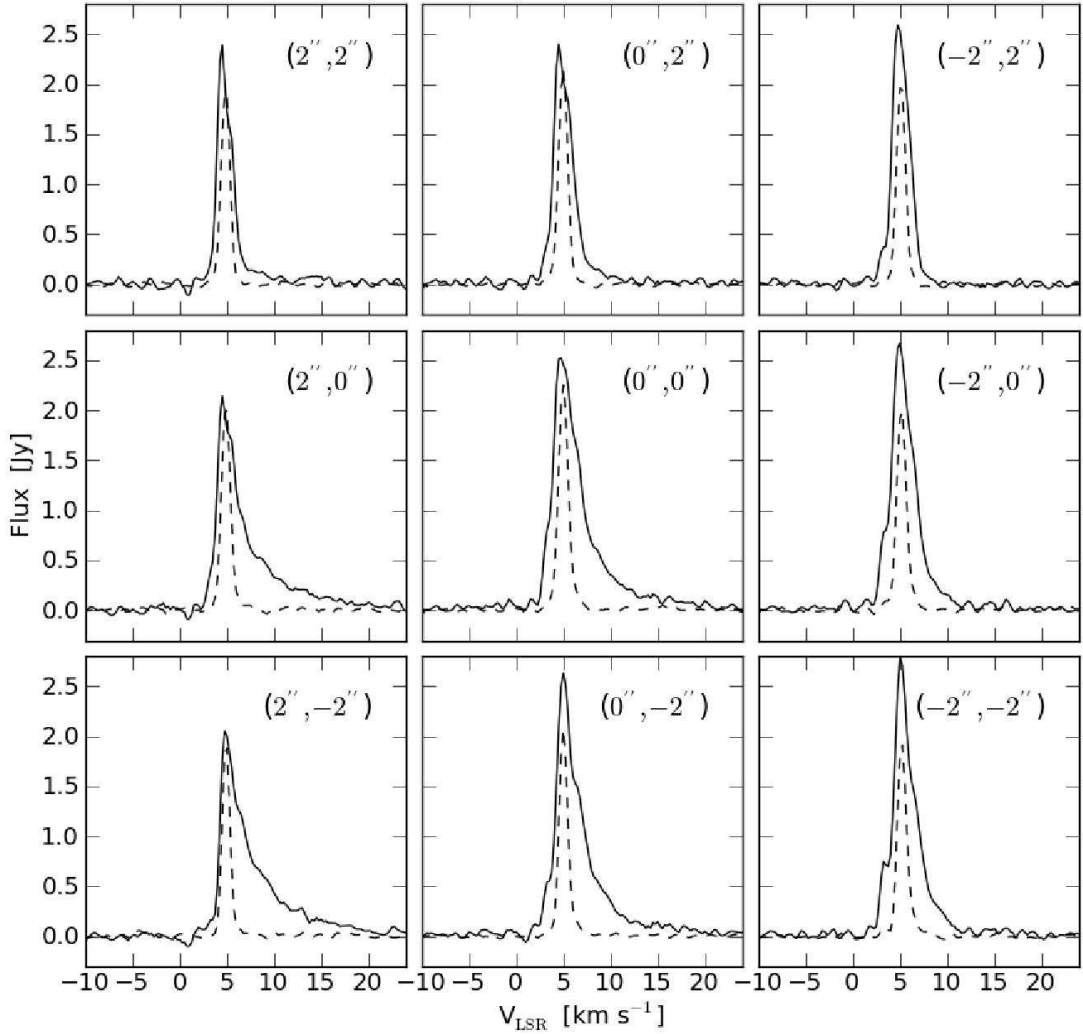}
\epsscale{1.0}
\caption{
$^{12}$CO (solid) and $^{13}$CO (dashed) J=1-0 spectra in the vicinity of PTF 10nvg.
Panels are labeled with positional offsets in arcseconds relative to the position of
PTF 10nvg. Both the $^{12}$CO and $^{13}$CO emission contain strong emission at the 
cloud systemic velocity that peaks at the location of PTF 10nvg (see Figure~\ref{fig:co}). The
$^{12}$CO emission also displays reshifted emission that extends to the south 
of PTF 10nvg (see right panel of Figure~\ref{fig:contoutflow}) that represents a redshifted outflow.
}
\label{fig:cospec}
\end{figure}
\clearpage

Figure~\ref{fig:cospec} shows the $^{12}$CO and $^{13}$CO spectra within $\pm2''$ of PTF 10nvg.
Both $^{12}$CO and $^{13}$CO contain strong emission at a velocity of $\sim$ 5 kms$^{-1}$,
which corresponds to the systematic velocity of the cloud. As shown in 
Figure~\ref{fig:co}, the
molecular emission at these velocities contains a local maximum at the position of PTF 10nvg.
The $^{12}$CO line profiles in Figure~\ref{fig:cospec} also display weaker emission that is redshifted
up to 10-15 km s$^{-1}$ from the systemic cloud velocity. The right panel in Figure~\ref{fig:contoutflow}
shows an image of the redshifted image constructed from the C-configuration data. The
$^{12}$CO emission at these velocities traces an arc-like structure extending to the south,
with PTF 10nvg at the apex. The morphology and the redshifted velocities indicate an
outflow from the young (proto)star. No evidence is seen in the $^{12}$CO spectra for a
corresponding blueshifted component of an extended flow.

\subsection{Adaptive Optics Imaging}

To assess the multiplicity of PTF~10nvg at wide orbital separations, we
acquired high spatial resolution
near-infrared images on 2010 September 25 UT using NIRC2 (K. Matthews) 
and the Keck II adaptive
optics system at Mauna Kea. The source was sufficiently bright during its
outburst phase to serve as its own
natural guide star at the time of our observations. We obtained 18
unocculted, dithered frames in the $K_s$
band totaling 135 sec of on-source integration time. The NIRC2 narrow camera
mode provided fine spatial
sampling (10 mas/pix) of the stellar point-spread function 
(FWHM$\approx 65$ mas for these observations).  Raw frames were
processed by cleaning hot pixels,
subtracting background noise, and aligning and coadding the results.
As discussed in \S~\ref{sec:mult}, no companions are found in these observations.

\subsection{New Spectroscopy}

Tables 1 and 2 describe the optical and infrared spectroscopy obtained
by our collaboration of PTF~10nvg since its photometric outburst in 2010.
Low resolution optical spectroscopy generally covering the red optical
spectral region ($>$6200 \AA) were acquired at over 20 epochs 
with the facility spectrographs at one of: the Keck I 10m 
\citep[LRIS;][]{Oke1995,McCarthy1998,Steidel2004}, 
Palomar 5m \citep[DBSP;][]{Oke1982}, or 
Lick 3m \citep[Kast;][]{Miller93} telescopes.
High dispersion optical data were also obtained at Keck I using 
HIRES \citep{Vogt94} on 6 occasions.  In the infrared, low resolution data 
covering approximately 1-2.5 $\mu$m were obtained at over 10 epochs 
using facility spectrographs at: the 3m InfraRed Telescope Facility (SPeX; 
\citep{Rayner03}),
Palomar Observatory 5m (TripleSpec; 
\citep{Herter08}), and
3.5 m Apache Point Observatory (TripleSpec; 
\citep{Wilson04}).  High dispersion infrared data in the Y-band
(1 $\mu$m atmospheric window) were also obtained, at Keck II
using NIRSPEC \citep{McLean98} at 3 epochs.

The data acquisition strategy included sensitivity 
to obtaining observations at the parallactic angle, 
and to proper acquisition of flux calibration standards.
Spectral data processing followed standard procedures for bias level
correction, flat fielding, spectral extraction including background subtraction,
and flux calibration for the low dispersion data.  
Of note is the extended emission from a nebular arc region about 7" 
away from PTF~10nvg (see Figure~\ref{fig:contoutflow} or
~\ref{fig:optimages} for morphology)
that must be accounted for when perfoming background subtraction on 
spectra.  The absolute calibration 
of the low dispersion data is typically good to $\pm 20\%$ and slightly better 
under photometric conditions.   The resulting signal-to-noise ratios in the
optical spectra range from 15-45, but lower in the 2011, 
late June to early August time frame when the source was optically very faint.
For the moderate resolution infrared spectra, the signal-to-noise 
was typically $\sim$250 in the K-band and $\sim$40 in the J-band though
varied by factors of a few depending on the weather and source brightness,
and  color at any particular epoch.

\section{Photometric Analysis}\label{sec:phot}

\subsection{Optical and Near-Infrared Variability}

The light curves published in 
\citeauthor{Covey11} included data obtained through 2010 December\ 14 
(JD 2455545). Updated multi-wavelength light curves appear in 
Figure~\ref{fig:lightcurve} and show that the source continued its large
amplitude fluctuations throughout the 2011 and 2012 observing seasons. 
The data do not 
completely sample the photometric behavior, notably during the period 
when the field was too close to the Sun from approximately 
late December to early February of each year or during survey-designed
gaps in the R$_{PTF}$ observing, 
as well as when the source was optically faint and below
the PTF detection limits.  Our near-infrared observations, 
in addition to the gaps in coverage associated with solar
conjunction, have large annular gaps in August 
when Mt.\ Hopkins is closed due to the Arizona monsoon season.
Despite the incomplete sampling, 
the photometric evolution of PTF~10nvg is obviously dominated by large amplitude
variations on several month time scales, with superposed chaotic
behavior on shorter time scales. 

\begin{figure}
\includegraphics[angle=0,width=1.00\hsize]{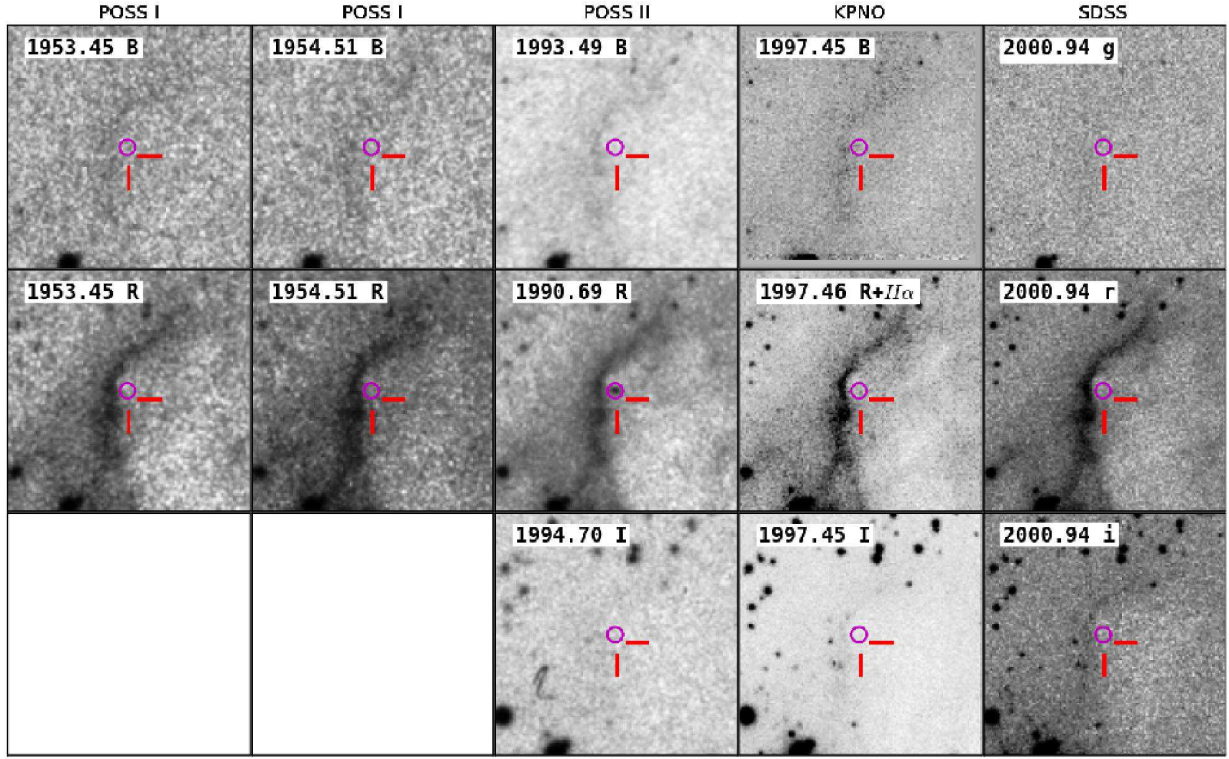}
\caption{
Optical images with north upward and east to the left covering 96'' 
of the PTF~10nvg vicinity, spanning many decades.
Rows contain blue, red, and far-red data from top to bottom, while
the columns contain different imaging surveys, as indicated above the figure.
The individual panels correspond to different epochs.
PTF~10nvg is located at the center of the
magenta circle and is not visible at most epochs, though was apparent on
the 1990 R band plate.  The intersection of the red cross hairs 
indicates the position of the USNO-B source identified 
in Aspin (2011) as the likely quiescent counterpart to PTF 10nvg but
no source is detected at this position in any of the POSS I plate scans. 
}
\label{fig:optimages}
\end{figure}

As described in \citeauthor{Covey11}, PTF~10nvg was not a recognized 
optical or near-infrared source prior to its 
2010 identification as an outbursting (possibly transient) source.
It is noted in Aspin (2011), however, that the source is reported at 
faint optical magnitudes in the USNO-B1 catalog in 1979, November\footnote{
We were unable to find a POSS image of the field taken in 1979. 
However, in the digitized scans of all POSS images including 
the position of PTF~10nvg, the source is detected in one red plate 
taken in 1990 (see Figure~\ref{fig:optimages}).
The $B$ and $R$ mag measurements reported in Aspin (2011) 
correspond to the USNO-B1.0 source 1340-0394684, which has a cataloged position 
that is $ \sim$3.4\arcsec\ away from the optically measured position 
of PTF~10nvg. 
The USNO-B1.0 catalog reports an epoch of 1966.2 for this source, 
which is the result of considering the B1 and R1 plates 
from the early 1950's (on which the reported photometry 
is in fact likely related to the nebulosity rather than the point source),
and the R2 plate from 1990 (on which there is a true source detection), with 
the final position given as the geometric mean of these three epochs
after taking a potential proper motion term into account 
(S. Levine, private communication).} 
as well as in a 1997, October 30 H$\alpha$ image (their Figure 2); 
it is also recorded in UKIDSS infrared data from 2006, June.
These detections led us to investigate the historical variability 
of PTF~10nvg via examination of images taken over several decades 
prior to the 2009 optical detection by PTF. In Figure~\ref{fig:optimages}
we show the vicinity of PTF~10nvg from several sources including  
the digitized plate scans of POSS-I and POSS-II, our own previously unpublished 
optical images from the KPNO-0.9m, and those from SDSS.
The apparitions of the source in 1990, 1997, and 2006 
combined with the large number of nondetections in other imaging data sets
spanning approximately 50 years (see Figure~\ref{fig:optimages})
hint that this young stellar object has displayed long-term, 
large amplitude variability. 
Notably, although the 1997, October H$\alpha$ detection of the point source
exhibited by \citet{Aspin11} is convincing, the source was not detected 
in BVRI or H$\alpha$ images taken only a few months earlier (1997, August).
Considering the body of evidence, the average faint-state magnitude 
in the red optical  is likely below about 22 mag, based on the
deepest digital data from the late 1990s. 
The source clearly has been intermittently detectable, however.

In August of 2009 the object was captured by PTF at around 
$R_{\rm PTF} \approx 17.2$ mag in what may have been the decline 
from an earlier local peak, with subsequent fading
over a roughly 1 month time scale to $\approx$20 mag.  
Our next observations were approximately 200 days later by which point PTF~10nvg 
had brightened by over four magnitudes.  It exhibited a first local
peak in July of 2010, then dropped in brightness by more than 
a magnitude and rose to its maximum recorded brightness 
of $R_{\rm PTF} \approx 13.5$ mag 
approximately 40 days later, in August of 2010. 
The light curve published in \citeauthor{Covey11}
demonstrate a third peak in early December of 2010.

During the 2011 and 2012 observing seasons, several subsequent peaks that were
typically narrower in time and fainter than the initial two 2010 peaks 
have been observed, as well as a broad, bright peak exhibited
in mid-2012. Notably, the declines from each peak were 
sharp, approximately 0.1 mag day$^{-1}$, while the rise times 
were about half as steep\footnote{The individually measured slopes 
for events lasting longer than a week are: 
0.14, 
0.14, and 0.14 mag day$^{-1}$ 
for the drops following the peaks that occurred on $\sim$JD 
2455678, 
2455884, and 2456086, respectively.
For the rise times, the measured slopes are 
$\sim$0.05, 0.05, and 0.07 mag day$^{-1}$ characterizing the events starting
on JD 2455652, 2455820, and 2456058, respectively.}.
The total optical variability amplitude observed during the 2011 and 2012 seasons
was more than 4.5 magnitudes, with PTF~10nvg as bright as 
$R_{\rm PTF} \sim 15$ mag in early 2011 and repeatedly reaching 
$R_{\rm PTF} \sim 16$ mag during the rest of 2011 and 2012. The source was
also repeatedly fainter than the P48 60-second detection limit 
($\approx$ 20.5 mag in this high background field).  Stacked images 
from these prolonged faint periods 
(e.g. between late June through late August in 2011 and again from late
December 2011 through mid-March 2012) 
indicate that the average flux from PTF~10nvg reached as faint as
$R_{\rm PTF} \sim 23$ mag (see \S~\ref{sec:P48-data}). 
Near-infrared detections continued through the periods of optical dimming
and at some epochs better sample the source fading 
and rising while it remained optically faint. 

\begin{figure}
\includegraphics[angle=0,width=0.90\hsize]{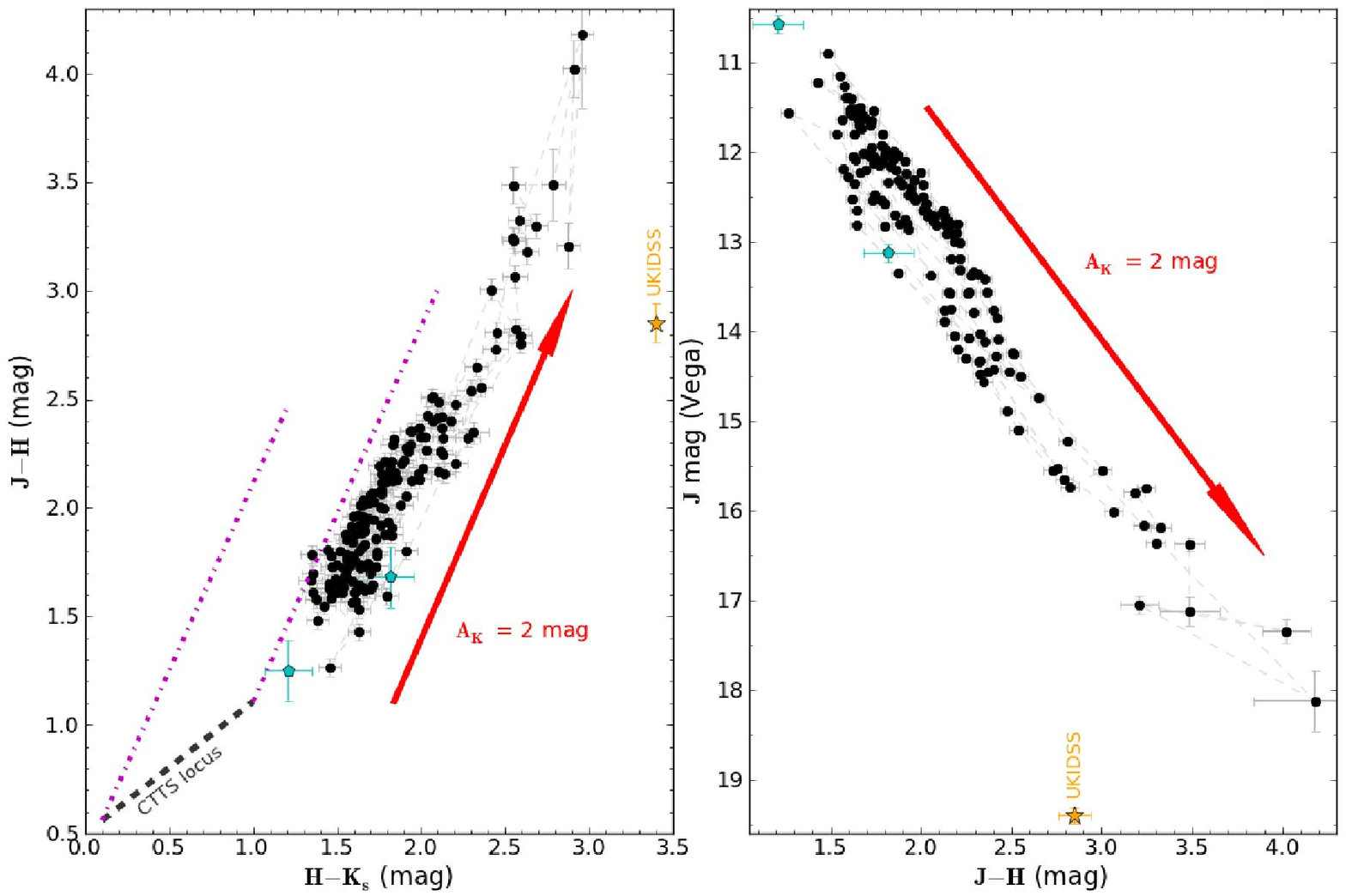}
\caption{Color-color and color-magnitude diagrams with the JHK$_s$ lightcurve
data exhibited in Figure~\ref{fig:lightcurve} shown as black points.  
Cyan pentagons show two measurements from \citet{Aspin11} 
while the gold star shows the 2003 measurement from UKIDSS. 
For reference, the dashed line indicates the locus of classical 
T Tauri stars as measured by \citet{Meyer97}
using the color transform of \citet{Carpenter01}. 
When the overall features of the dataset are considered,
the time variation of the black and cyan points is
consistent with the effects of extinction following an interstellar 
extinction law from \citet{Indebetouw05}, plotted in dash-dot lines.
The point-to-point
variations, however, may require additional thermal effects beyond the
dominant extinction effect.
}
\label{fig:color}
\end{figure}

The infrared light curves (Figure~\ref{fig:lightcurve}) generally follow 
the fading and peaking behavior seen in the optical, with 
no detectable time lag between the optical and infrared 
valleys and peaks seen within our coarse sampling. 
However, the amplitudes are different between the RJHK$_s$ bands, 
and indicate large color variations.  PAIRITEL obtains data
simultaneously in the $J$, $H$, and $K_s$ bands allowing for precise measurements of color change 
in the near-infrared. A color-color diagram (left side of Figure~\ref{fig:color}) indicates that 
during its large amplitude brightness changes,
the color evolution of PTF~10nvg does not deviate strongly from the  
behavior expected from variations in line-of-sight extinction. 
The observed scatter around the expected relationship for time-variable 
obscuration by standard dust grains is roughly consistent with 
the typical observational uncertainties.  A color-magnitude diagram 
(right side of Figure~\ref{fig:color}) reinforces the finding
that the source becomes redder when fainter in a manner consistent with 
extinction\footnote{We note that the same conclusion is not reached
in \citet{Aspin11} based on an analysis of the UKIDSS detection 
from 2006 and seven additional observations 
made between 2010 July--2010 September.
We find that indeed the 2006 detection by UKIDSS 
($J-H = 2.85$ mag; $H-K_s = 3.4$ mag) lies far from the ensemble of 
2010 outburst through 2012 points 
measured by PAIRITEL, as shown in Figure~\ref{fig:color}.
However, the deviation argued for in \citet{Aspin11} 
of the observations taken during outburst 
from standard reddening laws appears to be the result of an
assumption that the unreddened origin for PTF~10nvg lies 
along the ``classical T Tauri locus."  
As shown in both \citeauthor{Covey11} and \citet{Aspin11}, however,
PTF~10nvg {\it was not} a classical T Tauri star prior to outburst, but rather 
a Class I-type source.  As Class I sources have both envelope and disk
emission contributing to the near-infrared colors, there is no reason 
to expect that the origin of the reddening vector should lie 
on the classical T Tauri star locus.  This realization does not, however,
explain the discrepant colors reported by UKIDSS. One possibility is that
in its faint state the source becomes more dominated by scattered light
which could explain the turn blueward in the J-H color.  However, the
H-K discrepancy seems to require redder colors at this epoch not bluer. 
}. 

Assuming a standard interstellar extinction law and that all of the observed near-infrared color variations are the 
result of changes in extinction along the line of sight 
(rather than to an outburst), we find that the 
observed excursions in near-infrared color-color and color-magnitude 
diagrams correspond to variations in extinction of $\Delta A_V \gtrsim$ 30 mag
(see also discussion in \S~\ref{sec:spec}
related to Figure~\ref{fig:av_vs_time}). 
Extending the analysis, if we adopt the position that all of 
the historical photometric variation oberved for this source 
can be attributed to changes in line-of-sight extinction, the 2MASS 
non-detection of the source at K$_s$-band relative to the peak brightness
observed with PAIRITEL implies $\Delta A_V \approx$ 70 mag.

We sought next to quantify the time scales associated with the observed 
large amplitude systematic changes in source brightness. 
Some general challenges for interpreting the lightcurve of PTF~10nvg include: 
(i) the inhomogeneous sampling that fails to resolve all the peaks and valleys in the light curve,
(ii) the limited time baseline of the observations that may not properly 
resolve long-term trends, and (iii) the likely possibility 
that the system is dynamic, meaning that periods, amplitudes, and 
relative phases could be changing cycle to cycle. 

In our search for periodic signals in the brightness variations 
of PTF~10nvg we use a generalized Lomb-Scargle periodogram 
(\citealt{Lomb76,Scargle82,Zechmeister09}) and
analyze each photometric band separately (see 
\citealt{Richards11} for more details on our implementation). 
The extended period of non-detections in the 
$R_{\rm PTF}$ band would lead to a biased periodogram 
since the framework for the Lomb-Scargle analysis does not account for  
upper limits. As a result, we focus on the three near-infrared bands. 
The form of the long-term secular changes in 
the light curve is difficult to identify over the relatively limited timescales that we probe. Thus, 
we perform our Lomb Scargle analysis by simultaneously fitting for the periodic and long-term linear trend in the data. The observations do not clearly justify the use of a higher order polynomial, yet we note that simultaneously fitting for either a quadradic or cubic long-term trend does not significantly alter the results described below.

\begin{figure}
\includegraphics[angle=0,width=0.90\hsize]{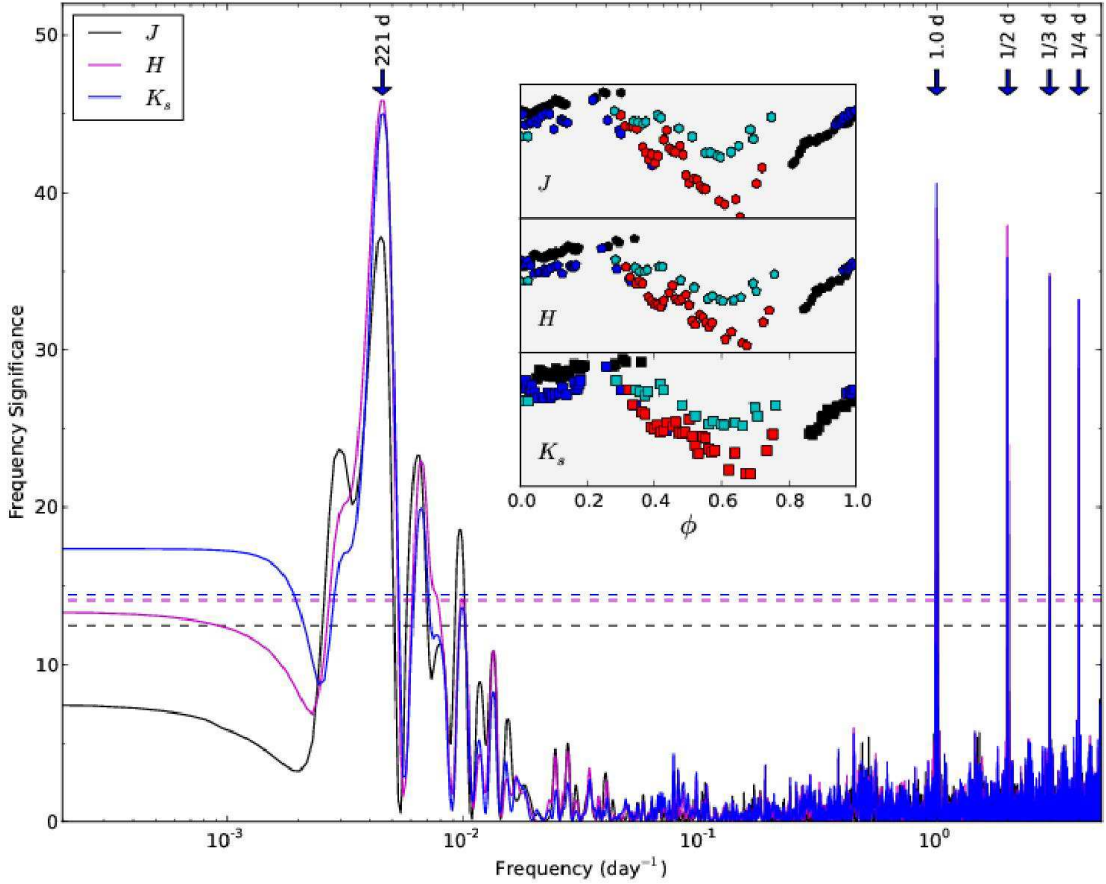}
\caption{Analysis of time scales inferred from the photometric time series.
Lomb-Scargle periodogram for each of the JHK$_s$ passbands.
A period of around 221 days, though with a broad frequency peak,
is apparent. It is well above the signal level of the calculated
0.1\% false alarm probability signal, shown as the horizontal dashed lines
and calculated using 10$^4$ random re-shuffled instances of the data.
Insets show the phased data, with the simultaneously fit linear trend removed 
for clarity of presentation.  The colors indicate sequential observing
seasons or other gaps in the data as follows:  
cyan = 2010, red = first part of 2011, blue = second part of 2011,
black = 2012 prior to June.
}
\label{fig:timescales}
\end{figure}

In Figure~~\ref{fig:timescales}
we show the periodogram for each of the three near-infrared bands, including
observations taken prior to 2012 June 16 only.  The periodogram exhibits 
significant power on long time scales.  In each of the J, H, and K$_s$ bands
we find a large peak in the periodogram corresponding to a period 
$P \sim 221$ days. The power at the 1 day parasite frequency, 
an artifact produced by the fact that we observe nightly, 
is strong, but weaker than the power inferred for the astrophysically
significant period at 221 days.  The peak is somewhat broad in frequency,
perhaps indicating only semi-periodic behavior from cycle to cycle.

To quantify the significance of the observed peak in the periodogram we randomly resample the PAIRITEL flux measurements while fixing the cadence to match our observations. Following this we repeat the Lomb-Scargle analysis described above and measure the peak value in the resulting periodogram, while excluding the 1, 1/2, 1/3, and 1/4 d overtones of the 1 d alias. 
This procedure is repeated 10,000 times, which allows us to determine a ``false alarm probability'' that the observed peak in the periodogram is the result of noise by examining the rank ordered values of the peak significance from the 10,000 trials. In Figure~\ref{fig:timescales} we show the 0.1\% false alarm probability, meaning that a peak in the periodogram greater than this value has $<$0.1\% chance of being the result of random fluctuations. In each filter the peak is well above this value, which clearly shows that the measured periodicity is significant. This can visually be confirmed based on the phase-folded trend-removed light curves shown in the inset of Figure~\ref{fig:timescales}. We interpret these results as tentative evidence that the emission from PTF~10nvg varies periodically on timescales of roughly 221 d; however, we caution that the signal may be driven by the particular peaks and minima that we do observe, specifically the two deep minima observed near JD 2455520 and 2455760. 
With higher cadence observations taken over longer time baselines we will be able to confirm if the periodic signal observed in the 2010-2012 data set is fundamental to the system, or a peculiarity of the observing window through which we have measured the source.

Summarizing our findings, we see evidence that the large-amplitude 
$JHK_s$ near-infrared variations are semi-periodic, 
with a roughly 220 day timescale. 
Local peaks in the $R_{\rm PTF}$ optical light curve 
on JD $\approx$ 2455679, 2455884, and 2456084 reinforce this time scale. 
We note, however, that during the 2010 season in particular,
the R$_{PTF}$ photometry does not phase well with the 
period derived from the near-infrared (JHK$_s$) photometry.
Further, beginning in the middle of June 2012, the $R_{\rm PTF}$ light curve 
clearly deviates from the previously observed semi-periodic behavior; 
rather than fading as expected based on the derived period,
PTF~10nvg instead reached its brightest optical state since 2010. 
PAIRITEL sampling during this timeframe is sparse due to 
the Arizona monsoon season, and corresponding telescope closure; 
nevertheless, the near-infrared observations confirm the deviation 
from the previously observed semi-periodic behavior. 
This suggests that while extinction clearly plays a prominent role
in the variability, other processes such as variable accretion (which would
affect optical photometry moreso than infrared photometry)
could also be important in driving the photometric evolution of this source.
The correlation of enhanced optical TiO/VO band emission 
(requiring heating and then cooling of molecular material) 
and permitted line emission (requiring accretion) 
with the bright state that we discuss in \S~\ref{sec:spec}, 
supports this explanation.  
Concerning the historical imaging of this source, while it was
not visible to the limits of the 2MASS survey, not once over the
past three years of our photometric monitoring has it been fainter
than K$_s \approx$ 11 mag.  Thus, although discussed above in the context
of possible variations in extinction of $A_V \approx$ 70 mag, the recently
bright K$_s$-band magnitude could instead support the importance of 
an outburst component, despite variable extinction 
clearly being part of the picture.

Should the periodic signal be confirmed, that in conjunction with 
the evidence that the photometric variations are largely due to changes 
in extinction along the line of sight (Figure~\ref{fig:color}) would 
naturally suggest that the recently observed long period
variability arises in a dusty structure 
characterized by a dynamical time $\tau_{\rm dyn} \sim$221 day. 
If this is a rotating disk around a 0.5-1 $M_\odot$ star, it would indicate 
higher opacity material in a clump or warp located
around 0.5-0.7 AU from the star.
The implied orbital velocity for a circular orbit (25 km/s) can be
compared to the characteristic time scale for the photometric 
rises and declines that would correspond to repeated clearing and 
obscuration events.  Refering to Figure~\ref{fig:lightcurve},
the relevant rise and fall times are a few tens of days.  This time scale
relative to the observed period is similar to the ratio between the
derived location and the orbital circumference (2$\pi$),
resulting in a size scale for the obscuring material that is thus similar
to its distance from the star, namely $\sim$0.5 AU.

\subsection{Mid-Infrared Variability}

The data acquisition strategy for the WISE mission included multiple scans 
of each point on the sky, which provides time series data on time scales
that vary with ecliptic latitude.  For the scans covering PTF~10nvg, 
the time series data are rather concentrated, with several observations 
obtained within just a few several-day-long time frames.
The observations took place around the time of the first peak and
before the absolute peak brightness exhibited in the optical light curve 
(Figure ~\ref{fig:lightcurve}).
PTF~10nvg is flagged in the WISE catalogs as a potential variable 
in the W3 and W4 bands based on 26 individual measurements.  Examination 
of the photometry from these individual scans reveals nothing systematic about 
the behavior in time between e.g. the 2010, May 27 and 28 WISE data 
vs the 2010, June 4 WISE data.  Scatter among the many observations 
taken so close in time that is larger than the quoted errors likely
leads to the variable source flag; it is not clear whether the source
is actually varying on sub-day time scales, however.  

Comparing the WISE photometry to previously reported measurements 
at similar wavelengths from Spitzer, MSX, and IRAS, 
illustrates that the source brightened in 2010 at mid-infrared 
wavelengths as well as at the near-infrared and optical wavelengths
highlighted in \citeauthor{Covey11} and discussed in detail above. 
Specifically, the Spitzer fluxes from 2006/2007 reported by 
Rebull et al. (2010) at both 3.6 $\mu$m 
and 4.5 $\mu$m 
were a factor of more than ten lower than those recorded by WISE in 2010
at 3.35 $\mu$m and 4.6 $\mu$m.
The WISE 22 $\mu$m measurment can be compared with data from
not only Spitzer but also from earlier MSX (21.3 $\mu$m) 
and IRAS (25 $\mu$m) measurements.  
While there are wavelength and beam size 
differences between these various missions to consider, it is clear that 
the 2010 WISE photometry is brighter by a factor of $\sim$4 compared to 
the 2006 Spitzer 24 $\mu$m data, with the earlier missions 
all reporting photometry at intermediate flux levels.

\section{Spectral Energy Distribution and Variability}\label{sec:sed}

\begin{figure}
\includegraphics[angle=0,width=0.80\hsize]{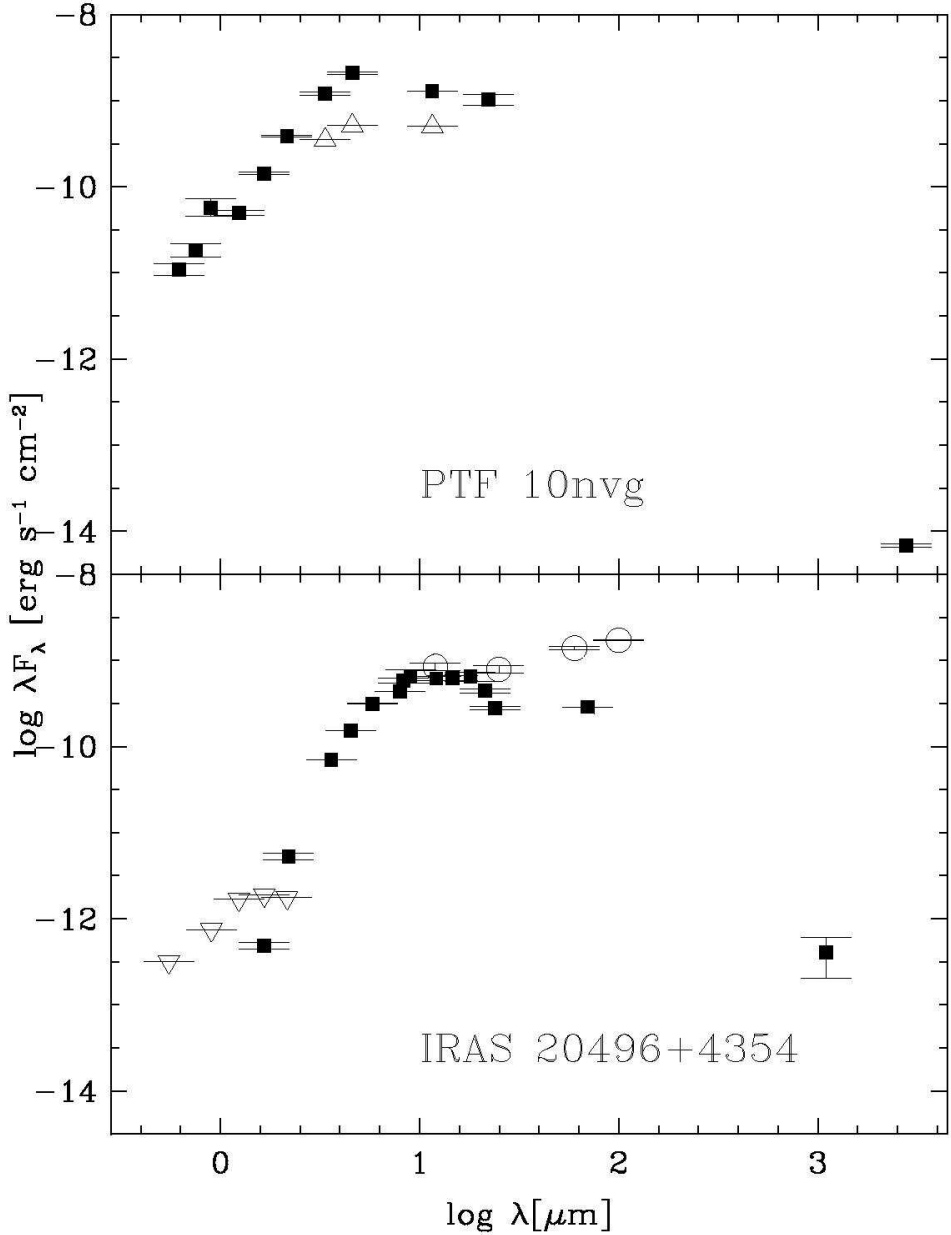}
\caption{Spectral energy distributions updated from those presented in
\citeauthor{Covey11} to include WISE and CARMA photometry. Bottom panel
shows literature data from 1980-2009 while upper panel shows outburst
photometry from 2010.  
Optical and near-infrared photometry is selected to be closest in time to
plotted mid-infrared and millimeter photometry.
At WISE wavelengths, both the saturated and
the profile-fit corrected magnitudes are shown.
Downward facing triangles are upper limits 
and upward facing triangles are lower limits
while squares and circles (large beam IRAS measurements in the left panel) 
are detections.  
As discussed in the text, the 2010 brightening of
the source had mid-infrared as well as near-infrared and optical manifestations.
}
\label{fig:sed}
\end{figure}

In Figure~\ref{fig:sed}, the outburst spectral energy distribution 
presented by \citeauthor{Covey11} is updated to include the photometry from 
WISE and CARMA.   

Using the CARMA continuum flux measurement of 5.6 mJy at 2.7 mm,
a total mass (dust $+$ gas assuming a ratio of 1:100) 
of 0.06 M$_\odot$ is calculated.  This assumes a distance of 520 pc, 
a dust opacity of 0.01~cm$^2$~g$^{-1}$ at 107~GHz (Beckwith et al. 1990),
that the continuum emission
is optically thin and isothermal with a temperature of 20~K,
and that free-free emission is a negligible contribution. 
This value is on the high end of circumstellar masses measured
for young stars and consistent with the Class I-type SED of the source.

The post-outburst CARMA measurements at 2.7~mm can be compared with the 
pre-outburst data from the Bolocam Galactic Plane Survey at 
1.1~mm (Rosolowsky et al. 2009) to search for variability at
millimeter wavelengths. The Bolocam survey reported a 2$\sigma$ detection at
the position of PTF 10nvg with an integrated flux of $148\pm74$~mJy, although
the integrated flux may be underestimated by 50\% (Rosolowsky et al. 2009). The
1.1~mm detection is consistent with a point-source at the 33$''$ angular
resolution of Bolocam. The higher resolution CARMA 2.7~mm continuum image
resolves the continuum emission into two compact sources and a diffuse extended
component (see Figure~\ref{fig:contoutflow}). The integrated flux of these
three components measured with aperture photometry is $11.3\pm1.4$~mJy, where 
a 10\% calibration uncertainty is assumed. While additional extended emission may
be ``resolved out'' with CARMA, we nonetheless derive a spectral millimeter
spectral slope ($F_\nu \propto \nu^\alpha$) of $\alpha=2.9\pm0.6$. 
Assuming the
millimeter continuum emission is optically thin, the expected slope is
$\alpha\approx4$ for interstellar medium grains, and $\alpha\approx3$ if 
grains have grown to approximately millimeter in size. Thus within the
considerable measurement uncertainty, the observed spectral slope is 
consistent with that observed in circumstellar disks and envelopes, and we
find no evidence for variation in the millimeter flux toward PTF 10nvg.

As demonstrated above, however, PTF~10nvg has
shown significant variability at mid-infrared wavelengths.
The $2-25 \mu$m spectral slope, originally defined for
ground-based K-band and IRAS data by \citet{Lada87}
as $\alpha$ = d log $\lambda F_\lambda$ / d log$\lambda$,
can be computed in the pre-outburst and post-outburst stages.
Using the data shown in Figure~\ref{fig:sed} and with the
caveat that there are not ideal matches in the timing between
available K-band and mid-infrared photometry, we find indeed
a dramatic change in the spectral slope of the source from
pre-outburst to post-outburst.  Regardless of whether IRAS 25
$\mu$m, MSX 21 $\mu$m, or Spitzer 24 $\mu$m photometry
(taken over a time span of three decades) is compared to the
2MASS 2 $\mu$m upper limits from 2000 or the UKIDSS
measurement from 2007, we find the pre-2010 spectral slope is indicative of
a ``Class I"-type spectral energy distribution: specifically, we
find $\alpha >$ 2 (in the range $>$2.1 to $>$2.5) when computed
with the 2MASS upper limit, and 1.6 $< \alpha <$ 2.0 when computed
from the UKIDSS measurement.

Computing an $\alpha$ value from the post-outburst PAIRITEL
K-band observations and the similarly post-outburst WISE
22 $\mu$m observations produces a much flatter spectral slope
of $\alpha \sim$0.3.  This $\alpha$ value lies at the low end of
the typically defined ``Class I" range, but the dramatic change
in spectral slope from the pre-2010 measurements indicates
that, while PTF10~nvg demonstrably brightened by factors of
several or more during and following the 2010 outburst, the
near-infrared and optical brightening significantly
exceeded that in the mid-infrared.

While the data in hand do establish that PTF~10nvg's
pre-outburst and post-outburst mid-infrared spectral slopes
were notably different, we have
little leverage for establishing the physical mechanism that drove the
observed change. We consider three mechanisms: 
1) direct extinction of the mid-infrared emitting region, as we invoke 
to explain much of the optical and near-infrared variability; 
2) shadowing of the outer disk by a vertical perturbation in the 
inner disk; 
3) an overall increase in the luminosity of the inner disk, due to internal 
or external heating related to an increase in the disk or stellar
accretion rate.  The current data does not clearly identify 
one of these possibilities as the favored interpretation.

\begin{itemize}
\item{ {\em Extinction}: 
As noted above,
PTF~10nvg's 2010 brightening included a four-fold increase in its $\sim$22 $\mu$m
flux. Interestingly, this change is equivalent to a difference of 1.5
mag in extinction at
this wavelength, and corresponds to a difference in A$_V \sim$ 30 mag according
to standard extinction laws. This is close to the maximum
extinction change inferred from variations in the near-infrared colors
(see Figures~\ref{fig:color} and~\ref{fig:av_vs_time}). Explaining the observed
mid-infrared flux variability as an extinction event does require, however, the
obscuration of a significantly larger portion of the inner disk than
is required to
explain the extinction experienced by the source's optical and near-infrared emission,
which presumably arises directly from the central protostar and the inner disk
wall.  To understand the spatial scale an obscuring clump would need to possess
to occult the full 22$\mu$m emitting region, we consulted a range of
disk models calculated as in Isella et al. (2012) spanning a range 
of flaring angles (h/R).
These models indicate that nearly all of the 2$\mu$m flux emits from 
the disk's innermost radii, $r < 0.2$ AU, with the bulk of it at
$r <$ 0.1 AU, regardless of geometry.  The 22 $\mu$m flux, by
contrast, does show a greater dependence on the disk's flaring angle;
for flat and moderately flared disks, the bulk of the 22 $\mu$m flux arises from
the disk's inner 0.5--0.7 AU region, within the radii implied for the location
of the obscuring clump based on the characteristic timescale of the extinction 
events detected at shorter wavelengths.  For disks with larger flaring
angles, as might be expected for a typical Class I source, 
the 22 $\mu$m emitting region extends over a larger radial range and
well beyond 1 AU, requiring a larger obscuring clump to explain the
observed mid-infrared variability.  }
\item{ {\em Shadowing events}: Models in which an increase in the
apparent scale height of the inner disk (due either to e.g. dynamical
changes in the disk structure, or the apparent position with
respect to the line of sight of a persistent orbiting perturbation)
decreases the illumination of the outer disk have recently been
developed to explain correlated near- and mid-infrared variability observed
from young disked stars 
(Muzerolle et al. 2009, Espaillat et al. 2011, Flaherty et al. 2012). These
models, by radiatively coupling the inner disk to more distant disk regions,
provide a viable mechanism for inducing changes in the 24 $\mu$m flux 
on timescales shorter than the local dynamical timescale.
However, these models typically have been invoked to explain changes
in 22$\mu$m flux at the 0.1-0.2 mag level, not the 1.5 mag level
observed from PTF~10nvg. In modelling mid-infrared  variability in T Tau S, 
van Boekel et al. (2010) note that detailed radiative transfer
models of this mechanism suggest that mid-infrared brightness changes larger
than 0.2 mag require remarkably large perturbations in the inner disk.
These models also predict a `pivot point' at 5-10 $\mu$m, where the
spectral changes on either side of the pivot are anti-correlated, due
to the increased near-infrared flux emitted by the inner disk as it shadows the
outer disk, and vice versa. The brightening exhibited by PTF~10nvg
provides no evidence for such a pivot point, but given the numerous
components which are likely contributing to PTF~10nvg's spectral
evolution, and the lack of simultaneous data over a broad wavelength
range, the existence of a $\sim$8$\mu$m pivot point cannot be ruled
out either.  }
\item{ {\em Accretion-related increase in disk luminosity}: If
PTF~10nvg's brightening is associated with a major accretion outburst,
the increase in external illumination from the stellar accretion shock
and/or a similar increase in viscous heating due to material flowing
through the circumstellar disk could lead to a significant increase in
the system's mid-infrared flux. This mechanism has provided the foundation
for increasingly detailed models of the spectral morphology of FU Ori
objects, over a range of wavelengths from the optical into the far-infrared
(Zhu et al. 2008), and is also consistent with observations of the
wavelength dependent timescales for the decay of the near- and mid-infrared
emission from V1057 Cyg, a post-outburst FU Ori object (Simon \& Joyce
1988).  Intriguingly, van Boekel et al. (2010) also demonstrate that
their models of accretion-driven disk emission predict near- and mid-infrared
color-mag changes which are broadly consistent with expectations from
standard extinction models: if true, this suggests that even
simultaneous broadband photometry spanning the near- to mid-infrared during a
distinct brightening and fading event may not suffice to establish the
true nature of PTF~10nvg's mid-infrared variations.}
\end{itemize}

\section{Spectroscopic Analysis}\label{sec:spec}

Our new optical and near-infrared spectra at low dispersion 
can be considered along with the spectroscopic data 
published by \citeauthor{Covey11} that sampled the source both before 
and after its absolute brightness maximum in 2010, August. 
The 2011 and 2012 spectra sample epochs in the lightcurve of PTF~10nvg 
when the source was near local maxima in brightness as well as when it was 
deep in extended faint states e.g. during summer/fall of 2011 
and fall/winter of 2012 (Figure~\ref{fig:lightcurve}). 
The high--dispersion spectra 
also sample the lightcurve at relatively bright as well as much fainter 
photometric states.  

\begin{figure}
\includegraphics[angle=0,width=0.90\hsize]{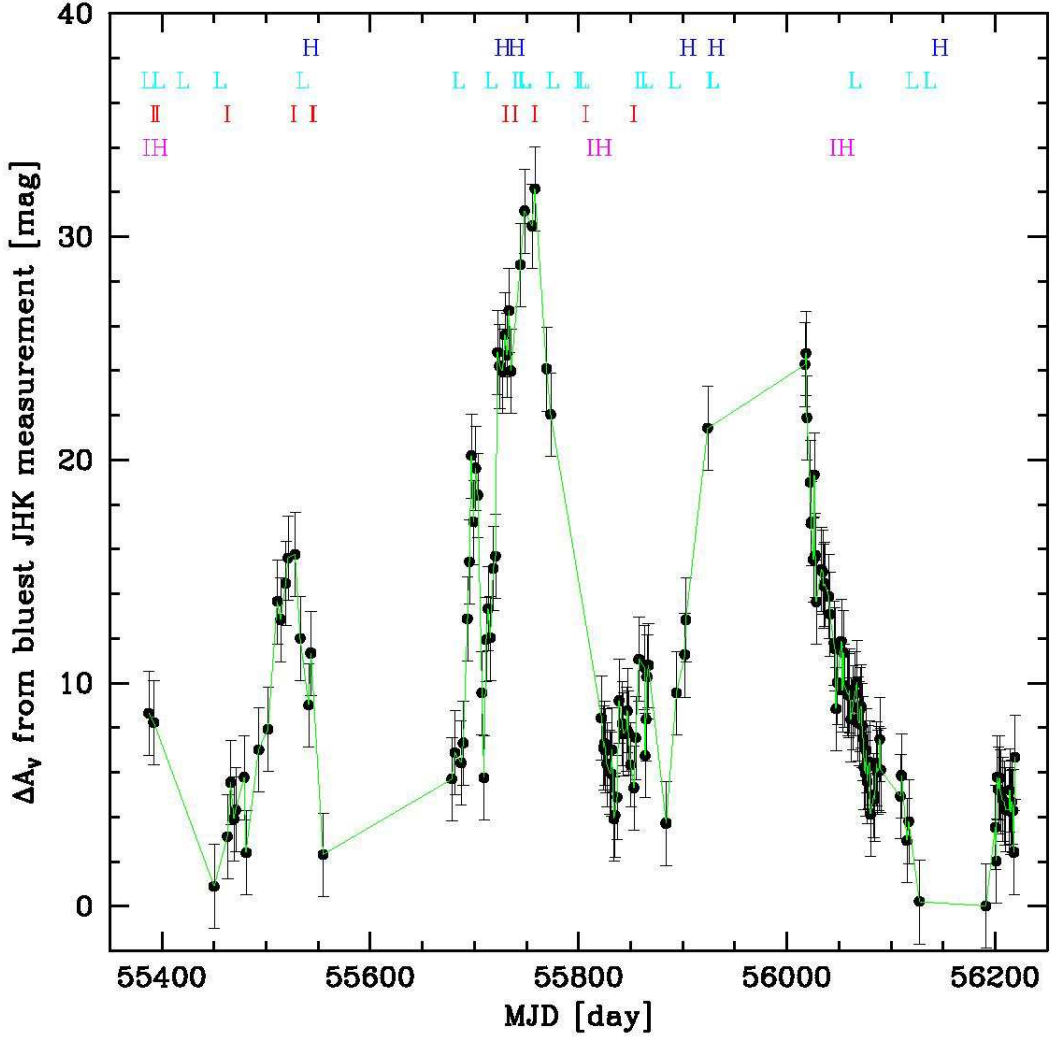}
\caption{Differential optical extinction, $\Delta A_V$, measured from 
the color changes in Figure~\ref{fig:color} relative to the bluest photometry captured 
in our near-infrared time series and an assumed extinction law.
Error bars are calculated assuming the typical photometric errors in 
Table~\ref{tab:PTEL}. 
The alphabetical labels indicate the timing of our spectroscopic follow-up 
of the 2010 brighting of PTF~10nvg, with ``H" representing
high--dispersion optical data, ``L" low dispersion optical data, 
``I" indicating low dispersion near-infrared data
and ``IH" high--dispersion near-infrared data, all as described in the text.
}
\label{fig:av_vs_time}
\end{figure}

Figure~\ref{fig:av_vs_time} illustrates the ensemble of
spectral epochs relative to the estimated optical extinction variation
over the time series. The latter comes from calculating
the color excess from the bluest observed near-infrared color
and applying the \citet{Indebetouw05} extinction law.  We note 
the high density of spectral coverage obtained during the faintest, reddest
state of the PTF~10nvg light curve in 2011, summer.
Given the sparse spectral sampling we do not consider
the evidence for spectroscopic periodicity. 
However, we do observe a seemingly repeating pattern in several spectroscopic
signatures. Specifically, 
the forbidden lines are most prominent during the high extinction epochs
represented in Figure~\ref{fig:av_vs_time}.  Further, spectra taken 
when PTF~10nvg was in one photometric cycle are very similar
to those taken in earlier photometric cycles at similar phase; for example,
repeatedly
when brightening, the forbidden lines are clearly present but quite weak 
relative to their dominance in the faint photometric states.

Our discussion of the spectral time series data is divided into 
presentation of the continuum behavior 
(illustrated for the optical in Figure~\ref{fig:optspec} 
and for the infrared in
Figures~\ref{fig:ircontinuum} and ~\ref{fig:irspec}) in \S 5.1, 
and then a description of the lines.
The line discussion is subdivided into wind-indicating
absorption lines in \S 5.2 and then accretion- and outflow-diagnostic
emission lines in \S 5.3. The various emission species are further
separated into molecular (\S 5.3.1), permitted atomic (\S 5.3.2), and 
forbidden atomic (\S 5.3.3) lines.  Evidence for spatially extended 
forbidden line emission is presented in \S 5.3.4.  
A wide range of thermal and mechanical processes are probed 
with the array of available spectral diagnostics.  
The molecular lines measure gas having temperatures up to
several thousand Kelvin.  The atomic lines diagnose
temperatures usually in the range $\sim$5000 -- 20,000 K (but up to 40,000 K), 
and densities $n_e \approx$ 10$^{4}$ -- 10$^9$ cm$^{-3}$ (forbidden lines)
and $n_H \approx$ 10$^{9}$ -- 10$^{12}$ cm$^{-3}$ (permitted lines).
Our main focus
is on describing the evolution in time of the line presence, line ratios,
and line profiles in the context of mass accretion and outflow. 

\subsection{Evolution of the Optical and Infrared Spectral Continuum}

\begin{figure}
\vskip-1truein
\includegraphics[angle=-90,width=0.72\hsize]{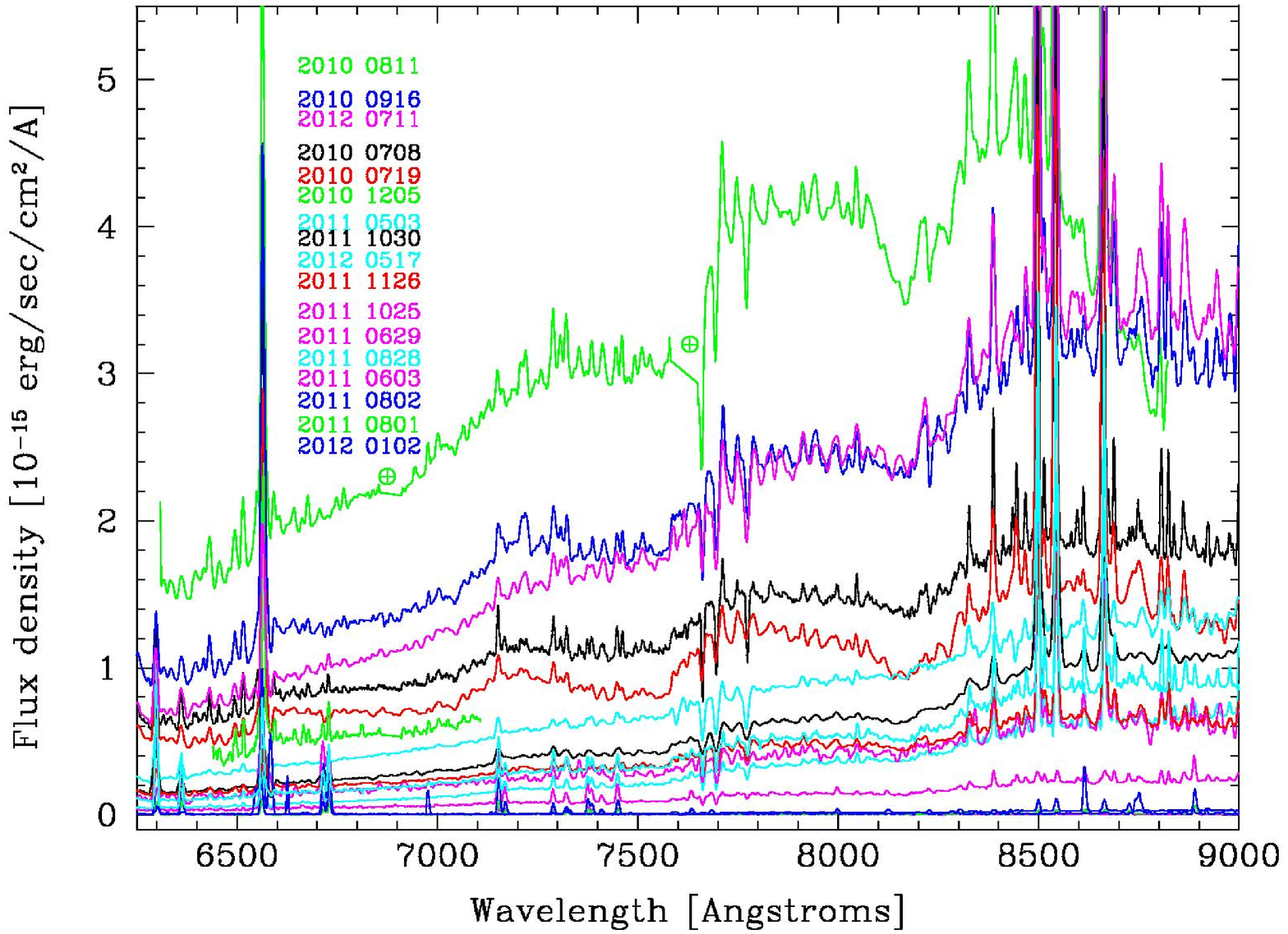}             
\includegraphics[angle=-90,width=0.72\hsize]{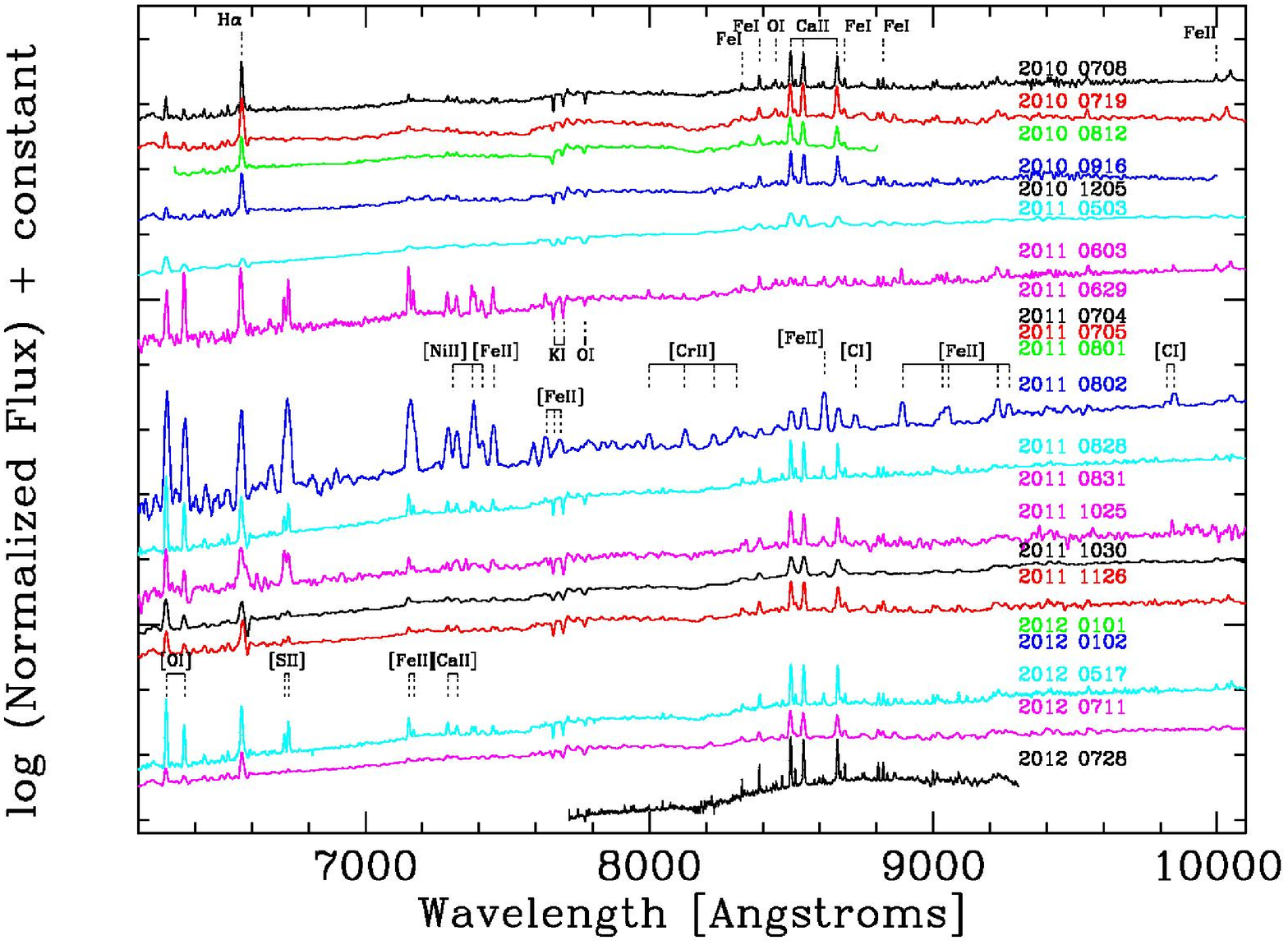}             
\caption{{\it Top panel:} Flux-calibrated optical spectra plotted
to highlight variable continuum brightness.
Data from 2011, July 4 and July 5, and from 2012, January 1 are excluded 
due to significant noise.  
The spectrum from 2010, August 11 is not corrected for telluric absorption
(the prominent atmospheric B and A bands have been smoothed over). 
The spectrum from 2010, August 12 is not shown but has 2.5 times the
flux of the previous night's spectrum, consistent with the lightcurve
in this period.
{\it Bottom panel:}
Normalized spectral time series, plotted on a log scale to
emphasize variable emission line strengths.  
Unmarked weaker emission lines are generally low excitation transitions 
of \ion{Fe}{1} or \ion{Fe}{2}.
Note the slightly wider spectral range than in the top panel, 
so as to feature additional forbidden lines.
Spectra are ordered by date; some dates have
no accompanying spectra plotted due to significant continuum noise
or, in the case of 2010, August 12, redundancy with an adjacent spectrum.
During brighter photometric states the H$\alpha$ and \ion{Ca}{2} lines are prominent,
while in fainter states, the forbidden emission lines are more prominent, 
as also shown in Figures~\ref{fig:optlines3} and ~\ref{fig:optlines4}.
}
\label{fig:optspec}
\end{figure}

\begin{figure}
\includegraphics[angle=90,width=1.00\hsize]{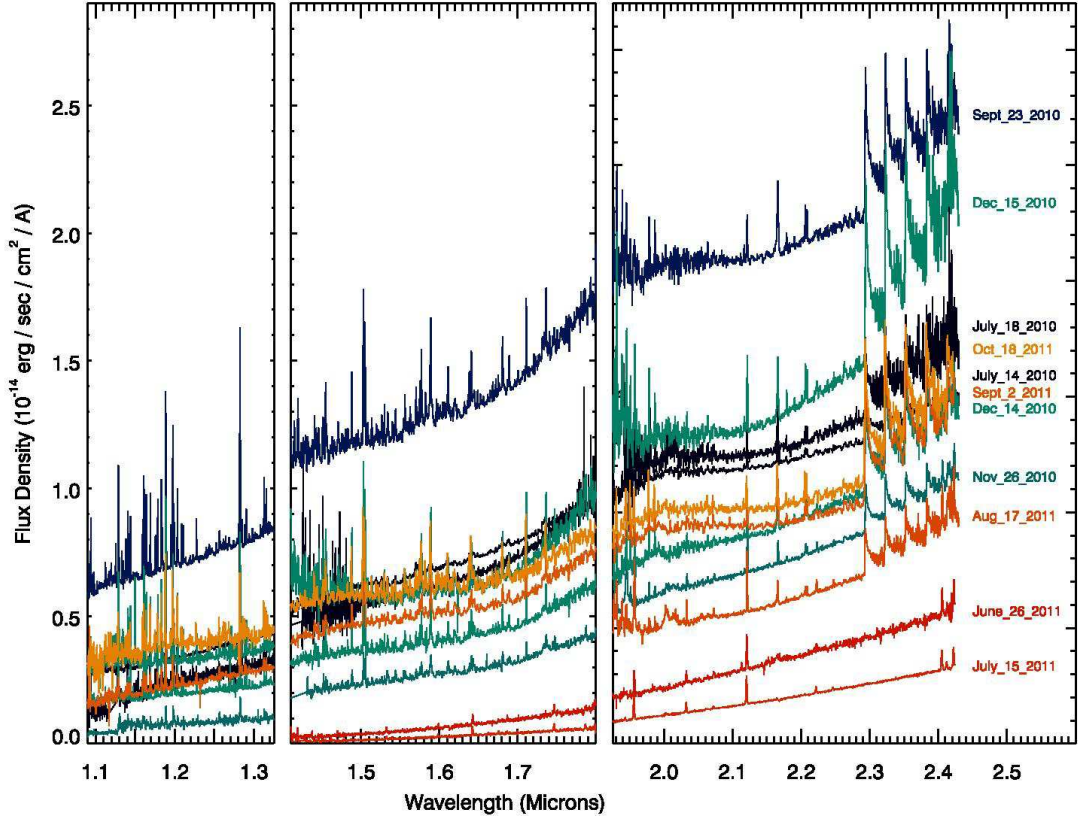}
\caption{
Flux-calibrated 
near-infrared spectra of PTF~10nvg.  
While the absolute flux densities observed in the time series
vary by a factor of $\sim$10, once the spectra have been corrected
with a standard R$_V$ = 3.1 extinction law for the photometrically
derived reddening estimates shown in Figure~\ref{fig:av_vs_time}, 
the spread in flux reduces to a factor of 2 -- within expectations
given the non-simultaneity of the
spectra and the photometric extinction estimates.  Spectra
obtained on 2011, June 26 and July 15 when PTF~10nvg was in a
particularly faint state, lack the numerous atomic
(e.g., H, Na, Ca, Mg, etc.) and molecular (e.g., CO, H$_2$O) emission
features that dominate the brighter spectra.  The few emission
features that remain visible in the faint state 
(e.g., H$_2$, [\ion{Fe}{2}]) likely arise in the jet/outflow.
}
\label{fig:ircontinuum}
\end{figure}

The large amplitude photometric changes exhibited by PTF~10nvg
are mirrored by large changes
in the spectral continuum. This is illustrated in
Figure~\ref{fig:optspec} showing the time series of optical spectroscopy
within the R and I-band regions and in 
Figure~\ref{fig:ircontinuum} showing the infrared spectra in the
Y, J, H, and K-band atmospheric windows.

Flux-calibrated optical spectra (top panel in Figure~\ref{fig:optspec}) 
exhibit spectral continuum slope changes that accompany 
the photometric brightness changes, as discussed in \citeauthor{Covey11}.  
Notably, the TiO/VO emission features (which are broad in wavelength 
and thus act like a continuum) have varied in strength relative to the 
nearby out-of-band continuum regions. 
Normalized optical spectra (bottom panel of Figure~\ref{fig:optspec}) 
illustrate line-to-continuum changes. 
During epochs when the source was fainter the signal-to-noise is lower,
but this does not preclude the detection of numerous prominent 
forbidden lines that are not seen in the brighter
photometric states when the continuum emission is stronger.  
Conversely, when the source is bright, permitted lines are more prominent than forbidden lines.
We discuss these and other spectral line changes in more detail in \S~\ref{sec:spec}.3.

\begin{figure}
\includegraphics[angle=90,width=1.0\hsize]{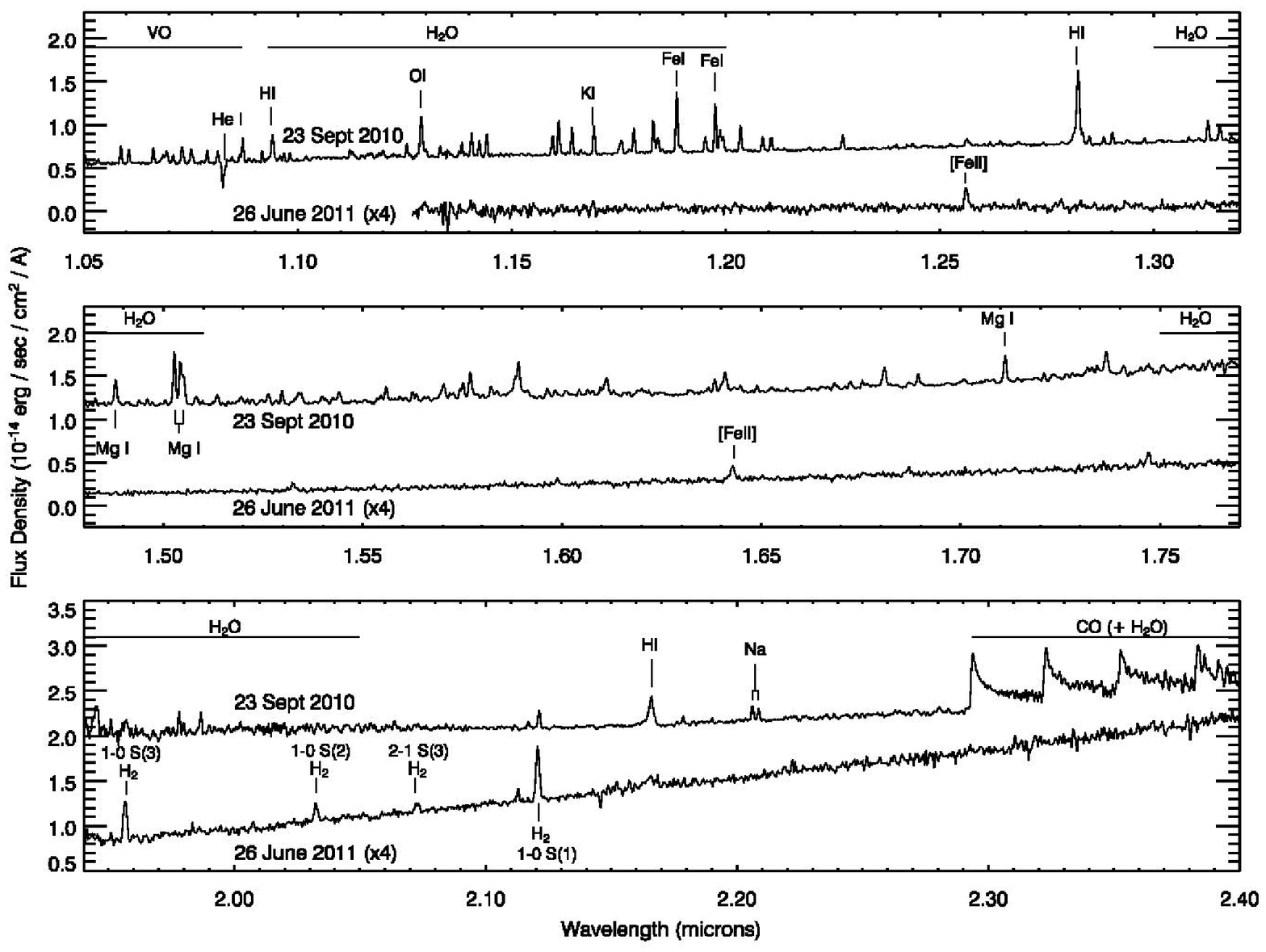}
\caption{
Comparison of PTF~10nvg near-infrared spectra during the brightest 
(2010 September 23) and second faintest (2011 June 26, adopted
because it is morphologically similar to the faintest 2011 July 15
spectrum but extends to bluer wavelengths) epochs 
captured in our moderate resolution near-infrared spectral
monitoring.  Numerous permitted atomic lines 
(e.g. \ion{Fe}{1} cluster at $\sim$1.15-1.2 $\mu$m) 
are prominent in the brighter spectrum
but absent in the fainter spectrum, which exhibits primarily 
H$_2$ and [\ion{Fe}{2}]. 
}
\label{fig:irspec}
\end{figure}

The infrared spectral continuum also exhibited differential brightening and
fading.  Along with the time series shown in Figure~\ref{fig:ircontinuum},
we highlight the near-infrared spectra from
the brightest and the (nearly) faintest epochs in Figure~\ref{fig:irspec}.
Variations in PTF~10nvg's near-infrared spectral continuum 
appear consistent with the extinction variations discussed
above based on the near-infrared photometric monitoring data, 
and illustrated in 
Figures~\ref{fig:color} and ~\ref{fig:av_vs_time}.  To verify this
assertion, we used an R$_V$=3.1 \citet{Fitzpatrick99} extinction law 
\footnote{
The \citet{Fitzpatrick99} reddening presecription
(employed here to the spectra because it is smoothly varying with wavelength) 
and the \citet{Indebetouw05} extinction (used earlier with the 
photometry since it is determined empirically using the same filters) 
are the same to within 5\% in the near-infrared; 
this difference is roughly the same as the estimated error 
in the flux calibration of our near-infrared spectra. 
}
to de-redden each moderate resolution near-infrared spectrum according to
the photometrically-derived A$_V$ estimates in Figure~\ref{fig:av_vs_time}, as
interpolated onto each spectroscopic epoch.  While the flux densities
of the observed spectra differ by more than an order of magnitude, the
flux densities of the de-reddened spectra agree to within a factor of
$\sim$2.  Similarly, the spectral slopes of the de-reddened spectra
also agree well generally, 
with the exceptions of the spectra acquired
on 2010 November 26, 2011 June 26, and 2011 July 15, when the source was
deep in (two different) photometric minima.
Considering the uncertainties imposed by the non-simultaneity of the
spectra and the photometric data from which the extinction estimates
are derived, this level of
agreement suggests that extinction variations account for much of the
near-infrared photometric and spectroscopic continuum variability within our time-series data.

\begin{figure}
\includegraphics[angle=0,width=0.90\hsize]{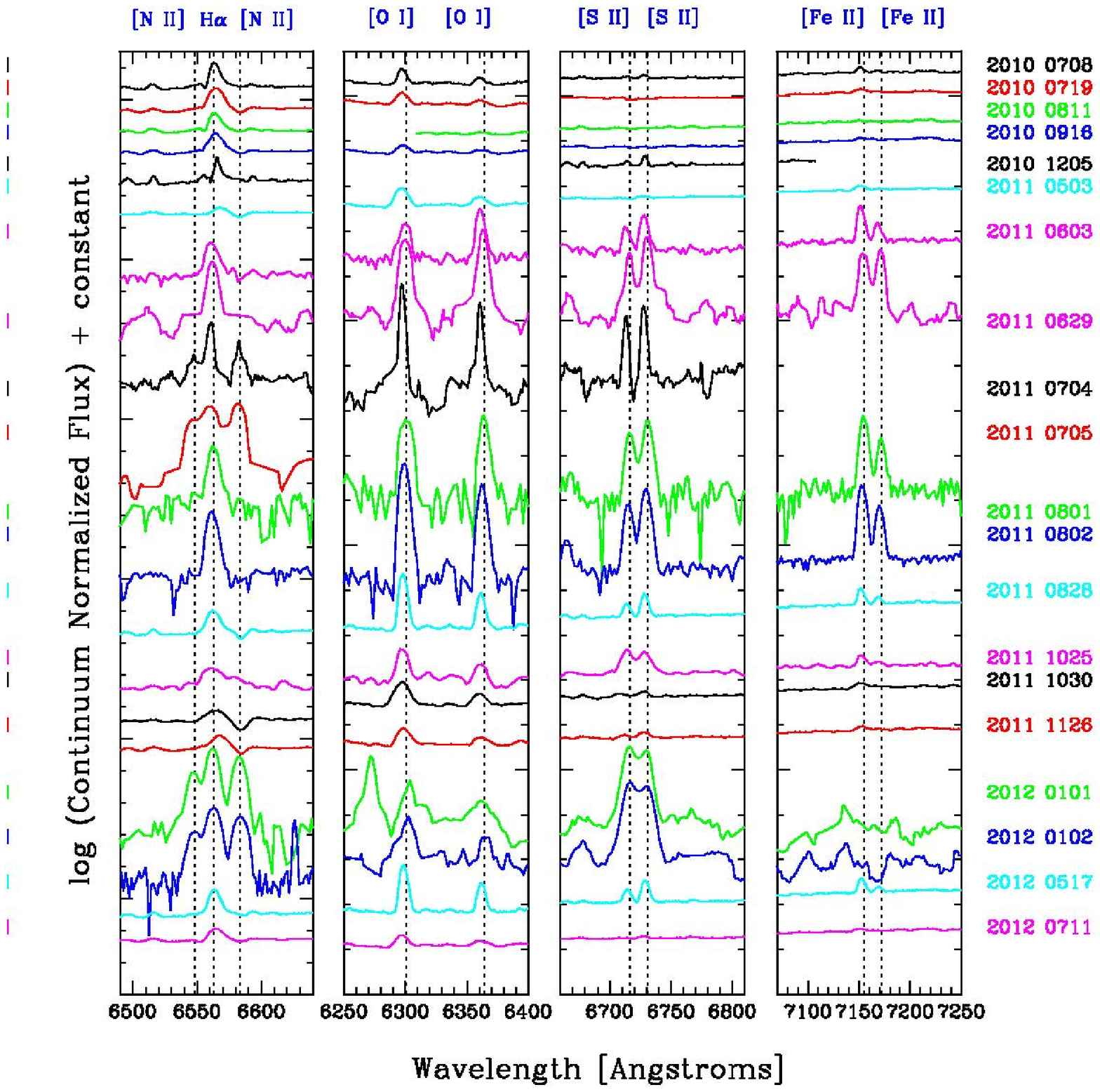}
\caption{Time series optical spectra in the same color scheme and order as 
the bottom panel of Figure~\ref{fig:optspec}, showing H$\alpha$ as well as
the [\ion{N}{2}], [\ion{O}{1}], [\ion{S}{2}], and [\ion{Fe}{2}] forbidden line doublets.
Lower signal-to-noise spectra have been smoothed.
Given the large variations in line-to-continuum ratio over the
time series, the ordinate is on a log scale.  Note the variation in the
doublet line ratios and the slightly blueshifted line centers which are
apparent even at low resolution.}
\label{fig:optlines3}
\end{figure}

\begin{figure}
\includegraphics[angle=0,width=0.90\hsize]{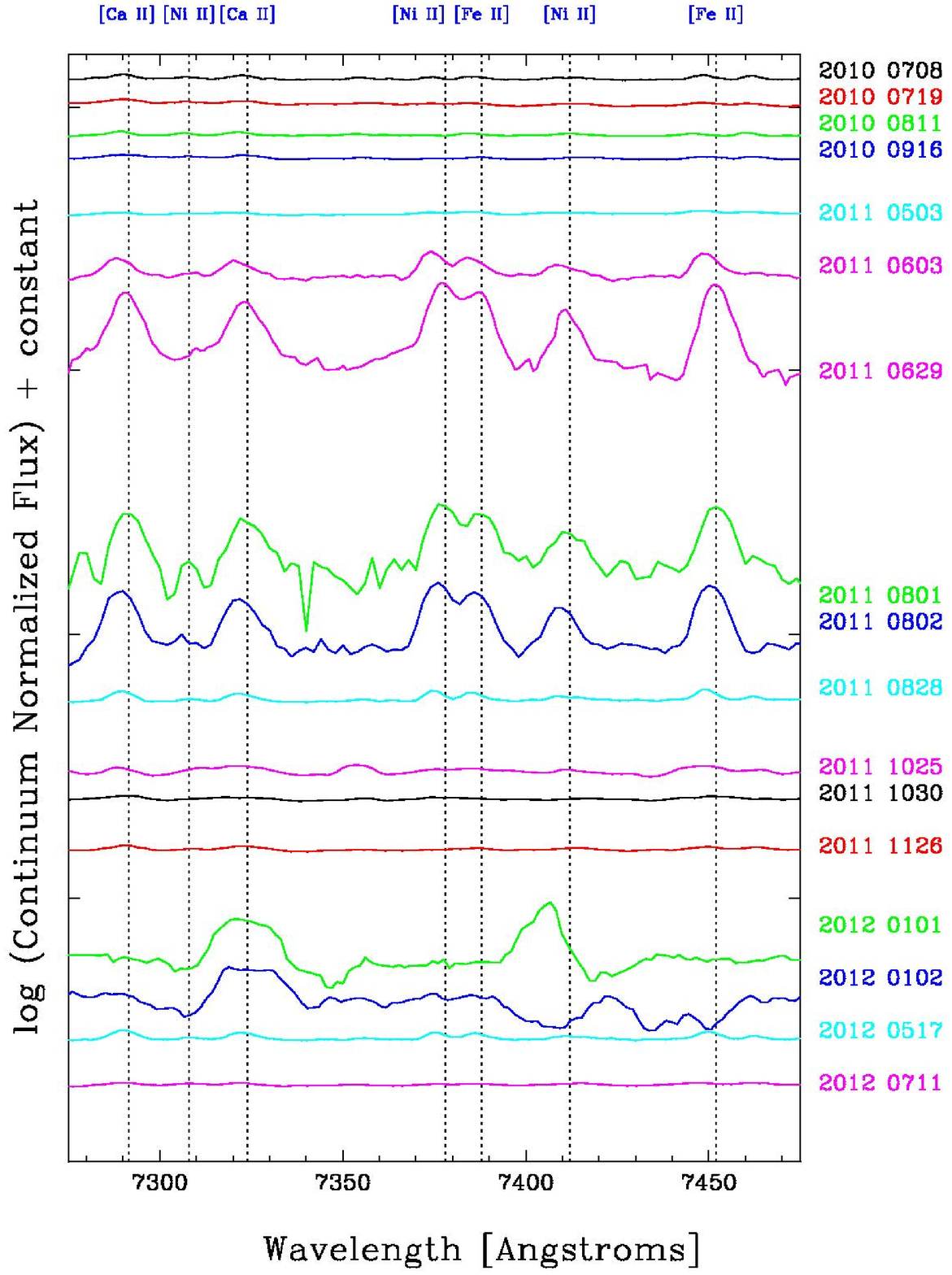}
\caption{
Time series spectra in the 7275-7474 \AA\ region showing
the [\ion{Ca}{2}] doublet as well as [\ion{Ni}{2}], and [\ion{Fe}{2}] lines
that are prominent during faint photometric states.
Lower signal-to-noise spectra have been smoothed, and neither of
the 2011 July 4 or 5 spectra are shown.}
\label{fig:optlines4}
\end{figure}

\subsection{Evolution of the Wind-Indicative Absorption Lines}

Our spectra contain a number of lines sensitive to outflowing material.
The early optical spectrum of PTF~10nvg shown by \citeauthor{Covey11} 
exhibited only the following lines in absorption
\ion{Na}{1} D, \ion{K}{1} $\lambda$ 7665, 7699, and 
\ion{O}{1} $\lambda$ 7774, all with
blueshifted velocities relative to the expected line center wavelengths.
These are either resonance lines or lines having a metastable lower level,
so they readily go into absorption.  Absorption was seen against the 
broad TiO/VO molecular emission continuum mentioned above, and 
there were no emission components to the profiles.
As the source varied photometrically over the 2010, 2011, and 2012 seasons, 
the blueshifted absorption remained relatively constant with respect to 
a normalized continuum, i.e. in equivalent width; values hovered 
around 2--3 \AA\ over most of this time period 
(though possibly somewhat higher, 3--4 \AA, towards the end of 2011).  
Between 2011 June 3 and 29, these absorptions  
disappeared as the continuum faded; this could be 
as they became undetectable in the lower signal-to-noise ratios 
characteristic of these spectra, or perhaps intrinsically if the
wind strength decreased.  By 2011, late August, when the source 
was again near a peak in brightness, line absorption 
from winds was again apparent at the same equivalent width level
and is seen in all subsequent ``bright state" spectra. 
The lack of change in absorption equivalent width is consistent with 
the extinction variation interpretation of the continuum behavior,  implying that the
large extinction variations arise exterior to the wind absorption region.

\begin{figure}
\includegraphics[angle=0,width=0.90\hsize]{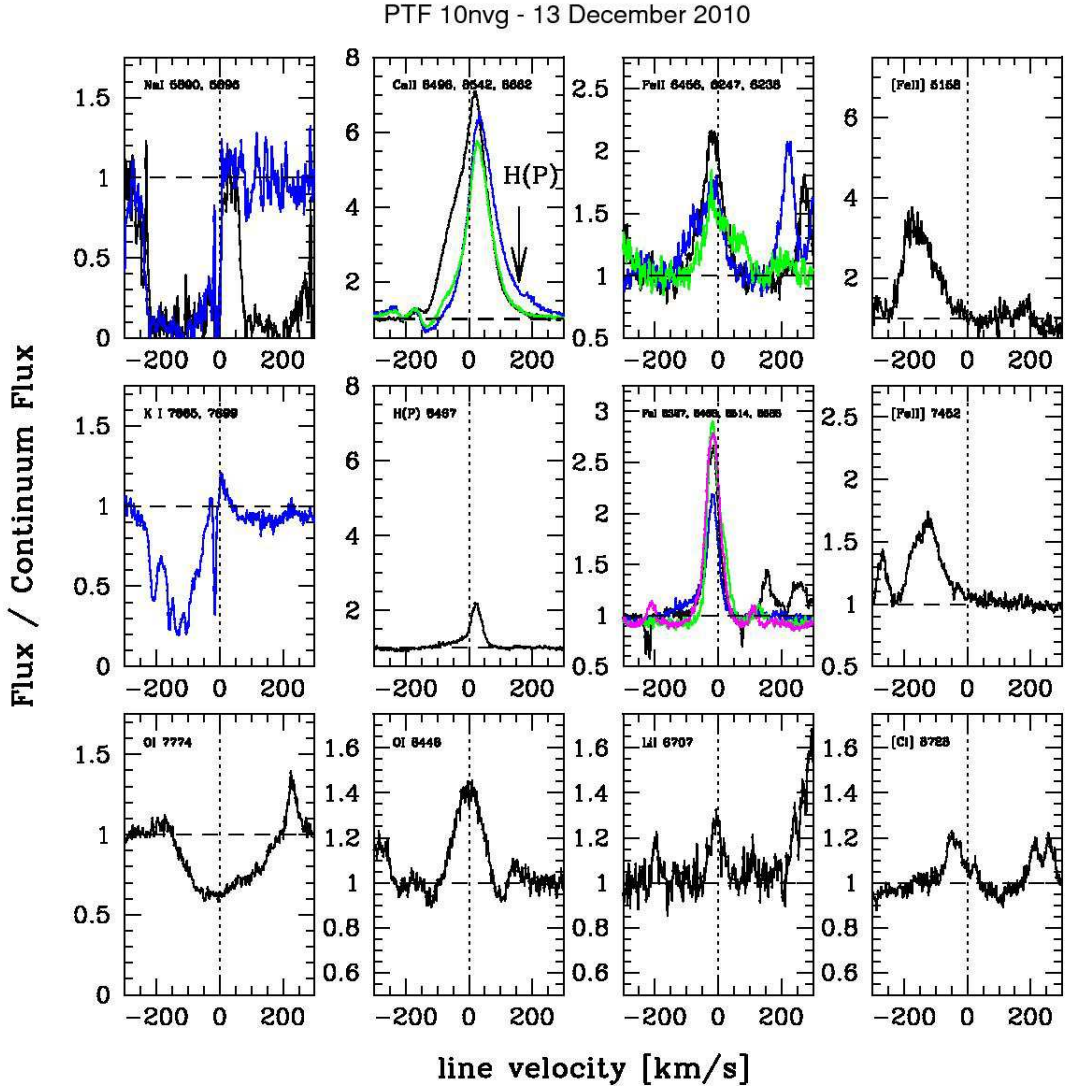}
\caption{Optical line profiles from 2010, December with vertical dotted lines
indicating zero velocity and the abscissa covering $-$300 to $+$300 km/s.
In each panel the black profile represents the line indicated first
in the panel legend; blue/green/magenta profiles, if present, indicate 
the second/third/fourth line in the panel legend 
(for example, in the \ion{Ca}{2} panel the black, blue, and green
profiles are those of the $\lambda\lambda$8498, 8542, and 8662 transitions).
The \ion{Ca}{2} triplet lines are broader than other atomic emission lines and
likely have both accretion-related and outflow-related components; note that
the red sides of these lines are each contaminated by \ion{H}{1} 
Paschen lines, with the panel immediately below the \ion{Ca}{2} 
panel showing another Paschen line on the same scale.
Highly and moderately blueshifted emission is apparent in forbidden [\ion{Fe}{2}]
and [\ion{C}{1}], respectively (see also Figure~\ref{fig:profforbidden}). 
In contrast, permitted \ion{Fe}{1}, \ion{Fe}{2}, and \ion{O}{1} emission 
is centered at low blueshifts or nearly zero velocity with a range in profile widths
including rather narrow profiles.  Blueshifted absorption is seen
in the wind-sensitive absorption lines of \ion{K}{1} and \ion{Na}{1}, with
the blue sides of the \ion{Ca}{2} triplet lines also exhibiting some
(sub-continuum) absorption.
The  \ion{O}{1} $\lambda$7774 triplet is seen in absorption as well.
Unusually, \ion{Li}{1} $\lambda$6707 appears to be in emission (see text).
} 
\label{fig:hires}
\end{figure}

\begin{figure}
\vskip-1truein
\includegraphics[angle=0,width=0.90\hsize]{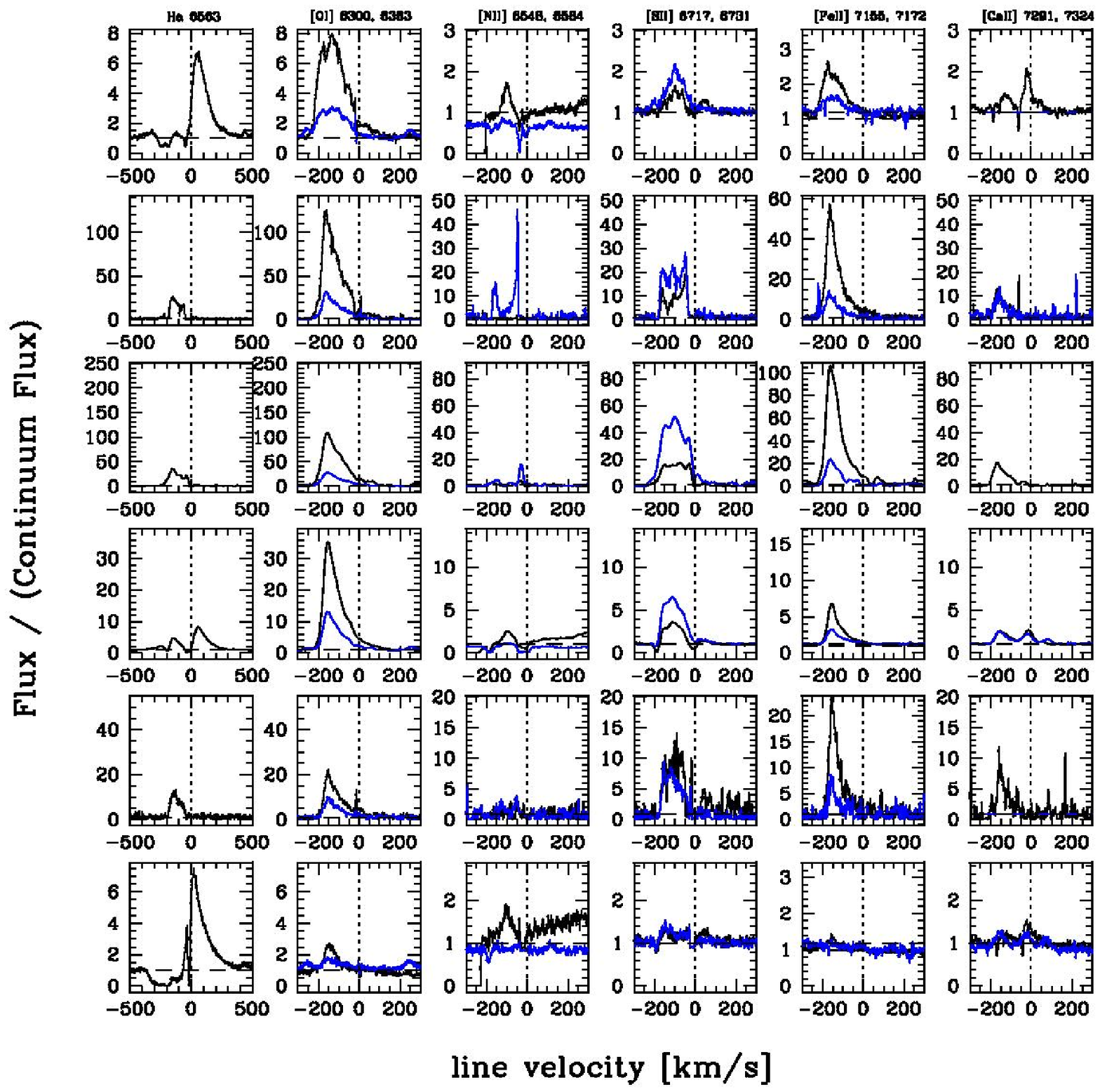}
\caption{Comparison of normalized optical H$\alpha$ and forbidden line profiles 
as identified above the top row.  
The velocity range along the abscissa is $-/+$500 km/s in the H$\alpha$ panels
and $-/+$300 km/s for all other lines.  Rows from top to bottom show data taken on 
2010, December 13 (ordinate axis scaled by 1.05), 
2011 June 15 (scale 18), 2011 June 28 (scale 32), 
2011 December 9 (scale 5), 2012 January 6 (scale 7), and 2012 August 7 (scale 1).
The ordinate range changes for the different dates (rows) by the scale factors
given above, set according to the strength of the strongest emission line. 
However, the relative scaling among panels along each row is the same, so that 
comparison of line ratios among dates (rows) is meaningful;
for example, [\ion{O}{1}] to [\ion{Fe}{2}] is high
when the source is bright in 2010 December and 2011 December, but
low in 2011 June and 2012 January when the source is faint. 
Black and blue lines indicate, respectively, the first and second wavelengths
as labelled at top.
Balmer line emission is visible over the same range of blueshifted velocities 
as the forbidden lines during faint photometric states, but there is an 
additional redshifted emission component during bright photometric states.
Variation is seen in the ratio of the [\ion{O}{1}] 6300 and 6363 lines 
and the [\ion{S}{2}] 6717 and 6731 lines;
for reference, the ratios are roughly as expected for optically thin emission 
in the top row.
} 
\label{fig:profforbidden}
\end{figure}

At high--dispersion, the bright state data (e.g. in Figure ~\ref{fig:hires}) 
show that the \ion{Na}{1} D doublet absorption is saturated at line center, 
with a boxcar like profile at zero flux extending blueward to $\sim$ -275 km/s. 
The \ion{K}{1} $\lambda$ 7665, 7699 doublet also has a broad blueshifted absorption
profile, extending from approximately $-50$ to $-250$ km/s.  Its depth is approximately 
40\% of the continuum level, shallower than the saturated \ion{Na}{1} D doublet absorption. 
Both lines have, in addition to their broad blueshifted absorptions, 
a very narrow very slightly blueshifted absorption component\footnote
{While this might be attributed to an interstellar contribution, the 
appearance of the low velocity narrow \ion{K}{1} and narrow \ion{Na}{1} 
is similar to a non-interstellar low velocity narrow component in 
the \ion{He}{1} $\lambda$ 10830 profile discussed below. In the bright state there is also 
a narrow absorption at this velocity in H$\alpha$ profiles
(see Figure~\ref{fig:profforbidden}).}, 
The \ion{O}{1} $\lambda$ 7774 triplet absorption profile is hard to
decipher due to the component blending, but it appears approximately neutral in
velocity and extends to appoximately 60\% of continuum depth; approximately
130 km/s of the line width is due to the triplet blending.  
We note that it is claimed in \citet{Aspin11} that 
the \ion{O}{1} $\lambda$ 8446 line also exhibited a blueshifted 
absorption profile in 2010, September, but we see only emission from our 
low resolution data taken in August and September (Figure ~\ref{fig:optspec})
and in our high resolution data taken in December
(Figure ~\ref{fig:hires}) of that year.

The time evolution of the optical wind-indicative absorption lines can not be
well-documented given that source faintness precluded the detection 
of continuum and thus any absorption against it in many of our high dispersion 
spectra.  In the HIRES data from 2011, December, however (taken after the
source had undergone several episodes of photometric fading and brightening), 
the \ion{K}{1} $\lambda$ 7665, 7699 doublet and the \ion{Na}{1} D doublet 
absorption profiles appear to have approximately the same form
and terminal velocities as in the 2010, December bright state spectrum.  
Another bright-state HIRES spectrum from 2012, August shows roughly similar 
\ion{K}{1} and \ion{Na}{1} features, but even higher velocity blueshifted
absorption in the \ion{O}{1} $\lambda$ 7774 line -- out to $-$250 km/s compared to
only $-$150 in Figure~\ref{fig:hires}.  This broadening of blueshifted absorption
by mid-2012 is also apparent at H$\alpha$; see below.

\begin{figure}
\includegraphics[angle=90,width=1.00\hsize]{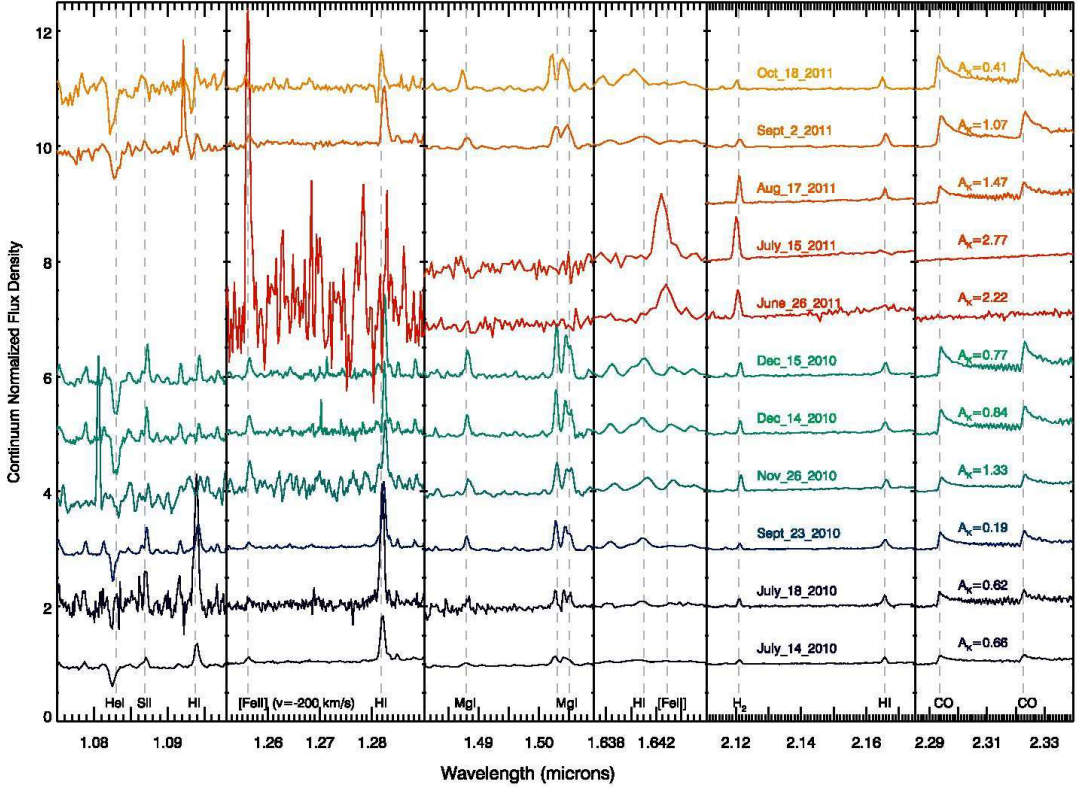}
\caption{
Time series near-infrared spectra, isolating representative permitted and forbidden
transitions. Extinction estimates based on the photometry are indicated 
in the right panel.  
Emission lines that are thought to trace hot, dense circumstellar material
(i.e., \ion{H}{1}, CO, \ion{Mg}{1}) are weaker in higher extinction states
whereas lines thought to trace shocks (H$_2$) or photo-excitation ([\ion{Fe}{2}]) in
lower-density circumstellar material are stronger.  
Absolute line flux measurements reveal that the lower-density tracers 
maintained a near-constant brightness, suggesting that they are formed
exterior to the extincted region, while the lines tracing hotter, denser
material faded along with PTF~10nvg's continuum, suggesting that they
are formed near the extincted photosphere or inner disk regions.  
}
\label{fig:irspeclines}
\end{figure}

\citet{Edwards06} have discussed the utility of the  
\ion{He}{1} $\lambda$ 10830 line (also with a metastable lower level)
as a wind indicator, demonstrating that its
sub-continuum absorption over a range of velocities probes both disk winds
and stellar/polar winds in young stars.  \citet{KF11} derived the
physical conditions traced by \ion{He}{1} (density $n_H \sim 10^9$  cm$^{-3}$, 
temperature $T \sim$ 10,000 K, excitation through UV photoionization) 
as well as other accretion/outflow tracing lines.  \citet{KR12} demonstrated in 
``observation" of MHD accretion/outflow radiative transfer
models the sensitivity of this line to the high energy radiation field,
namely $L_X$ and $T_X$, and were able to produce line profiles having 
redshifted and blueshifted absorption and emission features consistent with
observations of young accreting stars. 
The early infrared spectrum of PTF~10nvg shown in \citeauthor{Covey11} 
exhibited only \ion{He}{1} $\lambda$ 10830 in absorption.  The line 
was blueshifted relative to line center.  In the left panel of
Figure ~\ref{fig:irspeclines} we show the time series behavior of this line 
at low dispersion.  While the range
of resolutions of the spectra in Figure~\ref{fig:irspeclines}
do influence the details of the observed profile somewhat, 
the He I line appears to strengthen systematically 
over time, independent of the photometric state of PTF~10nvg.
The line increases from $\sim$3 \AA\ in equivalent width and a
maximum depth of 60\% of the continuum in 2010 July to $\sim$8 \AA\
equivalent width and a maximum depth of $\sim$30\% of the continuum in
2011 October.

\begin{figure}
\includegraphics[angle=0,width=0.90\hsize]{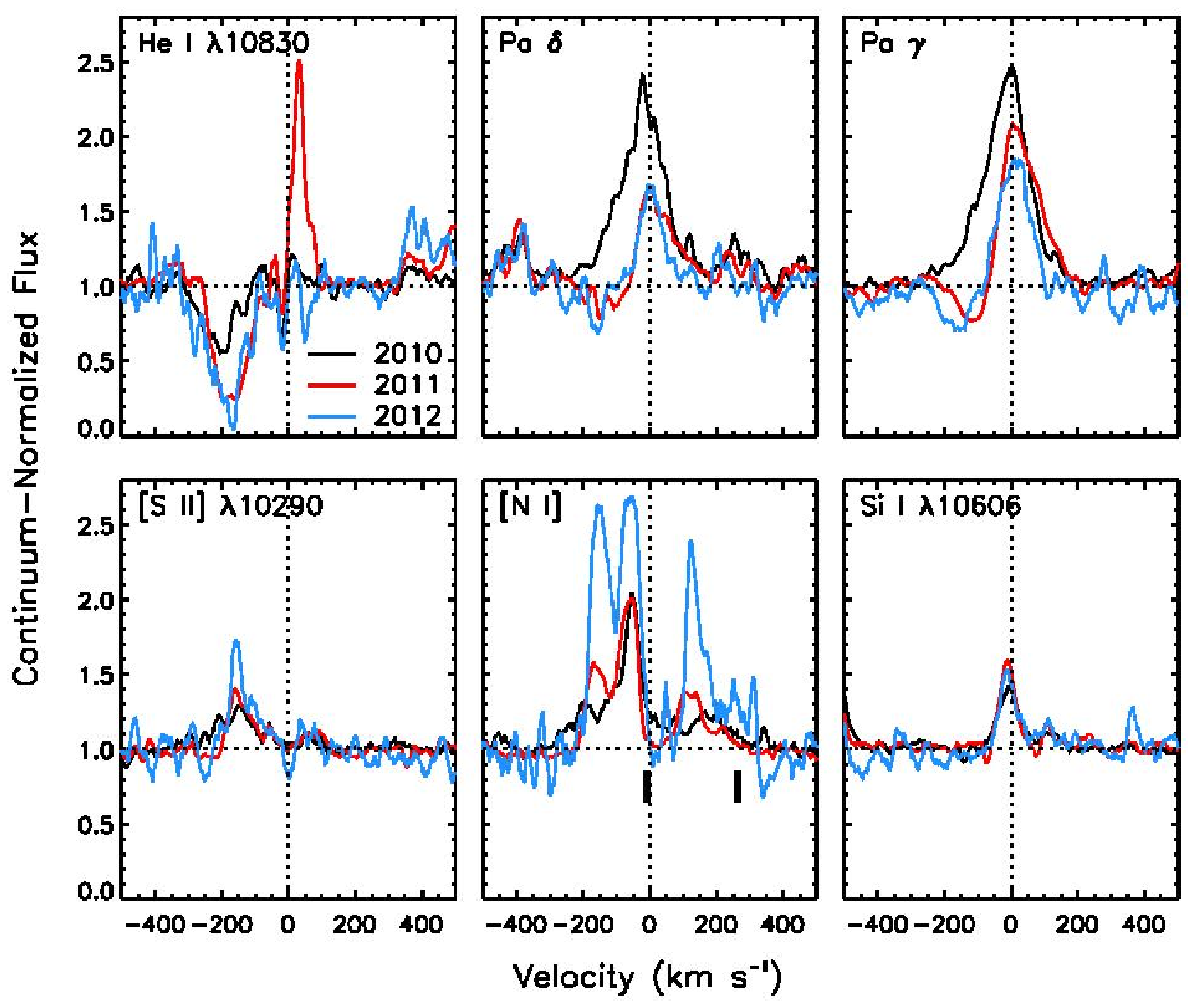}
\caption{Line profiles in the 1 $\mu$m region
as they appeared in 2010 July (black), 2011 September (red), 
and 2012 May (blue).  Top row shows a \ion{He}{1} and two \ion{H}{1} lines. 
Bottom row shows metallic forbidden lines (two [\ion{N}{1}] doublets as marked near
1.0401 $\mu$m and a [\ion{S}{2}] line at 1.029 $\mu$m)
and a permitted line (\ion{Si}{1} 1.061 $\mu$m). As in
the high dispersion optical spectra, the permitted emission is centered at
zero velocity while the forbidden emission is blueshifted
along with the wind-indicative \ion{He}{1} 1.083 $\mu$m absorption.
Variability over the time series is apparent, with the forbidden lines
stronger relative to the continuum
during epochs of higher extinction and the \ion{H}{1} lines
strong at lower extinction; however, little variability is apparent in the
permitted metallic lines.
}
\label{fig:1umprofiles} 
\end{figure}

These dramatic changes are also evident in the high--dispersion
\ion{He}{1} $\lambda$ 10830 profiles shown in Figure~\ref{fig:1umprofiles}. 
The NIRSPEC spectra were obtained in 2010, July,
when PTF~10nvg was near its absolute brightness maximum 
(which occurred in 2010, August), in 2011, September, when
the source was in its prolonged optical minimum state, and in 2012, May
as the source was rising towards a photometric peak. The change
exhibited is consistent with the trend observed in the
lower-resolution spectra. The faint-state 2011 spectrum 
is close to a classical P Cygni profile, with some sub-structure 
in each of the redshifted emission and the blueshifted
absorption components, along with a narrow, slightly blueshifted 
absorption. This low velocity narrow absorption component to 
the \ion{He}{1} $\lambda$ 10830 profile is present in all three spectra 
and may be the same as that seen in the \ion{K}{1} and \ion{Na}{1} profiles. 
In both brighter state spectra, the redshifted emission component  
at about $+$50 km/s that was present in the faint state is gone. 
However, the two brighter state spectra are different in their
broad blueshifted absorption components.
Relative to the 2010 bright state spectrum, the 2011 faint state
and 2012 bright state blueshifted absorptions both 
are deeper (25\% of continuum vs 60\% of continuum) 
and broader (about $-$100 to $-$300 km/s versus $-$50 to $-$250 km/s). 

In summary, the blueshifted absorption exhibited in
the \ion{Na}{1} D, \ion{K}{1} $\lambda$ 7665, 7699, 
\ion{O}{1} $\lambda$ 7774 lines and \ion{He}{1} $\lambda$ 10830 line,
as well as blueshifted absorption components to the broad H$\alpha$ and
\ion{Ca}{2} $\lambda\lambda$ 8498, 8542, and 8662 emission profiles
(discussed further in \S~\ref{sec:spec}.3 below)
are indicative of a strong outflow/wind.  The absorption must occur in a region
outside of the optical and near-infrared continuum in the case of the
pure absorption features, and outside the region producing the 
broad H$\alpha$ and \ion{Ca}{2} line emission (mainly the accretion column
but also the wind itself).

\subsection{Evolution of the Accretion and Outflow Diagnosing Emission Lines}

At low resolution,
Figures~\ref{fig:optlines3} and \ref{fig:optlines4} show detail relative 
to Figure~\ref{fig:optspec} of some of the 
optical emission line behavior of PTF~10nvg during 2010 and 2011.  
Among the rich optical emission line spectrum are permitted atomic species of:
\ion{H}{1},
\ion{O}{1},
\ion{Fe}{1},
\ion{Fe}{2},
\ion{Ca}{1},
and
\ion{Ca}{2}.
Notably there is no evidence of optical \ion{He}{1}.
There are also (strong) forbidden emission lines in the optical from:
[\ion{C}{1}],
[\ion{O}{1}],
[\ion{Fe}{2}],
[\ion{Ni}{2}],
[\ion{Cr}{2}],
[\ion{Ca}{2}],
and
[\ion{S}{2}].
Several of these species are rarely seen in young star spectra,
even those which are emission-line dominated.
Figure~\ref{fig:irspeclines} illustrates the infrared time series
of emission line behavior.  
In the infrared, we see permitted atomic species of:
\ion{Al}{1},
\ion{Ca}{1},
\ion{Fe}{1},
\ion{K}{1},
\ion{Mg}{1},
\ion{Na}{1},
\ion{O}{1},
\ion{Si}{1},
and
\ion{Ti}{1}
as well as forbidden
[\ion{Fe}{2}],
[\ion{N}{1}], and
[\ion{S}{2}].

In both the optical and infrared, there was
notable weakening of permitted atomic line emission when the continuum faded.  
The forbidden atomic line species, on the other hand,
became significantly more prominent in the faint state spectra; 
this prominence is due not only to better contrast relative to the
continuum as it faded, but also to variation 
(by up to a factor of two in both the optical and near-infrared) 
in the forbidden line absolute fluxes.  Infrared H$_2$ lines 
behaved in a manner similar to the forbidden atomic lines.  
As the permitted atomic emission is generally associated with the accretion
flow (with some lines influenced by outflow physics), and the forbidden
atomic and the H$_2$ lines are generally associated with outflow,
we suggest below that their respective correlation and anti-correlation
with the photometric state of PTF~10nvg implies that the forbidden and
H$_2$ lines are formed outside the extinction region.  
Other molecular lines, notably infrared CO and optical TiO/VO, 
however, followed the emission behavior of the permitted atomic lines,
suggesting that these lines are also formed in the inner disk region.

\begin{figure}
\includegraphics[angle=0,width=1.00\hsize]{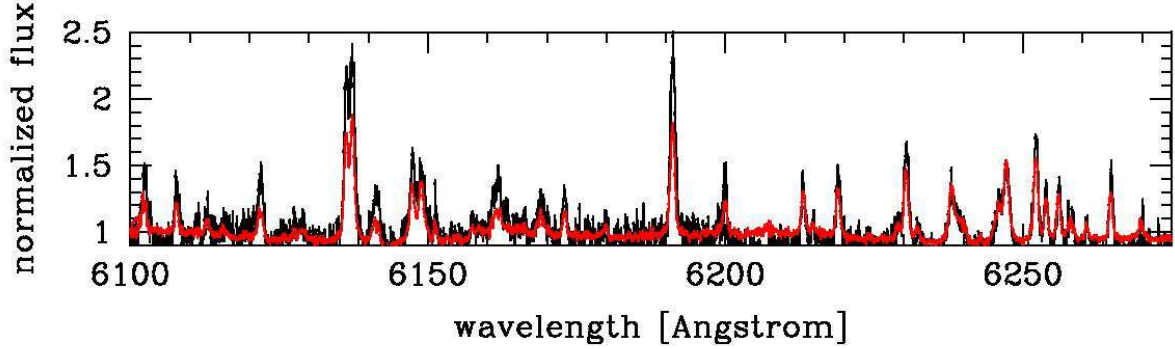}
\vskip-3.5truein
\includegraphics[angle=0,width=1.02\hsize]{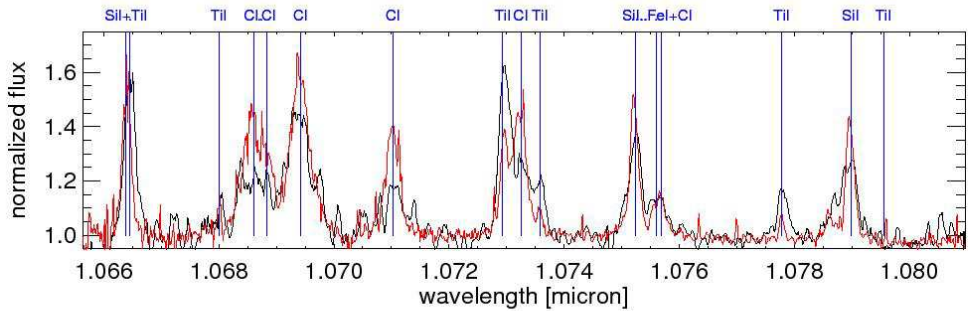}
\vskip-5.5truein
\caption{Portions of our high--dispersion spectra for PTF~10nvg (in black)
compared to the well-known emission line dominated object V1331 Cyg (in red).
There is a clear spectral similarity between these two young stars.
{\it Top panel:}
Keck/HIRES data in
the red optical spectral region, showing PTF~10nvg in a bright state in 2010, December.
{\it Bottom panel:}
Keck/NIRSPEC data in
the Y-band, showing PTF~10nvg on 2010 July 16, near peak outburst. 
Blue lines identify various permitted metallic emission lines,
which are similar in the two stars.
\ion{H}{1}, \ion{He}{1}, and forbidden metallic lines 
(located in other spectral regions) are also present in both stars but
exhibit different kinematic signatures.
}
\label{fig:1umspec}
\end{figure}

At high--dispersion, the emission line spectrum of PTF~10nvg
bears significant resemblance to that of V1331 Cyg in terms
of the species present and the relative line strengths.
Direct comparison of small portions of the red optical and 
of the 1 $\mu$m region high--dispersion spectra are
provided in Figure~\ref{fig:1umspec}.  The PTF~10nvg spectrum
is more ``extreme" in terms of emission line
presence and strength than the spectra of other well-known 
``continuum plus emission" sources.  Specifically, compared to 
DR Tau, Z CMa, or SVS 13, for example, 
PTF~10nvg exhibits a larger number of emission line species 
and with larger blueshifts. 
However, at 1 $\mu$m the blueshifted absorption trough 
in \ion{He}{1} $\lambda$ 10830 for PTF~10nvg (Figure~\ref{fig:1umprofiles}) 
is not as deep or as wide as in these objects (see Figure 1 
of \citealt{Edwards03} for profiles) despite the similar 
line terminal velocities.  V1331 Cyg similarly has a \ion{He}{1} profile 
that is much deeper and broader than that of PTF~10nvg.
But unlike the spectra of the above sources around 1 $\mu$m, 
V1331 presents a similarly rich emission line spectrum as PTF~10nvg 
in both the permitted and some of the forbidden line species  
(bottom panel of Figure~\ref{fig:1umspec}).  
In the optical, the PTF~10nvg spectrum (from 2010, December, 
the only epoch in our high--dispersion data set that is not limited 
by signal-to-noise) is again well-matched to V1331 in the permitted line
emission (top panel of Figure~\ref{fig:1umspec})  
though PTF~10nvg has stronger and more blueshifted forbidden line
emission.

In the sub-sections below we discuss the details and time series behavior of 
the molecular (\S~\ref{sec:spec}.3.1), permitted atomic (\S~\ref{sec:spec}.3.2),
and forbidden atomic (\S~\ref{sec:spec}.3.3) lines.  Summarizing our findings,
the evolution of PTF~10nvg's spectroscopic properties support our interpretation
from the photometry that the source undergoes
large variations in line-of-sight extinction ($A_v > 30^m$) 
towards the origin of the optical/near-infrared continuum and permitted
emission lines.  Because emission equivalent width increases (decreases) 
with continuum fades (rises) as $10^{0.4 \times\delta m}$, we do expect some
evolution in line strength.  However, the emission lines which are thought 
to arise from shocked circumstellar material do not appear
to be subject to the bulk of these extinction variations; the fluxes
of these lines are relatively stable throughout the monitoring period,
and therefore appear significantly more prominent at epochs when the
other components of the spectrum appear heavily extincted. 

\subsubsection{Molecular Emission}

Above when describing the behavior of PTF~10nvg's optical continuum, 
we discussed that the broad TiO and VO molecular emission
at optical wavelengths is time variable. 
To quantify the effect, we measured the strength of the TiO emission
at 6250, 6760, 7100, 7800, 8455, and 8880 \AA\ as well as the VO emission
at 7445 and 7865 using band indices as described in \citet{Hillenbrand97}.
There is stronger optical molecular emission relative to the continuum 
during the 2010 outburst and in early 2012 when the source was bright, 
relative to the periods when the source was fainter, notably in late 2011.  
This is consistent with the inference drawn from direct examination 
of the spectra both on an absolute scale and
on a continuum-normalized scale (see Figure~\ref{fig:optspec}). 
In addition to the direct correlation between strength 
of the different molecular bands, they are all stronger when 
the permitted atomic emission is stronger.
TiO/VO emission is correlated as well with the optical spectral continuum slope.

Molecular emission is also seen in PTF~10nvg's near-infrared spectra, 
specifically in bands arising from warm H$_2$O, CO, and VO (Figure ~\ref{fig:irspec}). 
These molecular band are all prominent in PTF~10nvg's bright states and 
all weaken relative to the continuum as the continuum fades.  
As Figures ~\ref{fig:irspec} and ~\ref{fig:irspeclines} 
demonstrate, the strength of the CO and H$_2$O
emission features are highly correlated with both emission from
permitted atomic lines (\ion{H}{1}, \ion{Na}{1}, \ion{Ca}{1}, etc.) 
and with the brightness of the underlying continuum. 

Like the optical TiO and VO bands, the infrared H$_2$O, CO, and VO  
bands are commonly seen in absorption in M-type photospheres.  
CO overtone emission is not uncommon in embedded protostars. For example,
\citet{Doppmann2005} and \citet{Connelley10} detected CO emission from
$\sim$15\% of the Class I and $\sim$25\% Class I/II (i.e. flat spectrum)
sources in their samples.  Detailed analyses of the kinematic profiles of 
these CO bandheads in several particularly well-studied CO emission sources
suggest that the emission arises from the innermost regions
of circumstellar disks in Keplerian rotation \citep[e.g.,][]{Najita1996}.  
The prominence of several oxygen species molecular lines when PTF~10nvg 
is brighter could be due to simple heating and cooling
related to variation in optical and higher energy photon flux,
or perhaps to molecule formation following dissociation 
during the bright state; see \citet{Hillenbrand12} for further discussion.
Alternately, the correlation could be caused by obscuration of 
the molecular emitting region during the faint stages; this is
the same explanation we invoke for the time variable behavior
of the atomic line emission.

\begin{figure}
\includegraphics[angle=90,width=1.00\hsize]{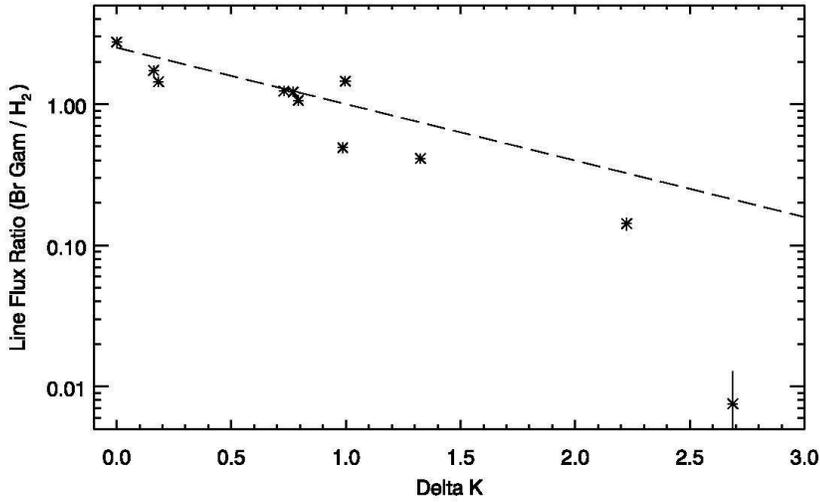}
\caption{
HI 2.16 $\mu$m - to - H$_2$ 2.12 $\mu$m line ratio measured from moderate
resolution near-infrared spectra (asterisks) as a function of PTF~10nvg's K$_s$
magnitude relative to peak brightness.  Error bars for the ratio
measurements were established by
adding synthetic noise to the spectrum then re-measuring the line
ratios. Overplotted for comparison as a dashed line are the
predictions of a simple model in which the H$_2$ line flux is constant
with time, while the HI 2.16 $\mu$m line flux diminishes as expected if the
differential K$_s$ magnitude is due entirely to extinction (i.e, HI =
HI$_o \times$ 10$^{-0.4 \times \Delta K_s}$).  The anomalous point at
high extinction corresponds to the 2011, July 15 epoch where the
brackett gamma line strength is very weak, and therefore may be more
susceptible to systematic errors introduced by mis-cancellation of the
HI profile of the telluric standard.  Except for this anomalous point,
the observed line ratios agree quite well with the quantitative
predictions of the extinction variation model.
}
\label{fig:linesvsak}
\end{figure}

Molecular emission from H$_2$, however, displays times series behavior 
with the opposite trend from the oxygen species (TiO, VO, CO).
While H$_2$ lines are present at all epochs, these lines appear
most prominent during PTF~10nvg's faint states. The absolute line flux
remains relatively constant, however, indicating that the increased
prominence of these features during faint states is due to improved contrast 
with the nearby continuum.  This behavior is qualitatively visible in Figure
~\ref{fig:irspeclines}, where the proximity of the 2.12 $\mu$m H$_2$ line 
to the 2.166 $\mu$m \ion{H}{1} line provides a clear demonstration 
that H$_2$ strengthens relative to \ion{H}{1} as the extinction (as inferred
from near-infrared colors) increases.  This behavior strongly suggests
that PTF~10nvg's variable extinction obscures the emission region
responsible for the \ion{H}{1} (and other permitted atomic emission 
and molecular emission) lines, but not the emission region 
from whence the H$_2$ emission arises.  To test this expectation in
more detail, we compare in Figure~\ref{fig:linesvsak}  the ratio of the
\ion{H}{1} and H$_2$ line strengths to a simple model in which H$_2$ line
strength is invariant while \ion{H}{1} line strength diminishes. 
This simple model accurately
replicates the dependence of the H$_2$-to-HI line ratio on A$_K$ for
all but the most heavily extincted epoch (for which increasing systematic 
errors may lead to underestimated errors based on photon statistics alone).

Shocks due to stellar winds and outflows, as well as 
photo-excitation by high--energy radiation, are both capable of producing
H$_2$ emission.  Previous studies of Classical T Tauri stars 
and Class I protostars have identified systems that appear to be
templates for each type of H$_2$ excitation mechanism, as well as numerous
systems where the excitation mechanism remains ambiguous
\citep{Beck2008,Greene10}.  From their reported 2010 July 14 spectrum, 
\citeauthor{Covey11} were able to measure confidently only
the 2.12 $\mu$m H$_2$ line, as well as state a tentative detection
of the 2.24 $\mu$m H$_2$ line.  From the ratio of these two lines,
it was concluded that shock excitation was the most likely
mechanism for exciting PTF~10nvg's H$_2$ emission, but noted that this
conclusion remained uncertain.   The 2011, July 15 spectrum, 
obtained when PTF~10nvg was near a photometric minimum,  
is by contrast nearly featureless save for a number of 
well detected H$_2$ and [\ion{Fe}{2}] emission features. 
We present line strength measurements in Table ~\ref{tab:h2table}.
for nine K-band H$_2$ emission lines. 
For comparision we also tabulate the line strengths expected for different 
physical emission mechanisms:
first models of J- and C-type shocks in stellar outflows 
\citep{Smith1995} and models of UV and/or X-ray photo-excitation in
the presence of circumstellar dust \citep{Nomura2007} and, second, 
empiricially measured values for spatially extended H$_2$ emission near RW Aur
\citep{Beck2008} and for a knot in the HH~54
outflow \citep{Giannini2006}.  Finally, Table ~\ref{tab:h2table} also
summarizes the observed and predicted values for specific line ratios
identified by \citet{Beck2008} as particularly useful for disentangling 
the underlying excitation
mechanism for H$_2$ emission.  

The PTF~10nvg line ratios
are extremely similar to those measured by \citet{Beck2008} using 
adaptive optics imaging that spatially resolves the 
circumstellar H$_2$ emission near RW Aur.
The line strengths measured by \citet{Giannini2006} from Knot B of HH54 
are also in reasonable agreement with our observations of PTF~10nvg.  
This similarity is consistent with the hypothesis that the origin 
of PTF~10nvg's H$_2$ emission is a region that is
well separated from the central protostellar source.  Models that 
reproduce the line ratios observed for PTF~10nvg and RW Aur include those 
calculated by \citet{Smith1995} for a $\sim$2000 K
C-type shock and the models produced by \citet{Nomura2007} for pure
X-ray photo-excitation.  

A final clue to the origin of PTF~10nvg's H$_2$ emission is provided by
the kinematics of the lines, which are uniformly blueshifted by
$\sim$200 km/s with respect to the rest velocity of the North
America Nebula as well as the observed velocities of PTF~10nvg's 
permitted atomic features.  This kinematic profile strongly suggests
that PTF~10nvg's H$_2$ emission originates from within a stellar jet. 
Interestingly, isolating the emission region location does not 
completely resolve the ambiguity between X-ray and shock heating as potential
excitation mechanisms, as {\it Chandra} observations of the HH 154 
protostellar jet revealed that shock regions can generate X-rays {\it in situ}
in emission regions with expansion velocities of $\sim$ 500 km/s.
Figure~\ref{fig:irspeclines}
demonstrates the behavior of H$_2$ (second panel from right) relative to 
[\ion{Fe}{2}] (third and fifth panels from right) which vary in a 
correlated manner over the time series,
and relative to CO (right-most panel) and various permitted
atomic lines (all panels) where there is anti-correlated behavior.

In summary, multiple lines of evidence support the conclusion
that PTF~10nvg's H$_2$ emission likely arises from a spatially extended 
outflow. First, the H$_2$ emission strength is insensitive
to the variable extinction that apparently occults the innermost disk 
and accretion region -- where the near-infrared continuum and the
other molecular (see above) emission as well as the permitted atomic 
(see below) emission lines form. Second, there is similarity
in H$_2$ line strengths to RW Aur's spatially extended
emission and to models of excitation due to mechanical shocks and
X-ray photoexcitation, both of which have been associated with other
protostellar jets.

\subsubsection{Permitted Atomic Emission}

Over most of the time series, the strongest emission lines 
in the optical spectra of PTF~10nvg are those of 
H$\alpha$ and \ion{Ca}{2} -- a typical situation for young stars.  
As discussed below, during fainter photometric stages, however, 
the forbidden lines became stronger than these atomic lines. 
The H$\alpha$ and \ion{Ca}{2} line strengths vary
both absolutely and relative to the red optical continuum
(top and bottom panels of Figure~\ref{fig:optspec}). 

The \ion{Ca}{2} triplet emission varied between $-$15 and $-$25 \AA\ 
but became much weaker in 2011, June, when the continuum faded.
For classical T Tauri stars
in Taurus, the line strengths among the three lines of the \ion{Ca}{2} triplet 
are typically $1.20\pm0.19$ for the 8542 \AA\ to 8498 \AA\ ratio and
$1.28\pm0.13$ for the 8542 \AA\ to 8662 \AA\ ratio
(based on analysis of our own collection of optical spectra for representative
samples).  In PTF~10nvg, however, the 8498 \AA\ line is particularly strong; 
the 8542 \AA\ to 8498 \AA\ ratio has a mean and dispersion of
$0.93 \pm 0.15$ over the time series (considering only spectra
for which the lines are measureable),
while the 8542 \AA\ to 8662 \AA\ ratio is consistent
with expectations at $1.14 \pm 0.18$.  Although highly unusual, these 
observed line ratios for PTF~10nvg are not unique.  
\citet{HP89} found for V645~Cyg
ratios among the triplet lines of 0.85 for 8542 \AA\ to 8498 \AA\ and
1.1 for 8542 \AA\ to 8662 \AA, similar to those in PTF~10nvg.
Notably the line ratios are far from their expected ratio of 1:9:5 for 
optically thin emission, which is typically the case for young accreting stars.
\citet{KF11} find that the physical conditions for producing 
strong \ion{Ca}{2} triplet emission include relatively
high density $n_H \sim 10^{12}$  cm$^{-3}$ and low temperature $T <$7,500 K,
with the line formation taking place in the accretion flow.

As illustrated in the high--dispersion profiles of
Figures~\ref{fig:hires}, 
the \ion{Ca}{2} triplet lines all peak to the red of zero velocity.
The velocity profiles are similar between the 2010, December and
2011, December spectra (when atomic emission was also prominent)
although the 2011 spectra are even narrower on the blueshifted side.
There are indications, however, at both epochs of differences among
the three lines of the triplet. 
The 8498 \AA\ line is the broadest of the three, with 
FWHM$\approx$200 km/s, has an approximately symmetric profile, 
and exhibits no evidence for sub-continuum absorption.  
The 8542 \AA\ and 8662 \AA\ lines, by contrast,
are somewhat narrower and asymmetric (narrower on the blue side than the red side),
with sub-continuum ``dips" in their blueshifted components 
that change between the two high dispersion spectra 
and indicate wind absorption.
The 8542 and 8662 \AA\ structure is similar to that seen 
in other strong outflow sources such as V1515 Cyg and V1057 Cyg.
On its red side, the 8498 \AA\ line traces well the 8662 \AA\ profile but the 8542 \AA\ line is broader.
on the blue side, though, it is the 8542 and 8662 \AA\ profiles that are 
quite similar.   While these three lines share the 2P$\rightarrow$2D 
transition, they differ in their spin with the 8542 and 8662 \AA\ electrons 
changing spin between the 4p and 3d energy levels, 
and the 8498 \AA\ electrons retaining the same spin, which is perhaps 
a clue to the apparent differences in line optical depth.

In H$\alpha$, the emission line equivalent widths varied at low dispersion 
between $-$20 and $-$60 \AA\ during brighter photometric states but
became quite large (many hundreds of \AA) as the source continuum 
level faded during late June through August of 2011.  
The strength and profile behavior as discerned from our collection of
low resolution spectra is illustrated in Figure ~\ref{fig:optlines3} while
the high--dispersion behavior is illustrated in Figure~\ref{fig:profforbidden}. 

The low resolution spectra from 2010 July, August, and September all show
a redshifted emission shoulder in H$\alpha$,
with the first July spectrum exhibiting a
clear blueshifted absorption component, which is then ambiguous in the 
August and September data.
The line profile as seen in the high--dispersion 2010, December spectrum 
shows more clearly both the redshifted emission component and 
a two-component structure to the blueshifted absorption, 
similar to the structure seen in the \ion{Ca}{2} lines 
described above\footnote{Based on optical spectra taken on
2010 September 5 and 2010, November 25, \citet{Aspin11} claim that the blueshifted 
H$\alpha$ absorption reported by \citeauthor{Covey11} had disappeared. Our data
show that the absorption weakened in the low dispersion data
but it clearly remained present.}.
In H$\beta$, the line structure has the same $-$250 km/s absorption component as
seen in the H$\alpha$ line, but a deeper $-$125 km/s absorption component 
that extends to lower velocities and even to the positive side of the line.
During the faint photometric state spectra (two in 2011, June and one in 2012, January), 
the H$\alpha$ profile lacks any redshifted emission 
and only strong blueshifted emission is observed correlating with strong
forbidden-line emission at these same velocities
(see Figure~\ref{fig:profforbidden} and disussion below). 
The brighter state (2012, August and 2010, December) profiles are also similar.
The 2011, December profile is intermediate, exhibiting
both strong blueshifted emission and strong redshifted emission components.
The blueshifted emission components in H$\beta$ and H$\alpha$ are similar
and exhibit similar time series behavior.
Specifically, while from 2011, May through to 2011, October,
the H$\alpha$ line is considerably weaker 
(Figure~\ref{fig:optspec}),
the even greater dimming of the red continuum produced enhanced 
line equivalent width during this period 
(left panel of Figure~\ref{fig:optlines3}).
The 2011, October and November profiles then revert back to the
redshifted emission shoulder, as the source brightened.

This time series behavior suggests that at least some of the blueshifted 
H$\alpha$ emission originates in the outflow or outflow shock region,
similar to the H$_2$ emission (dicussed above)
and to the optical/infrared forbidden line emission (discussed below), 
both of which are similarly blueshifted.  The blueshifted velocities
for all of these emission tracers at the position of the continuum source
are similar to the velocities reported in \S~\ref{sec:offset}
for the spatially offset blueshifted emission that is observed, 
and arises from the interaction of the outflow with either
ambient material in the local vicinity or itself.   

In both the \ion{Ca}{2} and the \ion{H}{1} profiles  
there is no apparent change over the existing time series 
in the velocity of the emission
line peaks (which, as described above, are slightly to the red
in both of these lines) or in the terminal velocity of the 
blueshifted absorption components.  That there is little 
change in the terminal velocity of the wind 
means that the launching velocity is not changing, even though
there is changing structure in the absorption part of the
P Cygni profiles.

Notably, no optical \ion{He}{1} $\lambda$ 5876 emission is 
seen\footnote{A narrow 6678 \AA\ emission line might be associated
with \ion{He}{1} but the absence of the related 5876 \AA\ line 
suggests that this is more likely \ion{Fe}{1} $\lambda$ 6678.},
even though this line is near-ubiquitously detected for classical 
T Tauri stars in high--dispersion spectra.
For strong emission at 5876 \AA, T$\approx$15,000 - 20,000K  
and $n_H > 10^{11}$ would be required \citep{KF11} and thus
 a lack of 5876 \AA\ emission while lower energy 10,830 \AA\ emission 
is observed could indicate
an upper limit on temperature and/or density in the emission region.  

Other prominent atomic emission lines in the optical include those of 
\ion{O}{1},
\ion{Fe}{1},  
\ion{Fe}{2},
\ion{Ca}{1}.
Those lines that are visible in the existing high--dispersion spectra
(see Figure~\ref{fig:hires}) 
are generally symmetric and centered at zero velocity, as well as narrow
in velocity with full-width half-maximum $\sim$45-50 km/s
(e.g. \ion{Fe}{1} lines) to $\sim$65-70 km/s (e.g. \ion{Fe}{2} lines).
Their origin may be on the stellar surface or in the disk.  
The \ion{O}{1} 8446 \AA\ line, however, though also centered at zero
velocity, is broadened with full-width
half-maximum $\sim$250 km/s.  This width is similar to that exhibited
by the \ion{Ca}{2} and H$\alpha$ lines, and suggests
origin in the inflowing and/or outflowing material.
In the low dispersion spectra we measure a ratio of \ion{Ca}{2} 8498 \AA\ 
to \ion{O}{1} 8446 \AA\ (for which the time variable extinction 
effects are not a factor given the small wavelength difference)
that does change somewhat over time but notably always exceeds 10.  
\citet{KF11} find that values larger than 7
for this ratio indicate temperatures $T <$7500 K and densities 
$n_H >10^{12}$cm$^{-3}$ for the emitting gas.  Only the most extreme accretors 
among well-studied Taurus stars have similar \ion{Ca}{2} 8498 \AA\ : 
\ion{O}{1} 8446 \AA\ line ratios.

Another notable atomic line seen in emission is \ion{Li}{1} $\lambda$ 6707.
Interpretation of the profile (Figure ~\ref{fig:hires}) should take
into account that there is a $y^3$F$^o$-e$^5$P term transition of \ion{Fe}{1}
reported by the NIST Atomic Spectra Database \citet{NIST}
at similar wavelength, specifically 6707.43 \AA.  No other lines 
of this \ion{Fe}{1} multiplet are within our spectral range 
and so the possible identification as such can not be supported.  
We note, however, that
another \ion{Li}{1} line at 6103 \AA\ may also appears weakly in emission 
in our spectra, possibly supporting the \ion{Li}{1} emission scenario.
Although this line also has a nearby \ion{Fe}{1} potential identification,
the $w^5$D$^o$ -5/2[7/2] term transition at 6103.64 \AA, this is a 
much higher excitation \ion{Fe}{1} line than those found otherwise in the
PTF~10nvg spectra.

Redward, as illustrated in Figure~\ref{fig:1umspec},
the atomic line emission apparent in the high--dispersion 1 $\mu$m
spectra includes zero velocity \ion{C}{1}, \ion{S}{1}, \ion{Si}{1}, 
\ion{Sr}{2}, and \ion{Ti}{1} multiplets.  There is no \ion{Fe}{1}
or \ion{Fe}{2} apparent.
The three epochs of NIRSPEC spectra, which cover
dramatically different brightness stages of PTF~10nvg, show little
difference in the metallic line emission profiles
(e.g. Figure~\ref{fig:1umprofiles}) which generally match quite well
over a range of line excitation potentials.  Changes over time are apparent, 
however, in the \ion{H}{1} and \ion{He}{1} profiles, with both lines exhibiting 
blueshifted absorption components as discussed above in the context of wind lines. 
Other prominent near-infrared emission lines seen in the low dispersion
spectra include \ion{Na}{1}, \ion{Mg}{1}, and \ion{Fe}{2}.  
As Figure~\ref{fig:irspeclines} demonstrates, 
these lines appear to vary in concert
with one another, and are anti-correlated with the extinction
estimates inferred from PTF~10nvg's near-infrared photometry.  

A common practice is to convert infrared hydrogen line fluxes into 
mass accretion rates on to the central star.  Following \citet{Muzerolle1998}
and assuming a mass of 0.5 M$_\odot$ and radius of 3 R$_\odot$ for the star,
we derive accretion rates of $3\times 10^{-7}$\msun\pyr\ in the
brighter states and $6\times 10^{-8}$\msun\pyr\ 
in the fainter states using Br$\gamma$ with an upper limit 
of $5\times 10^{-12}$\msun\pyr\ when the line was undetected during summer
of 2011.  Using Pa$\beta$, the values are about a factor of 4 lower.  
No extinction corrections are applied in these calculations, 
and thus the numbers are clearly lower limits in the faint state and probably
the bright state as well.  In addition to the extinction uncertainty,
it is not clear that the accretion model on which the line luminosity to
accretion rate calibration is based applies to PTF~10nvg.

\subsubsection{Forbidden Atomic Emission}

In young stars, forbidden lines of e.g. 
[\ion{O}{1}], [\ion{N}{2}], [\ion{S}{2}], and [\ion{Fe}{2}]
in the optical and near-infrared spectra
are attributed to few hundred km/s radiative shocks 
(those that cool by such radiation more quickly than they would 
by expansion) in outflowing material in jets associated with accretion disks. 
Unlike the permitted optical/infrared emission lines discussed above, 
the forbidden emission lines are (almost always) optically thin.
They trace plasma over a wide range in density 
$n_e \sim 10^3-10^8$  cm$^{-3}$ in the temperature range 
$T \sim$5,000--20,000 K --
for hydrogen ionization fraction $n_e/n_H \approx$ 1/10, which is appropriate
for the mostly neutral young star jets.

The moderate temperature, low density forbidden
lines are formed closest to the shock, 
with temperature decreasing and density increasing further from the shock 
in the pre-shock medium.  The infrared H$_2$ lines discussed above 
are generally thought to form further away from the shock front 
than the optical/infrared forbidden lines, probing
cooler temperatures in either post-shock gas along the jet axis that has sufficiently cooled,
or material perpendicular to the jet axis that is entrained/stirred by the flow.
As with the density and temperature gradients, the ionization fraction is likely
not constant along the flow.  The implied range of physical parameters in the outflow  
sampled in spectroscopic observations such as ours that do not spatially 
resolve (see, however, \S~\ref{sec:offset}) 
the cooling region/s in the outflow zone 
means that we should consider the forbidden line emission indicative 
but not diagnostic of jet/outflow conditions.  
See \citet{Hartigan00} for a review and \citet{BBM81}
for a classic study of forbidden lines in Herbig-Haro objects.

Indeed, the typical [\ion{O}{1}], [\ion{N}{2}], [\ion{S}{2}], and [\ion{Fe}{2}]
forbidden lines are present in PTF~10nvg spectra, but they display
significant variability over time (see Figure~\ref{fig:profforbidden}).  
The forbidden profiles are purely blueshifted, with a range of 
profile shapes.  Notable in the spectroscopic time series 
is the increasing prominence of these and other forbidden lines in 
2011, June -- as the broadband optical/infrared continuum faded to below detectable
levels in our PTF R-band photometry.  Specifically in the red optical spectra,
while the somewhat common forbidden line doublets of [\ion{O}{1}] 6300, 6363 \AA, 
and [\ion{S}{2}] 6717, 6731 \AA\ as well as [\ion{Fe}{2}] 7155, 7172 \AA\ 
and [\ion{Ca}{2}] 7291, 7324 \AA, 
had been weakly present in the 2010 spectra and in
the first 2011 (May) spectrum obtained, by 2011, June these lines had
signficantly increased in strength relative to the continuum.  
Over just a three week time frame, they along with other somewhat unusual 
forbidden lines (such as the [\ion{C}{1}] 9824, 9850 \AA\ doublet, 
[\ion{Cr}{2}] and [\ion{Ni}{2}]; 
see Figures~\ref{fig:optspec} and \ref{fig:optlines3}-\ref{fig:optlines4})  
rose in prominence to dominate the optical spectrum. 
Notably, emission in [\ion{N}{2}] is 
seen {\it only} during the very faintest epochs. 
This increasing prominence of the forbidden lines was accompanied by a
weakening of the permitted emission lines, most visibly the \ion{Ca}{2}
triplet and several \ion{Fe}{2} lines that previously were quite strong.
This behavior suggests that we are seeing, in the faint photometric states,
emission from a Herbig-Haro flow that is not visible when the
source is brighter.  

We can gain insight into the physical conditions necessary for these
forbidden line observations by taking guidance from the material assembled in
\citet{Dougados10} and references therein.  For the different forbidden line 
species we discuss first the line strengths and their variation, 
and then the line profiles and their variation.

For the [\ion{O}{1}] 6300 and 6363 \AA\ doublet, having the highest critical
density of the red optical [\ion{O}{1}], [\ion{N}{2}], and [\ion{S}{2}] 
lines, the evolution of the observed line ratio is truly remarkable. 
Typically observed [\ion{O}{1}] 6300 \AA\ : [\ion{O}{1}] 6363 \AA\ values are
in the range 2.5-4 : 1 with the theoretical line ratio 3:1 based on Einstein A values. 
Before and after the deep photometric fading episode of 2011, 
the expected 3:1 ratio for optically thin emission was
indeed consistently observed to within 10\% in PTF~10nvg spectra. 
However, as the continuum flux faded and the forbidden
line prominence increased, the [\ion{O}{1}] doublet ratio was variable 
between the typical 3:1 and values  
as extreme as 1:2 in several of the faint state 2011
June, July, and August low dispersion (see Figure~\ref{fig:optlines3}),  
as well as high--dispersion (see Figure~\ref{fig:profforbidden}) spectra.  
The variability in the doublet line ratio suggests dramatic changes 
in the optical depth of the [\ion{O}{1}] lines, from optically thin 
to a likely optically thick case where 1:1 is expected; we can not, however,
explain the observed situation that the 6363 \AA\ line is {\it stronger than}
the 6300 \AA\ line at some epochs.  
We note that \citet{LiMcCray92} developed a model for 
[\ion{O}{1}] 6300 \AA\ : [\ion{O}{1}] 6363 \AA\ doublet emission in
SN 1987A, which exhibited a systematically time variable ratio that increased from an
optically thick 1:1 to the usual optically thin 3:1. The model was based on an expanding shell 
with a small filling factor, i.e. clumpy distribution of [\ion{O}{1}], that is
not necessarily applicable here given differences in expected geometry between
a spherically expanding supernova shell and a more collimated young star outflow,
but does suggest that variable line optical depth is a plausible explanation for our observations.
Notably, the [\ion{O}{1}] changes observed in PTF~10nvg were not 
gradual and systematic as observed in SN 1987A; rather, they were abrubt 
with the 6300 \AA\ : 6363 \AA\ ratios changing literally from night to night, 
e.g. from 1:1 on 1 August, 2011 to 3:1 on 2 August, 2011 when the data derive
from the same instrument, configuration, observers, slit position angle, etc. 
so the variability can not be attributed to spatial alignment or other similar
obserational effects.
Plausible explanations for our observations, whether related to
optical depth effects or not, need to account not only for the truly 
peculiar 1:2 line ratio but also the short term variability of the ratio.

At all epochs, including during the photometric fading period when 
the [\ion{O}{1}] ratio is out of the expected 3:1 ratio, 
the much lower critical density [\ion{S}{2}] 6717, 6731 \AA\ doublet 
notably remained within the expected range for its ratio.
Typically observed values in young stars are 0.5-1 : 1, with the theoretical 
ratio in the high density limit 0.43.  The PTF data are consistent
with these values but with some variation over the time series. 
Most of the data exhibited line ratios close to the high density 
expectation (indicating $n_e >10^4$ cm$^{-3}$) 
but some epochs show more equal strength doublet lines, 
indicating lower density gas ($n_e \sim 10^3$ cm$^{-3}$).

Notably, neither [\ion{O}{3}] 4959 / 5007 
with a similar critical density to the strong [\ion{O}{1}] doublet, nor
[\ion{O}{2}] 3726 / 3728 with a similar critical density to the strong [SII] doublet,
are seen.  
Although our spectra become somewhat
signal-to-noise challenged in the blue, strong line emission comparable
to the red line emission would be detectable.
The [\ion{O}{3}] line absence provides a constraint on the electron temperature of
the emitting gas and is, indeed, 
rarely seen in well-studied HH objects. The [\ion{O}{2}] absence
can be taken to confirm the assumption of a low ionization fraction.  

The [\ion{N}{2}] 6548, 6583 \AA\ doublet is of intermediate critical density and has 
a typically observed ratio of 1-5 : 1 in young stars with a theoretical ratio of
3 based on Einstein A values.
These lines are present in just a few of our low resolution spectra during 
the faintest photometric states, but are hard to disentangle from the broad Halpha emission (see Figure~\ref{fig:optlines3}).
They were only weakly visible in our high--dispersion data.
We note that [\ion{N}{2}] 6548, 6583 \AA\ emission is seen, however, at positions
that are spatially offset from the PTF 10nvg point source, as discussed in \S~\ref{sec:offset}.  Some of the weak emission seen in  
Figure~\ref{fig:profforbidden} may arise in the extended emission region
closest to the trace (see Figure~\ref{fig:extended}).

Considering a higher density diagnostic, the infrared
[\ion{Fe}{2}] 1.533 and 1.644 $\mu$m lines, like the optical forbidden lines,
are more prominent when the source is faint in the continuum.  The 
ratio of these two lines is density sensitive and indicates $n_e >10^5$ cm$^{-3}$. 
The 1.257 and 1.644 $\mu$m line ratio can be used to measure extinction
since the lines share the same upper level (see \citet{Connelley10}). 
For PTF~10nvg the time series data indicate low extinction (A$_V < 2$ mag)
to the jet region -- in contrast to our earlier findings on the large and time
variable extinction towards to near-infrared continuum.
The above and other infrared \ion{Fe}{2} lines that are prominent 
during PTF~10nvg's faintest epochs are tabulated in Table ~\ref{tab:feiitable}
and can be compared in line strength ratio to that observed
by \citet{Giannini2006} from Knot B of HH54.  The two sets of
measurements agree moderately well, providing additional support for
the existence of a significant outflow within the PTF~10nvg system.
The optical [\ion{Fe}{2}] lines probe even higher density material.  
The measured 
7155 \AA, 7452 \AA, 8617 \AA, and 8892 \AA\ lines of [\ion{Fe}{2}] 
yield ratios that according to \citet{BP98}
indicate gas at densities of $n_e \sim 1-5 \times 10^6$ cm$^{-3}$
in the vicinity of PTF~10nvg.
The optical [\ion{Fe}{2}] doublet at 7155, 7172 \AA\ 
varies in our time series, strengthening during the 2011 faint epochs
but perplexingly failing to re-emerge in the late 2011 / early 2012 fade
like other forbidden lines.
These lines also exhibit a ratio both well above and below its expected 3:1 
ratio for optically thin emission.  Unlike the time series behavior of the
[\ion{O}{1}] doublet ratio, however,
the [\ion{Fe}{2}] anomalies are not restricted to the faint epochs.
There is no correlation of the [\ion{Fe}{2}] line ratio variations 
and the [\ion{O}{1}] line ratio peculiarities.  

High density gas is also probed by [\ion{Ni}{2}] and [\ion{Ca}{2}] lines. 
The observed ratio of [\ion{Ni}{2}] 7412 \AA\ to  7378 \AA\ in the 0.3-0.5
range indicates densities $n_e >5 \times 10^7$ cm$^{-3}$ according to \citet{Bautista96}. 
The observed ratio of \ion{Ca}{2} 8542 \AA\ to  [\ion{Ca}{2}] 7291 \AA\ 
varies between 8 and 15 during the bright photometric states and is
indicative of densities in the range $n_e \sim 3-5\times 10^8$ cm$^{-3}$. 
During the faint states this ratio becomes very small as the atomic emission
is not detected while the forbidden emission becomes quite strong
relative to the continuum.

As illustrated in Figure~\ref{fig:1umspec},
there is also forbidden line emission in the high--dispersion 1 $\mu$m
spectrum, with a blueshifted set of [\ion{N}{1}] doublets around 1.040 $\mu$m 
(specifically 1.040059, 1.040100 and 1.041002, 1.041044 $\mu$m) and 
a blueshifted [\ion{S}{2}] series located at 1.029, 1.032, 1.034, 1.037 $\mu$m. 
They are indicative of hotter temperature gas, 
in the 30,000--40,000 K range, than the optical forbidden lines 
which probe 5,000 to 20,000 K temperatures.  Like the optical lines,
the near-infrared forbidden line strength appears to correlate 
with the photometrically derived extinction
estimates (Figure~\ref{fig:av_vs_time}); specifically,
the 2012 spectrum exhibits stronger [\ion{N}{1}] and [\ion{S}{2}] lines
than the 2011 and 2010 spectra, both of which were taken when the extinction
was about half of the value for the 2012 spectrum.
The non-simultaneous optical-infrared spectroscopy in combination with 
the unknown extinction and evidence for large extinction variations 
over time precludes us from using infrared-optical line pairs 
of these species to probe either temperature 
(e.g. from optical 6730 \AA\ vs infrared 1.03 $\mu$m [\ion{S}{2}] 
at high temperature, or 
optical 8620 \AA\ vs infrared 1.64 $\mu$m [\ion{Fe}{2}] at lower temperature)
or ionization fraction 
(e.g. from optical [\ion{N}{2}] and infrared [\ion{N}{1}]).
We simply note the presence of these hotter lines in the 1 $\mu$m region spectra.

In addition to the line strength variations, there were variations in the
line profiles of the forbidden emission in PTF~10nvg.  These are 
illustrated by our high--dispersion optical data as shown in 
Figure~\ref{fig:profforbidden}.  
In all forbidden lines and at all epochs, the profiles are purely blueshifted with maximum emission velocities of $\sim$250 km/s.
At the earliest 2010, December epoch, the [\ion{O}{1}] doublet has a broad FWHM
and nearly square profile ranging from 0 to $-$250 km/s, 
peaking at about $-$130 km/s, 
whereas at the later epochs the profile becomes skewed with a broad base 
but increasing flux towards higher velocities, peaking near the terminal velocity 
which remains at the same $-$250 km/s terminus.
The skewed profile is present in both fainter and brighter continuum states
of PTF~10nvg following the initial 2010 December spectrum in which there was
more lower velocity emission relative to later times.  
Similar line profile morphology and behavior over time
is seen in the [\ion{Fe}{2}] doublet.
The [\ion{S}{2}] doublet seems to have the opposite trend, however,
with a similarly shaped but narrower and slightly lower velocity 
($-$100 km/s) peak relative to the [\ion{O}{1}] doublet early on, 
but becoming more boxy and broad as the [\ion{O}{1}] became more skewed.
The [\ion{N}{2}] doublet appears quite similar to the [\ion{S}{2}] doublet
but is much weaker, and essentially undetected in most of our high--dispersion
spectra.

The [\ion{Ca}{2}] doublet, unlike the above lines, is double-peaked, with
a higher velocity component centered at about -150 km/s and extending
to $-$200 km/s (similar to the [\ion{S}{2}] and [\ion{N}{2}] doublets),
as well as a second, nearly zero velocity but slightly blueshifted 
peak.  This double-peaked profile (see Figure~\ref{fig:profforbidden})
is seen in the 2010, December and 2011, December spectra, 
both taken near photometric peaks 
in the lightcurve.  During the fainter photometric states, however,
 only the blueshifted peak is seen, and presenting the same
skew presented by the [\ion{O}{1}] and [\ion{Fe}{2}] doublets.
A similar double-peaked and time-series behavior is exhibited by the
optical [\ion{C}{1}] lines.  The [\ion{Ca}{2}] doublet, unlike other
forbidden lines, has a possible origin in photon pumping through the permitted 
\ion{Ca}{2} 8498, 8542, and 8662 lines
rather than requiring collisional pumping 
(though collisions may still be the dominant mechanism 
in young star jets).  The 7324 \AA : 7291 \AA\ line ratio
of [\ion{Ca}{2}] is consistent with the expected equilibrium value of 0.67.

In summary, the forbidden emission line strengths in PTF~10nvg indicate
a wide range of gas densities, from $n_e \sim 10^3$ cm$^{-3}$ 
to $n_e > 10^8$ cm$^{-3}$.  Line strength variations,
some line ratio variations, and line profile variations were all seen
over the time series.

\subsubsection{Spatially Offset Emission}\label{sec:offset}

\begin{figure}
\includegraphics[angle=0,width=0.90\hsize]{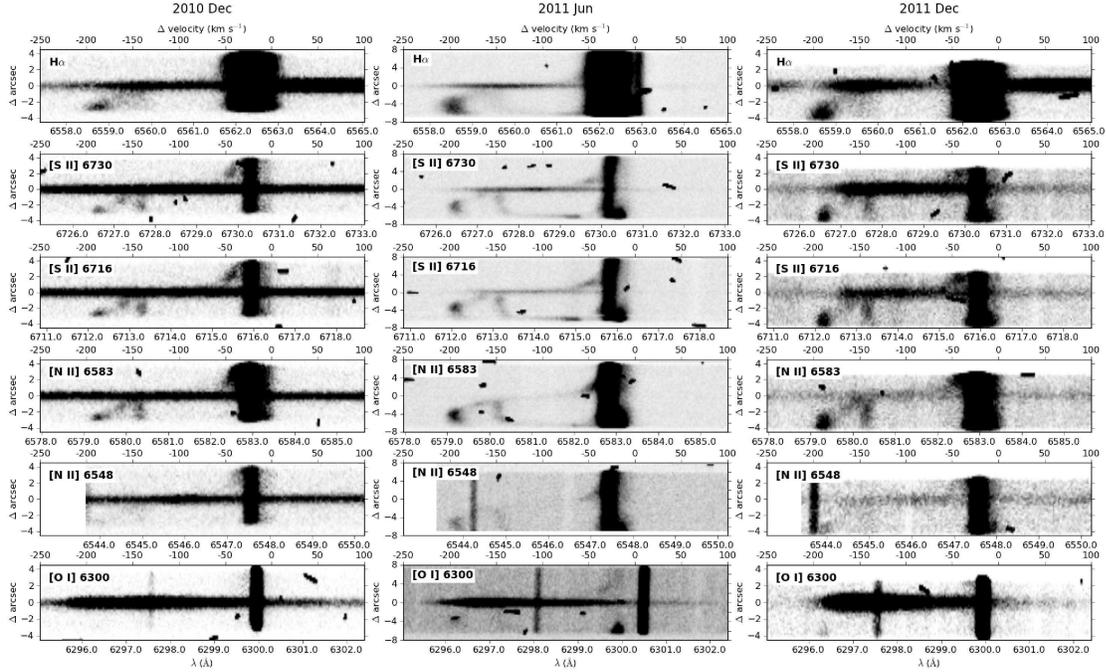}
\caption{Spatial-spectral diagrams centered around the optical
H$\alpha$, [\ion{O}{1}], [\ion{N}{2}], and [\ion{S}{2}] lines.
Units on the ordinate are arc seconds while those on the abscissa are given
in both wavelength (\AA) and velocity (km/s) with 
each panel showing a 350 km s$^{-1}$ portion of 
the emitting region.
The image intensity scaling is linear with the range set over a range
of counts selected for each feature to emphasize the spatially extended emission
(not the same in all panels). 
Cosmic rays have not been removed from the images. 
``Sky" emission fills the decker vertically at low velocity locations 
that depend on heliocentric correction and may not be the same from line to line
depending on whether the emission is dominated by telluric vs nebular 
background emission.
The left and right panels show the HIRES observations from
2010, December (PA = 110$^\circ$.8) and 2011, December (PA = 102$^\circ$.6),
respectively.  The middle panel shows data from 2011, July (PA = 176$^\circ$.0)
when the decker setting allowed twice the spatial coverage of the other two panels
but otherwise exhibits similar structure away from the main trace.
Blueshifted emission is seen below the trace at several velocities
(roughly -190, -160, -140, and -20 km/s in all columns plus at -75 
and blueward
in the middle panel only, with its larger extent in the spatial direction). 
Blueshifted emission is also seen above the trace at about -50 km/s.
}
\label{fig:extended}
\end{figure}

As discussed above, the velocity structure along the 
spectral trace as defined by the (sometimes weak) spectral continuum 
exhibits blueshifted emission in a large number of forbidden lines 
as well as in H$\alpha$, notably during the fainter photometric states
of PTF~10nvg's lightcurve. 
In addition, many of the HIRES images 
exhibit blueshifted emission that is spatially offset 
from the continuum source position.  This is illustrated via
position-velocity diagrams in the vicinity of H$\alpha$ and forbidden 
[\ion{O}{1}] 6300 \AA, [\ion{N}{2}] $\lambda\lambda$  6548, 6583, 
and [\ion{S}{2}] $\lambda\lambda$ 6716, 6730 in Figure~\ref{fig:extended}.  

We show the two-dimensional HIRES spectra at
three epochs: 2010 December when PTF~10nvg was relatively bright, 
2011 June when it was quite faint, and 2011 December when it was
again bright.  The two December spectra were obtained at 
similar position angle (PA), 110.8$^\circ$ and 102.6$^\circ$ for the 2010 and 2011 
observations, respectively, with a 7\arcsec\ long decker, while the 2011 June spectrum 
was obtained with a 14\arcsec\ long decker and a PA of 176.0$^\circ$.
Despite the differing PAs between the spectra, there are a few features that are 
commonly observed at all epochs. These include a bright knot of emission extending 
$\sim$2--6\arcsec\ from the stellar trace at $\sim -190$ km s$^{-1}$, which can be seen 
in all epochs in H$\alpha$, [S II], and [N II] $\lambda$ 6583 \AA\ 
(note that only the 2011 June spectrum had a long enough decker to see the emission 
6\arcsec\ from PTF~10nvg). This same knot is also visible in [N II] $\lambda$ 6548 \AA\ 
in the two 2011 spectra as well as [O I] $\lambda$6300 \AA\ in the 2011 December spectrum. 
In addition, there is plausible but weak offset emission in [\ion{Ca}{2}] 7291 \AA\ at this same $-$190 km/s velocity, but in the 7324 \AA\ line of the doublet it
is not obviously offset from a sky line that coincidentally appears at a similar velocity; these images are not shown.
Fainter knots, apparently extending from the position of the continuum source, can be seen 
a few arcsec below the trace at $\sim -140$ and $-160$ km s$^{-1}$. These features 
are present in all three epochs and are most easily seen in [\ion{S}{2}]. 
The final knot that can be seen in all epochs is $\sim$2\arcsec\ above the stellar trace 
at $\sim -50$ km s$^{-1}$, and it too is most readily identifiable in [\ion{S}{2}]. 

Interestingly, in the 2011 June spectrum there is virtually no 
[\ion{N}{2}] emission from the PTF~10nvg continuum position, 
and for $\lambda$6583 in particular, it can be seen that each of the aforementioned knots
are connected spatially. This structure is clear evidence that PTF~10nvg is driving 
a jet with a terminal shock ending $\sim$4\arcsec\ from the star and maximum velocity 
$\sim$200 km s$^{-1}$. The appearance of these features at all epochs, 
regardless of the PA, suggests that PTF~10nvg is driving a wide-angle outflow. 

There are additional spatially offset features which are not observed in each epoch. 
The most prominent is a knot seen in the 2011 June spectrum at $\sim -20$ km s$^{-1}$ 
located $\sim$4\arcsec\ below the trace of PTF~10nvg and extending to the edge of 
the decker $\sim$7\arcsec\ from PTF~10nvg. This feature is most easily identified 
in [\ion{O}{1}] since it is blended with diffuse nebular emission in all the other panels. 
This feature is too far from PTF~10nvg to be detected in the 2010 December spectrum 
as the decker did not extend 4\arcsec\ below the PTF~10nvg continuum, while the same emission feature 
may be present in the 2011 December spectrum, though blending with nebular emission 
(or the [\ion{O}{1}] night sky line) make a positive identification challenging. Finally, 
the last prominent feature, which is seen only in the 2011 June observation, 
is a point source-like trace seen $\sim$7\arcsec\ below PTF~10nvg. This trace 
is seen in H$\alpha$, [\ion{S}{2}], and [\ion{N}{2}] $\lambda$ 6583 \AA. Optical imaging does not 
reveal any point sources within $\sim$8\arcsec\ of PTF~10nvg and so the origin of this 
emission is currently unclear. High-angular resolution narrow-band imaging of the field 
around PTF~10nvg should elucidate if this feature is from another star or somehow related to 
PTF~10nvg itself.

No spatially offset emission is apparent in the
NIRSPEC 1 $\mu$m high--dispersion spectral imaging data, 
e.g. in [\ion{N}{1}] or [\ion{S}{2}],
or in the 1.083 $\mu$m \ion{He}{1} line.
For these observations the slit length was 12" and thus able
to sample the relevant spatial range where  
offset emission is seen in optical forbidden lines,
as discussed above.

\section{Constraints on Multiplicity}\label{sec:mult}

\begin{figure}
\includegraphics[angle=0,width=0.90\hsize]{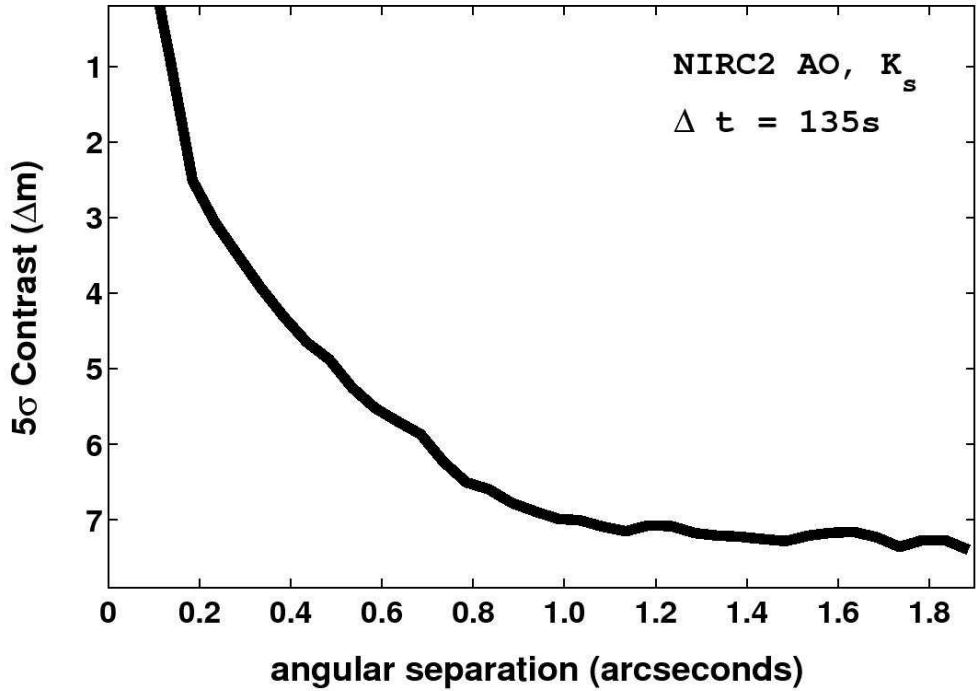}
\caption{Contrast versus angular separation curve for the field centered on PTF~10nvg.
A maximum contast of 7.5 mag is achieved in the K$_s$ band
beyond 1".0 separation, corresponding to
520-600 AU depending on the adopted source distance.  Detailed image analysis 
reveals that no stellar companion candidates are detected above these limits
though the observations can not rule out very low mass stellar or any
substellar companions. 
}
\label{fig:aocontrast}
\end{figure}

Neither the raw nor the reduced 
and coadded adaptive optics imaging frames from Keck/NIRC2
show evidence for point source companions within several arcseconds 
of PTF~10nvg. 
Figure~\ref{fig:aocontrast} shows the sensitivity of the observations 
to off-axis sources. 
The diffraction-limited data rule out the presence of companions down to
a level of $\Delta K_s = 3.2, 5.0, 7.0, 7.4$ mag at 0.25", 0.50", 1.0", and
2.0" ($5 \sigma$) respectively.
For comparison, the typical FWHM of our seeing-limited PTF
observations in the R-band and 
the beam size of our CARMA data at 2.7 mm are both 2.0\arcsec. 

With a primary source brightness of K$\approx$8.5 mag 
at the time of the AO observations,
the achieved contrast limits correspond to a lack of companions brighter than
K$\approx$16 mag outside of the optical and mm PSF.  At closer separations,
the flux limits can be translated into upper mass limits to any compansions 
of 1.14, 0.66, and 0.30 $M_\odot$ at separations of 130, 260, and 520 AU. 
These numbers use the isochrones of \citet{Girardi02} and 
assume an age of 1 Myr, a distance of 520 pc, but do not include a foreground
extinction estimate since this is unknown (but would raise the mass limits).
The companion upper mass limit sensitivity is not particularly deep despite
the large contrast achieved with Keck/NIRC2
since the 2 $\mu$m brightness of PTF~10nvg is dominated by
circumstellar excess rather than pure photosphere.

For reference, companions to several well-known FU Ori-type stars 
do reside within this separation and contrast range range:
at 196 AU and $\Delta K_s = 4.1$ mag for FU Ori \citep{Wang04,Pueyo12}
and at 115 AU and $\Delta K_s = 1.2$ mag for Z CMa \citep{Koresko91,Hinkley12}.  Although the companions have been claimed to have some relevance 
to the outbursting behavior of FU Ori stars, the ability for such
wide companions to instigate the inner disk instability is limited.
Companions are also invoked in several examples of presumed
disk-eclipsing young stars that, like PTF~10nvg, have long periods.   
Specifically, KH15D at 48 days \citep{Hamilton12},
WL4 at 131 days \citep{Plavchan08}, 
and YLW16A at 92 days \citep{Plavchan10} have extinction-like 
flux dips between a few tenths and several magnitudes in depth.  
Several of these stars have companions in the 25-50 AU region
that are hypothesized to induce warps in the disks.
Our existing AO observations are not sensitive to such close binaries.

\section{Discussion and Interpretation}\label{sec:disc}

\subsection{The Nature of the Large-Amplitude Variable PTF~10nvg}

In this section we provide broader context for the observations
of PTF~10nvg that we have described above, namely:
1) the large amplitude optical/near-infrared magnitude and color variations 
that are broadly consistent with expectations from extinction variations;
2) the semi-periodic nature of the multi-band lightcurves;
3) the respective correlation / anti-correlation of permitted / forbidden
atomic spectral line emission strength with continuum brightness;
4) the rare detection of molecular TiO/VO emission at optical wavelengths,
with corresponding molecular CO, H$_2$O and VO emission in the near-infrared;
5) the exhibition of strong P Cygni line profiles in several atomic
species that are sensitive to winds and outflows;
6) the lack of evidence for stellar multiplicity;
7) the Class I-type spectral energy distribution;
8) the spatially extended molecular outflow. 

Before the 2010 brightening that was detected by both 
Itagaki et al. and \citeauthor{Covey11}, 
the source discussed here as PTF~10nvg was
known as IRAS 20496+4354 but had not been well-studied.  
Although there was no previously reported near-infrared
or optical photometry for the object, \citet{Aspin11} and we in this paper have demonstrated
through investigation of archival images that the source has been sporadically
visible in several past observations.  The recent
photometric behavior included brightening on a several month
time scale followed by an overall fade that has been punctuated by semi-periodic
spikes in brightness over the past several years.

Both previous to and subsequent to its 2010 brightening, the source can be characterized as
a Class I (disk-plus-envelope) young stellar object.   
Spectroscopically, PTF~10nvg is
highly unusual for a Class I, or any other Class, young star.
In addition to the rare TiO/VO emission and more common
CO emission as well as permitted atomic emisison during bright 
photometric states, in its faint photometric state the source 
is dominated by forbidden emission lines, some of which are commonly seen
in young stars but others of which are rarely seen.

One interpretation is that the object could be in its initial
envelope-clearing stage such that the embedded optical/infrared point source
is in the process of becoming revealed.  The consistency of
both the photometric and spectroscopic behavior with variations 
in line-of-sight extinction supports this interpretation. Further, 
the molecular emission could be the result of heating and cooling
at high latitudes in the outflow cavity as stellar photons first reach
these areas of the circumstellar environment.
As an analogy, classical novae, which are also driven by accretion 
(on to a white dwarf of material from an unseen companion) frequently show  
``dust dips" in their outbursting light curves with several episodes of
fading and re-brightening ocurring on tens to hundreds of days time scales, similar to the situation we have found for PTF~10nvg.
Another interpretation is that, instead of repeated
geometric clearing of the line of sight to the central star
and subsequent obscuration of this line by dust, the source
is surrounded by dust that is being in situ
reformed in the nearby circumstellar environment and then either
destroyed or accelerated away though wind/outflow.  In this
scenario, an analogy to the deep occulting phenomenon 
of R Coronae Borealis stars \citep{Clayton12} could be appropriate.
The very large extinction variations, $\Delta A_V$ up to 30,
and the detected periodicity,
may be a challenge for this hypothesis, however.  The quasi-periodicity of the 
dust obscuration events in PTF~10nvg strongly suggest that the appropriate 
geometry is one of dense material along our line of sight to the
optical/near-infrared continuum, that is in orbit about the 
central star, and located approximately 
$0.7 ({M_\ast}/{M_\odot})^{1/3}$ AU from it.

The periodicity and the extinction-dominated nature of the
2011 and 2012 light curves do not, however extrapolate back to the
2010 (optical) data in a satisfactory manner, 
though the limited 2009 data do seem to phase well with
the derived period.  
During 2010 the source was much brighter than 
over the next two years, and in fall of 2012 the lightcurve again
reached a similar level of brightness 
(in a departure from the previously observed quasi-periodicity).  
These unusually bright periods perhaps indicate that at least some
of behavior is driven by accretion-related ``outburst" events. 
The recent outburst hypothesis may be further supported 
by the lack of detection in previous infrared imaging surveys,
notably 2MASS, but detection at not fainter than K$_s\approx$11 mag
over the past three years.
Brightening in the mid-infrared as well, specifically during 2010, suggests 
a bolometric luminosity increase; however, the scale of the mid-infrared 
brightening could plausibly be due to the inferred extinction changes. 
Dramatic photometric and spectroscopic changes in young stars are
often classified as outbursts and associated with large scale
(one or more orders of magnitude) increases
in the accretion rate of material from the disk on to the central star. 
Below we pursue the interpretation that the body of evidence 
in hand for PTF~10nvg is indeed indicative of an outbursting young star, 
though mediated by semi-periodic dust obscuration events.

Spectroscopically, PTF~10nvg bears strong resemblance during its brighter states 
in the atomic emission line pattern to the well-known extreme emission-line object 
V1331 Cyg.  The accompanying forbidden line emission in PTF~10nvg is explained if
the enhanced accretion episode of 2010 also drove an enhanced outflow, perhaps
with outflowing material reaching a shock-enabling surface approximately
one year post-outburst, when forbidden line prominence increased significantly.
However, given that we see a richer forbidden spectrum when
the source is in a faint state, and given the evidence for large variations
in extinction towards the central continuum source, another interpretation
is that it is only when the continuum (arising from the photosphere, accretion shock,
and inner disk regions) is suppressed that we are able
to see a much fainter, spatially distinct, but ever-present jet region.
This scenario requires an inclined disk to do the obscuring, but not
too inclined given the $>$200 km/s line widths of the [\ion{O}{1}],
[\ion{S}{2}], and [\ion{Fe}{2}] lines.
The situation would be similar to that advocated by \citet{WH04}: 
there is an orientation effect such that the spatially 
extended forbidden line emission is more readily observed
in systems without a direct view of the central continuum source.  This
explanation was prefered to invoking differences, or in our case variations in,
the mass outflow rate itself for explaining stronger forbidden line emission.
We note that \citet{Appenzeller05} also found enhanced line-to-continuum
ratios for forbidden lines in known edge-on disk systems relative to those
typically seen in accreting T Tauri stars.

Another peculiarity of PTF~10nvg is the rare observation of optical 
TiO/VO molecular bands in emission. 
 \citet{Hillenbrand12} reported several additional sources besides PTF~10nvg that have been observed to exhibit similar, broad TiO/VO emission at optical wavelengths.  
In common between the sources is the Class I-type spectral energy distribution and spectroscopic evidence for rapid accretion and strong outflow.
It is unclear at present whether an ``outburst" phase is required for detection of TiO/VO emission.
 Future study of such TiO/VO sources may help elucidate the relative molecular abundances in circumstellar disk/envelope environments, and perhaps the mechanisms for molecule formation and destruction. The TiO and VO species appear from the statistics to be not readily observable during more quiescent stages of disk evolution but notably they are present in planetary atmospheres and so must be present in the disks.  \citet{Banzatti12}, for example, have recently discussed in detail for EX Lup the H$_2$O and OH (increased emission during outburst), as well as several organic  (decreased emission during outburst) molecules during quiescence and in outburst. These authors discuss the photo-dissociative and photo-desorption effects of enhanced high energy radiation during an episode of enhanced accretion, and the ensuing broad implications for disk chemistry and structure evolution.  

In summary, a coherent physical model for PTF~10nvg would include
the following.  The optical TiO/VO and infrared CO emission comes
from the inner disk region. The broad permitted atomic lines
are formed in the accretion/outflow zone, with the narrower atomic
lines part of the rest-frame accretion shock.  The 2010 brightening
was indeed an ``outburst" event, driven by accretion, but
perhaps accompanied by dust clearing along the line-of-sight.
The photometric variability since 2010 is primarily due to large
variations in extinction, which periodically obscure the inner disk. 
This causes the dense gas tracers (molecular TiO/VO, CO and 
permitted atomic lines) to dissappear.  However, the low density tracers
(molecular H$_2$ and atomic forbidden lines) are coming from an
extended jet/outflow, which may include self-shocks. We can see
this faint emission only when the central continuum source is obscured.
The characteristic timescale of the extinction variations is few week
occulations, with $\sim$221 day periodicity.  If these events are linked
with an orbital timescales, it suggests that the occulting material 
is located at $a \approx 0.7 ({M_\ast}/{M_\odot})^{1/3}$ AU.
The departure from periodic behavior in late 2012
may be due to the dissipation of, or geometrical reconfiguration of,
the occulting material.  Alternately, it could be due to a 
recurring dominance of accretion variation behavior 
over extinction variation behavior, similar to the 2010
outburst that initially drew the attention of several groups worldwide to this source.

\subsection{PTF~10nvg in the Context of other Large-Amplitude Variable Young Stars}

In this section we discuss the characteristics 
of PTF~10nvg relative to those of other young stellar objects
that have exhibited outbursting behavior, finding some common elements
with known classes, but also some unique aspects of PTF~10nvg.  
We advocate the position that our knowledge of the diversity
of time-variable behavior in young stars is vastly incomplete,
as is our understanding of the underlying physical mechanisms.
Although FU Ori stars are certainly a distinguishable class once 
the candidates are suitably characterized, other variable categories 
such as EX Lup-type objects and V1647 Ori-type objects appear to
contain a range of behaviors among their members.  While all of these 
variable types are likely explained by time-variable accretion
and related outflow, young star ``eruptions" may manifest themselves 
in very different ways with the currently identified categories 
including somewhat heterogeneous objects at present.

The lightcurve of PTF~10nvg is dissimilar to those 
of other categories of young star outbursts.  Beyond the initial rise, 
the light curve is very different from those of FU Ori stars, which 
remain at or near the elevated brightness for decades.  The FU Ori stars
are inferred to be undergoing significantly enhanced accretion at rates
of roughly $10^{-4}$ or $10^{-5}$ \msun\pyr\ \citep{Zhu10}.  
The PTF~10nvg light curve is also dissimilar 
to the lightcurves of EX Lup-type objects, which have smaller 
amplitude rises and a clearly distinct quiescent vs outbursting state.   
The EX Lup star accretion rates are inferred to be lower than those
advocated for FU stars; the unusually large 2008 outburst of EX Lup 
itself was interpreted as an accretion enhancement from 
roughly $10^{-9}$\msun\pyr\ to roughly $10^{-7}$ \msun\pyr\
(Sipos et al. 2009; Aspin et al. 2010; Juhasz et al. 2011). 
The lightcurve for this outburst event (see e.g. Figure 1 of
Juhasz et al. 2012) indeed bears some similarity to the 
photometric behavior of PTF~10nvg, both in terms of the amplitude of the 
initial rapid rise and the multiple photometric peaks occuring
over the next 1/2 to 3/4 of a year before the source faded 
to quiescent levels. As described here, PTF~10nvg also exhibits
quasi-periodic brightness peaks at $\sim$7 month-long intervals.
However, the outbursting phase with superposed
large exinction variations has lasted almost 3 years now, 
longer than the events typically associated with EX Lup outbursts.
It remains to be seen whether a true quiescent stage 
will reveal itself in the future. 

The physical mechanism for the outbursting behavior in the several
classes of large amplitude variable young stars is thought to be related 
to an inner disk instability.  For FU Ori stars, thermal, gravitational,
 and magneto-rotational models have all been proposed
(Bonnell \& Bastien 1992; Bell \& Lin 1994; Clarke \& Syer 1996;
Kley \& Lin 1999; Armitage et al. 2001; Vorobyov \& Basu 2005;
Boley et al. 2006; Zhu et al. 2009b).   
For the EX Lup stars, the origin
is postulated to be essentially at the inner disk edge itself, 
and the ``instability" driven by a cyclic
interaction between the differentially rotating accretion disk and the
magnetosphere (roughly dipole in this region), which is rotating with the star.
Episodic and sometimes periodic accretion and outflow are naturally explained 
as a consequence of the oscillations in magnetic field topology  
\citep{GW99,Romanova05,Romanova09,DS10,DS12}.  The field structure alternates
between an expanded magnetospheric structure that diffuses into the disk and
leads eventually to field line opening and outflow.  Following reconnection, 
a more compact magnetospheric structure results and allows encroachment 
by the accumulated inner disk material.  
Enhanced accretion on to the star results when co-rotation is crossed, 
after which the magnetosphere again expands.
Alternately, the enhanced accretion could involve similar mechanisms
but operating in a more stochastic and rapid rather than a semi-cyclic manner
\citep[e.g.][]{Romanova08}.  In all of the above scenarios, however,
the predicted time scales are highly uncertain, ranging between several 
times the dynamical time scale and several orders of magnitude times 
the viscous time scale.
Thus it is not yet known if the postulated inner disk oscillations can be
associated with the interval between the young star outbursts that
have been observed.
At least qualitatively, there is some resemblance between the
accretion rate evolution during the outbursts described in e.g. Figure 4 of
\citet{DS10} and the light curve behavior of PTF~10nvg.  The same
interpretative conclusion was reached for 
the 2008 outburst of EX Lup by \citet{Juhasz12}.

\citeauthor{Covey11} made the analogy between the outburst behavior of 
PTF~10nvg and V1647 Ori.  This claim was before discovery of the
large amplitude near-infared variations as consistent with extinction
variations, and before discovery of the quasi-periodicity in the lightcurve.  
However, \citet{Aspin11} do claim that the near-infrared color
variations of V1647 Ori are consistent with variable extinction
(though they argue that those of PTF~10nvg are not consistent with
this interpretation, contrary to the evidence presented here).
Furthering the analogy, we have found a single outflow cavity in the
long wavelength molecular line observations of PTF~10nvg which could
also be argued for V1647 Ori based on the strong asymmetry in the 
scattered light nebula.

\citet{Lorenzetti12} have suggested based on the near-infrared
color differences between the quiescent and the outbursting states
of several large amplitude variables (including PTF~10nvg and V1647 Ori)
that the outbursts are characterized by an extra thermal component
that is not present during the quiescent stages.
In the case of PTF~10nvg, the extra radiation during the outburst
is estimated to have $\sim$1400 K and to be located at 0.3 AU 
(based on a fitted blackbody model), after accounting for 
extinction variations of 4.6 mag between a low state and a high state.
While our observations indicate that the observed photometric variations
(Figure~\ref{fig:color}) and also spectrophotometric variations
(considering a de-reddened version of Figure~\ref{fig:ircontinuum})
are roughly consistent with variations in extinction, there is some
stochasticity to the color evolution which could indeed necessitate underlying
changes in the source SED below the extinction variation which dominates.

If the \citet{Lorenzetti12} model is robust, the physical situation would be
that of changes in thermally emitting material located in the
inner region the disk.  Although \citet{Lorenzetti12} interpret this as
reprocessing of the radiation emitted on the stellar surface as
the accretion shock becomes hotter due to the enhanced accretion rate,
one can also draw analogy to the above theories invoking
inner disk instabilities wherein the inner disk heating is due to
viscous and dynamical process rather than radiative reprocessing.  
Notably, the $\sim$0.3 AU scale derived in the \citet{Lorenzetti12}
blackbody model is not so different from the 
$a \approx 0.7 ({M_\ast}/{M_\odot})^{1/3}$ AU scale that
was found above in our discussion of the semi-periodicity of the repeated
photometric peaks.  However, perhaps inconsistent between the
postulation of an extra, cool emission component and the instability theory,
is that there does not appear to be any lag between the infrared brightening
and the optical brightening in PTF~10nvg.  This is true for the initially
recorded outburst in 2010 as well as all subsequent brightenings
that we have observed from 2010 to 2012.
\citet{Lorenzetti12} do note that the blackbody model becomes less good 
of a fit as their EX Lup-type sources evolve beyond the initial 
photometric outburst towards quiescence.

\subsection{The Rate of PTF~10nvg-like events}

As detailed both here and in previous work \citep{Covey11,Aspin11}, 
PTF~10nvg does not belong to either the FU Ori-like or the EX Lup-like 
classes of YSO outbursts.  It may be an intermediate timescale and amplitude
outburst class, perhaps analogous to V1647~Ori as discussed in
\citeauthor{Covey11}.  Or it may be a more extreme version of the 
extinction-driven phenomena seen in UX~Ori or GM~Cep. We advocate that
both phenomena (accretion and extinction) play some role in the 
interpretation of PTF~10nvg photometry and spectroscopy.  
At present, estimates of the absolute rates of these types of large
amplitude events (whether accretion-driven, extinction-driven, or both) 
remain highly uncertain.  Nevertheless, we attempt a rough estimate 
from our ongoing photometric monitoring survey of the North America Nebula 
of the relative rate of PTF~10nvg-type phenomena. 

To our knowledge, the P48 survey
of the North America Nebula region is unique in both its depth and cadence.
During the past $\sim$3 yr the P48 has observed the same 
$\sim$7.3 deg$^2$ field with roughly a nightly 
cadence while it was visible from Palomar mountain to a depth of $\sim$20 mag per observation.
In this time there has been one FU Ori-like outburst (HBC 722/PTF~10qpf; 
\citealt{Semkov10,Miller11}) and one PTF~10nvg-like outburst event, suggesting that
the rate of the two phenomena is similar to within the accuracy afforded 
by low number statistics.  A more general discussion of smaller-amplitude but significant
outbursting and extinction dominated behavior among young members of
the North America Nebula will be presented by Findeisen \& Hillenbrand (2013, in preparation).

Overall, with $\sim$12 FU Ori-like eruptions 
observed to date \citep{Reipurth10} and only one PTF~10nvg-like event 
it is also possible that PTF~10nvg-like events are far more rare 
and only $\sim$one-tenth as common as FU Ori outbursts. 
With $\sim$10 FU Ori-like outbursts expected for each low mass star 
\citep{Hartmann96}, even this lower limit would suggest $\sim$1 PTF~10nvg-like 
outburst per low mass star.  Likewise, in the interpretation that this event
corresponds to the initial cloud clearing and source revelation, we would 
expect $\sim$1 event per star.  

As future surveys continue to study
star forming regions with high cadence observations over long time baselines, 
the importance of PTF~10nvg-like events in our global understanding 
of the low mass star formation process will be elucidated.  
History shows that astronomy is rich
with examples of new phenomena being discovered as surveys push to explore
undersampled regions of parameter space.  
 
\section{Summary and Conclusions}\label{sec:summ}

1. Continued photometric monitoring of PTF~10nvg 
by the P48 and PAIRITEL telescopes throughout 2011 and 2012 
detected magnitude variations of 
$\Delta$R $\sim$ 10 mag and $\Delta$K $\sim$ 3 mag, and color variations of  
$\Delta$J-K $\sim$ 3 mag.  The time series behavior 
in color-color and color-magnitude diagrams indicates
that these photometric variations are consistent with those expected
due to changes in line-of-sight extinction amounting to $\Delta$A$_V >$ 30 mag.

2. Over the long term, attributing PTF~10nvg's non-detection in 2MASS 
to similar extinction changes suggests the source has historically 
experienced a range of $\Delta$A$_V \sim$ 70 mag.  Alternately, the
non-detection in 2MASS may reflect a true outburst scenario for the
2010 brightening and subsequent photometric evolution.

3.  Time scales of several weeks can be associated with dramatic 
flare-like photometric changes and $\sim$7 months with the repeated maxima  
observed in the time series over the past several years. 
While the initial 2010 maxima were rounded/smooth, later maxima
in 2011 and early 2012 were characterized by narrow/sharp photometric peaks. 
A late-2012 maximum was again broad and rounded/smooth. 

4.  The Lomb-Scargle periodogram calculated from 
JHK$_s$ light curves between 2010 and mid-2012 features a prominent, 
statistically significant peak at $\sim$221 days.  
Under the assumption that the period is the Keplerian orbital
timescale, the dust obscuration governing PTF~10nvg's
photometric behavior is located at about 
$0.7 ({M_\ast}/{M_\odot})^{1/3}$ AU from the central star.
Continued monitoring will be essential to confirm the reality of the
periodic signal; the existing light curves span $<$1000 days in
total, and include several 100$+$ day gaps.  Further, 
the late-2012 lightcurve appears to depart from the previously derived periodic trend
and may reflect renewed dominance of accretion over extinction effects.

5. WISE detected PTF~10nvg over multiple epochs in 2010. Comparison of
these mid-infrared detections with prior detections at similar wavelengths by
IRAS, Spitzer, and MSX indicate that PTF~10nvg brightened in the mid-infrared
by factors of a few during its summer 2010 brightening.

6. CARMA observations reveal an extended millimeter continuum source 
having a total (dust plus gas) mass of 0.06 M$_\odot$.  CARMA also detects
spatially unresolved emission in both $^{12}$CO and $^{13}$CO centered on 
PTF~10nvg's near-infrared position and systemic velocity.  
Spatially extended, redshifted $^{12}$CO emission is 
detected to the south of PTF~10nvg, which we interpret as revealing the
presence of an outflow cavity.  Over the wider field, the molecular emission
is well-aligned with the H$\alpha$ emission arc.

7. The bright-state optical and near-infrared spectrum of PTF~10nvg 
is that of a ``continuum-plus-emission" object, similar to V1331~Cyg but 
with significant molecular emission contributions to the continuum.
The faint-state spectrum is that of a Herbig-Haro object.
Both neutral and singly ionized line species probing a range of
densities and temperatures are observed.  The lines that characterize 
PTF~10nvg's
faint-state spectrum are present at all epochs, however, suggesting
that the spectral evolution is primarily due to the supression of the
bright state spectral features during epochs of enhanced extinction.

8. Spectral monitoring reveals that the strong wind signatures 
detected in the 2010 outburst persisted through 2011 and 2012.  The depth
and terminal velocities of optical wind absorption features (i.e., the
Na I D and K I 7665/7699 doublets, the O I 7774 triplet) appeared
consistent -- or perhaps slightly increased over time in some lines 
-- across all epochs where the optical continuum was
sufficiently strong to enable these features to be measured.
In the near-infrared, \ion{He}{1} $\lambda$ 10830 
absorption increased in both depth and equivalent width by nearly a
factor of two from 2010 through 2011.  A strong redshifted emission component 
to \ion{He}{1} $\lambda$ 10830 
likely also traces the outflow rate of the inner wind.

9. The evolution of the strengths and profiles of permitted atomic
emission lines reveal the presence of time variability in PTF~10nvg's
line-of-sight accretion and outflow activity.  The source
exhibits optically thick Ca II triplet emission with line
ratios that are unusual even for young stars.  The H${\alpha}$
emission also demonstrates significant variations in both overall
strength and kinematic profile, with the line weakening and shifting
blueward as the source fades.  Strong redshifted
H${\alpha}$ emission disappeared in June 2011 as PTF~10nvg approached
its photometric minimum, while the blueshifted H${\alpha}$ emission
remained distinctly visible.  These observations suggest that
PTF~10nvg's \ion{Ca}{2} and \ion{H}{1} lines 
include contributions from both accretion and outflow activity. 
While the accretion component may dominate during the brightest
states, during fainter states only the outflow signature remains.
Notably absent from the spectra at any
epoch is the \ion{He}{1} $\lambda$ 5876 line, a high temperature
(T $>$ 15,000 K) tracer seen near-ubiquitously in classical T Tauri stars.

10. Forbidden line emission measuring $\sim$5000-20,0000 K gas
reveals the presence of a likely Herbig-Haro jet
with a wide range of densities, $n\sim10^3-10^7$ cm$^{-3}$.
Some line ratios such as [\ion{O}{1}] 6300 \AA\ : 6363 \AA\
and [\ion{Fe}{2}] 7155 \AA\ : 7172 \AA\ show extreme and unusually 
time-dependent ratios that are not easily explained in the standard
optically thin assumption for forbidden line emission.  Instead,
these lines may be partially optically thick.

10. Spatially extended blueshifted emission is visible in several
forbidden line species as well as H$\alpha$
in two-dimensional HIRES images.  These position-velocity diagrams provide
further evidence that PTF~10nvg is driving a jet with a line-of-sight
velocity of about $-200$ km/s and exciting a strong shock region
$\sim$4" projected distance from the source (2000AU at 520 pc). 

11. The moderate resolution near-infrared spectra obtained in June/July 2011,
when PTF~10nvg was near its faintest state, provide an opportunity to
study the properties of the jet and extended outflow.
PTF~10nvg's H$_2$ line strengths strongly resemble theoretical
predictions for C-type shocks and pure X-ray photo-excitation, as well
as empirical observations of H$_2$ emission from RW Aur and Knot B of
the HH54 protostellar jet; the same is true for the near-infrared [\ion{Fe}{2}]
emission.

12. From near-infrared adaptive optics observations we can exclude
the presence of stellar companions within several hundred AU of PTF~10nvg.

Overall, we conclude that
the recent time series data on PTF~10nvg will not be easy to interpret without 
a good deal of hindsight.  Further observational and theoretical
study of this enigmatic source is warranted. 

\facility{Facilities: Palomar 48", PAIRITEL, Lick 3m, Palomar 200" (two instruments), APO, Keck (four instruments), IRTF, CARMA, DSS, KPNO 0.9m, USNO, 2MASS, UKIDSS, Spitzer, WISE} 

\acknowledgements
{\it Acknowledgements:}
We commend PTF and PAIRITEL critical personnel Peter Nugent, Eran Ofek, and
Dan Starr, Cullen Blake for effective facility operation and data distribution.  
Mark Sullivan ran his PTF-specific PSF fitting code for us. 
In addition, we thank Joey Richards for helpful conversations and advice
about the lightcurve analysis.
We are also grateful to the many observers who kindly helped with
the acquisition and/or reduction of spectra that are reported 
either in this paper or in Covey et al. (2011), including 
Katie Hamren, Everett Schlwain,
Aaron Barth, Evan Kirby, Melissa Graham, Eric Hsiao,  Peter Blanchard, 
Antonio Cucchiara, Adam Morgan, Daniel Perley, Christopher Griffith, 
Michael Kandrashoff, Luisa Rebull, Wilson Liu, John Angione, 
Shriharsh Tendulkar,  Aaron Ofer, Kelsey Clubb, 
Greg Herczeg, Adam Kraus, Nick Law, Andrew Howard, Dan DeFelippis, 
Kunal Mooley, Trevor David, Geoff Marcy, Howard Isaacson.
CARMA Summer School students Isaac Shivvers, Che-Yu Chen, and Aaron Juarez
participated in some of the CARMA observations and data reduction
described herein. 
We are also very grateful for the assistance of staff members 
at the observatories where data were obtained. 
Funding acknowledgements are as follows.
1. K.R.C. acknowledges support for this work from the Hubble Fellowship
Program, provided by NASA through Hubble Fellowship grant
HST-HF-51253.01-A awarded by the STScI, which is operated by the AURA,
Inc., for NASA, under contract NAS 5-26555.
2. A.V.F.'s group at UC Berkeley acknowledges generous
financial assistance from Gary and Cynthia Bengier, the
Christopher R. Redlich Fund, the Richard and Rhoda Goldman Fund, 
the TABASGO Foundation, 
and US National Science Foundation (NSF) grant AST-0908886.
3. PAIRITEL has been supported by a Swift Guest Investigator grant NASA/NNX12AE67G.
4. Some of the data presented herein were obtained at the W. M. Keck
Observatory, which is operated as a scientific partnership among the
California Institute of Technology, the University of California, and
NASA; the observatory was made possible by the generous financial
support of the W. M. Keck Foundation.
5. The Infrared Telescope Facility is
operated by the University of Hawaii under Cooperative Agreement no.
NNX-08AE38A with the National Aeronautics and Space Administration,
Science Mission Directorate, Planetary Astronomy Program
6. Support for CARMA construction was derived from the Gordon and Betty Moore
Foundation, the Kenneth T. and Eileen L. Norris Foundation, the James S.
McDonnell Foundation, the Associates of the California Institute of Technology,
the University of Chicago, the states of California, Illinois, and Maryland,
and the National Science Foundation. Ongoing CARMA development and operations
are supported by the National Science Foundation under a cooperative agreement
(grant number AST-0838260) and by the CARMA partner universities.

\clearpage

\begin{deluxetable}{lcc}
\tablewidth{0pt}
\tabletypesize{\tiny}
\tablecaption{P48 $R$-Band Photometry of PTF~10nvg\tablenotemark{1}\label{tab:P48}}
\tablehead{
\colhead{Epoch (MJD)}  & 
\colhead{R$_{PTF}$ (mag)} & 
\colhead{err} \\ 
}
\startdata
55056.221 & 17.831 & 0.019 \\
55056.295 & 17.772 & 0.020 \\
55059.262 & 17.328 & 0.017 \\
55059.357 & 17.380 & 0.017 \\
55061.316 & 17.520 & 0.023 \\
\nodata & \nodata & \nodata \\
\nodata & \nodata & \nodata \\
\nodata & \nodata & \nodata \\
\nodata & \nodata & \nodata \\
\nodata & \nodata & \nodata \\
\nodata & \nodata & \nodata \\
\nodata & \nodata & \nodata \\
\nodata & \nodata & \nodata \\
\enddata
\tablenotetext{1}{ The table will appear in its entirety in the electronic edition of the journal.}
\end{deluxetable}

\begin{deluxetable}{lcccc}
\tablewidth{0pt}
\tabletypesize{\tiny}
\tablecaption{Stacked R$_{PTF}$ Photometry Representing Faint Epochs\label{tab:P48-stacked}}
\tablehead{
\colhead{Mean Epoch (MJD)}  & 
\colhead{R$_{PTF}$ (mag)} & 
\colhead{err} &
\colhead{Start of Stack (MJD)} &
\colhead{End of Stack (MJD)} \\ 
}
\startdata
55729.67 & 22.7 & 0.7 & 55721.32 & 55735.46 \\
55763.68 & 22.6 & 0.2 & 55750.18 & 55775.31 \\
55788.93 & 21.0 & 0.2 & 55781.21 & 55794.51 \\
55999.68 & 21.8 & 0.4 & 55922.10 & 56041.48 \\
\enddata
\end{deluxetable}

\begin{deluxetable}{lcccccc}
\tablewidth{0pt}
\tabletypesize{\tiny}
\tablecaption{PAIRITEL JHK-Band Photometry of PTF~10nvg\tablenotemark{1}\label{tab:PTEL}}
\tablehead{
\colhead{Epoch (MJD)}  & 
\colhead{$J$ mag} & 
\colhead{$J$ err} & 
\colhead{$H$ mag} & 
\colhead{$H$ err} & 
\colhead{$K$ mag} & 
\colhead{$K$ err} }
\startdata
55387.376 & 12.229 & 0.025 & 10.232 & 0.030 &  8.456\tablenotemark{s} & 0.060\tablenotemark{s} \\
55392.294 & 12.231 & 0.026 & 10.231 & 0.030 &  8.475\tablenotemark{s} & 0.063\tablenotemark{s} \\
55450.191 & 10.896 & 0.025 &  9.414\tablenotemark{s} & 0.031\tablenotemark{s} &  8.029\tablenotemark{s} & 0.060\tablenotemark{s} \\
55463.151 & 11.538 & 0.025 &  9.801 & 0.030 &  8.303\tablenotemark{s} & 0.060\tablenotemark{s} \\
55466.126 & 11.585 & 0.025 &  9.966 & 0.030 &  8.347\tablenotemark{s} & 0.060\tablenotemark{s} \\
\nodata & \nodata & \nodata & \nodata & \nodata & \nodata & \nodata \\
\nodata & \nodata & \nodata & \nodata & \nodata & \nodata & \nodata \\
\nodata & \nodata & \nodata & \nodata & \nodata & \nodata & \nodata \\
\nodata & \nodata & \nodata & \nodata & \nodata & \nodata & \nodata \\
\nodata & \nodata & \nodata & \nodata & \nodata & \nodata & \nodata \\
\nodata & \nodata & \nodata & \nodata & \nodata & \nodata & \nodata \\
\nodata & \nodata & \nodata & \nodata & \nodata & \nodata & \nodata \\
\nodata & \nodata & \nodata & \nodata & \nodata & \nodata & \nodata \\
\enddata
\tablenotetext{1}{The table will appear in its entirety in the electronic edition of the journal.}
\tablenotetext{s}{Photometry measured on short read PAIRITEL images (see text).}
\tablenotetext{n}{No reliable photometric measurements are available due to large thermal backgrounds (see text).}
\end{deluxetable}

\clearpage

\begin{deluxetable}{lccccc}
\tabletypesize{\tiny}
\label{tab:h2table}
\rotate
\tablewidth{0pt}
\tabletypesize{\tiny}
\tablecaption{Summary of Optical Spectra \label{tab:opt_spec_summary}}
\tablehead{
\colhead{UT} &
\colhead{Wavelength} &
\colhead{Spectral} &
\colhead{Telescope/} &
\colhead{Slit} &
\colhead{} \\ 
\colhead{UT Date} &
\colhead{Range (\AA)} &
\colhead{Resolution} &
\colhead{Instrument} &
\colhead{PA} &
\colhead{Observers} }
\startdata
\\
2010 July 8  & 3010--10200 & 4/7 \AA & Keck/LRIS-Blue+Red & 113.0&  Bloom, Cucchiara, Morgan, Perley \\  2010 July 19  & 3452--10800 & 4/10 \AA  & Lick/Kast-Blue+Red & 117.6&  Griffith, Kandrashoff \\  2010 August 12  & 3400--5500 / 6300--8800 & 2/5 \AA & Palomar 200\"/DoubleSpec-Blue$+$Red & 268.0&  Rebull \\ 2010 September 16  & 3500--10000 & 4/10 \AA  & Lick/Kast-Blue+Red &186.4 &  Cenko \\  \hline
2010 December 5  & 6440--7105 & 2.6  \AA & Palomar 200\"/DoubleSpec-Red & 97.0& Hillenbrand, Mooley \\ 
2010 December 13 & 4310--8770 &HIRES  & Keck/HIRES         & 110.8&  Hillenbrand, Kraus, Law\\
2011 May 3  & 3440--10700 & 4/10 \AA  & Lick/Kast-Blue+Red &90.8 & Barth \\
2011 June 3  & 3370--10200 & 4/10 \AA  & Keck/LRIS-Blue+Red & 166.0& Silverman, Filippenko, Cenko\\   2011 June 15  & 3645--7975 &HIRES & Keck/HIRES         & 176.0 &  Hillenbrand \\
2011 June 28  & 4315--8775 &HIRES & Keck/HIRES         & 16.9 & Kraus, Law \\
2011 June 29  & 3324--10150 & 4/10 \AA  & Keck/LRIS-Blue+Red & 124.0 &  Silverman, Cenko, Nugent \\ 
2011 July 4  & 6165--6832 & 2.6 \AA & Palomar 200\"/DoubleSpec-Red & 198.0&  Kirby \\  2011 July 5  & 3500--9816  & 4/10 \AA & Lick/Kast-Blue+Red & 132.8 & Graham, Hsiao \\   2011 August 1  & 3298--10200 & 4/10 \AA  & Keck/LRIS-Blue+Red  &125.0& Silverman, Cenko \\ 
2011 August 2  & 3276--10200 & 4/10 \AA  & Keck/LRIS-Blue+Red &125.0 &  Silverman, Cenko \\ 
2011 August 28  & 3184--10200 & 4/10 \AA  & Keck/LRIS-Blue+Red & 107.0&  Silverman, Cenko, Miller \\ 
2011 August 31  & 3300--5200 / 6100--8250 & 8/10 \AA  & Palomar 200\"/DoubleSpec-Blue$+$Red  & 210.0&  Liu \\  2011 October 25& 3432--10246  & 4/10 \AA  & Lick/Kast &133.6  & Kandrashoff, Blanchard, Silverman\\
2011 October 30 & 3500--10185 & 4/10 \AA & Palomar 200\"/DoubleSpec-Blue$+$Red & unk.&  Tendulkar, Ofer\\  2011 November 26 & 4350--10260 & 4/6 \AA  & Keck/LRIS-Blue+Red & 93.6 &  Silverman, Clubb\\ 
2011 December 9 & 4775--9220 &HIRES & Keck/HIRES         &  102.6 & Hillenbrand \\
2012 January 1  & 4800--11000 \AA & 6 \AA & Palomar 200\"/DoubleSpec-Red & 82.0&  Hillenbrand\\  2012 January 2  & 4800--11000 \AA& 6 \AA & Palomar 200\"/DoubleSpec-Red & 81.&  Hillenbrand\\
2012 January 6 & 4420--8770&HIRES & Keck/HIRES         & 86.6 &  Kraus, Law, Hillenbrand\\  2012 May 17 & 3230--10200 & 4/6 \AA  & Keck/LRIS-Blue+Red & 220.0 &  Clubb, Silverman, Cenko, Filippenko, Miller\\ 
2012 July 11& 3460--10395  & 4/10 \AA  & Lick/Kast-Blue+Red &133 & Clubb, Silverman\\
 2012 July 28& 4000--6000 /7750-9250 & 4/10 \AA & Palomar 200\"/DoubleSpec-Blue & 263.0&  Hillenbrand, David\\ 2012 August 7 & 3640--7985 &HIRES & Keck/HIRES         &  91.3 & Marcy, Isaacson \\
\enddata
\tablecomments{Divider separates spectra published in Covey et al. (2011)
from those newly presented here.}
\end{deluxetable}

\clearpage

\begin{deluxetable}{lcrcc}
\rotate
\tablewidth{0pt}
\tabletypesize{\tiny}
\tablecaption{Summary of Near-Infrared Spectra \label{tab:nir_spec_summary}}
\tablehead{
\colhead{UT} &
\colhead{Wavelength} &
\colhead{Spectral} &
\colhead{Telescope/} &
\colhead{} \\
\colhead{UT Date} &
\colhead{Range ($\mu$m)} &
\colhead{Resolution} &
\colhead{Instrument} &
\colhead{Observers} }
\startdata
\\
2010 July 14  & 0.8-2.5 & R $\sim$ 2000  & IRTF/SpeX & Rayner \\
2010 July 14  & 3-4.2, 4.5-5.0 & R $\sim$ 2500  & IRTF/SpeX & Rayner \\
2010 July 16 & 0.95-1.12 & R$\sim$25000 & Keck/NIRSPEC & Hillenbrand \\
2010 July 18  & 1-2.5 & R $\sim$ 5000 & APO/Triplespec & Covey \\
\hline
2010 September 23  & 1-2.5 & R $\sim$ 2700 & Palomar/TripleSpec & Covey\\
2010 November 26  & 1-2.5 & R $\sim$ 2700 & Palomar/TripleSpec & Liu, Angione \\
2010 December 14  & 1-2.5 & R $\sim$ 2700 & Palomar/TripleSpec & Muirhead \\
2010 December 15  & 1-2.5 & R $\sim$ 2700 & Palomar/TripleSpec & Muirhead \\
2011 June 26  & 0.8-2.5 & R $\sim$ 1200  & IRTF/SpeX & Covey \\
2011 July 15  & 0.8-2.5 & R $\sim$ 1200  & IRTF/SpeX & Covey \\
2011 August 17  & 1-2.5 & R $\sim$ 2700 & Palomar/TripleSpec & Muirhead \\
2011 September 2  & 0.8-2.5 & R $\sim$ 1200  & IRTF/SpeX & Covey \\
2011 September 15 & 0.95-1.12 & R$\sim$25000 & Keck/NIRSPEC & Hillenbrand \\
2011 October 18  & 0.8-2.5 & R $\sim$ 2000  & IRTF/SpeX & Covey \\
2012 May 5  & 0.95-1.12 & R$\sim$25000 & Keck/NIRSPEC & Herczeg, Gong\\
\enddata
\tablecomments{Divider separates spectra published in Covey et al. (2011)
from those newly presented here.}
\end{deluxetable}

\clearpage

\begin{deluxetable}{lcc}
\label{tab:feiitable}
\tablewidth{0pt}
\tabletypesize{\tiny}
\tablecaption{PTF~10nvg [Fe II] Line Fluxes  \label{tab:FeII}}
\tablehead{
\colhead{Line ID} &
\colhead{PTF 10nvg} & 
\colhead{G06 HH54} \\ }
\startdata
1.257 $\mu$m & 4.5 & 8.7 \\
1.644 $\mu$m & 5.06 & 8.2 \\
\\ \tableline \\
1.257 / 1.644 & 0.89 & 1.06 \\
\enddata
\end{deluxetable}

\begin{deluxetable}{lcccccccccc}
\tablewidth{0pt}
\tabletypesize{\tiny}
\tablecaption{PTF~10nvg H$_2$ Line Fluxes  \label{tab:H2}}
\tablehead{
\colhead{Line ID} &
\colhead{PTF 10nvg} & 
\colhead{B08 RW Aur (spatially extended)} & 
\colhead{S95 C Shock} &
\colhead{S95 J Shock} &
\colhead{G06 HH54} &
\colhead{N07 X-ray (10 $\mu$m)} &
\colhead{N07 UV (10 $\mu$m)} &
\colhead{N07 X+UV (10 $\mu$m)} &
\colhead{N07 X+UV (1 mm)} &
\colhead{N07 X+UV (10 cm)} }
\startdata
1.958 $\mu$m [1-0 S(3)] & 1.8 & \nodata & 2.68 & 0.0341 & 15.4 & 1.326 & 8.096 & 14.475 & 1.472 & 0.079 \\
2.034 $\mu$m [1-0 S(2)] & 0.75 & 0.25 & 1.06 & 0.011 & 4.2 & 0.533 & 3.857 & 6.477 & 0.719 & 0.041 \\
2.073 $\mu$m [2-1 S(3)] & 0.192 & 0.12 & 0.139 & 0.00868 & 1.63 & \nodata & \nodata & \nodata & \nodata & \nodata \\
2.123 $\mu$m [1-0 S(1)] & 2.045 & 1.0 & 3.04 & 0.0266 & 9.59 & 1.565 & 12.968 & 20.725 & 2.526 & 0.142 \\
2.154 $\mu$m [2-1 S(2)] & 0.041 & $<$0.09 & 0.0539 & 0.0027 & 0.51 & \nodata & \nodata & \nodata & \nodata & \nodata \\
2.222 $\mu$m [1-0 S(0)] & 0.476 & 0.17 & 0.679 & 0.00526 & 2.58 & 0.357 & 3.256 & 5.023 & 0.667 & 0.04 \\
2.248 $\mu$m [2-1 S(1)] & 0.206 & 0.11 & 0.151 & 0.00635 & 1.00 & 0.09 & 0.242 & 0.583 & 0.068 & 0.022 \\
2.407 $\mu$m [1-0 Q(1)] & 1.63 & 0.76 & 2.34 & 0.0167 & 10.9 & 1.27 & 12.299 & 18.528 & 2.707 & 0.355 \\
2.413 $\mu$m [1-0 Q(2)] & 0.557 & 0.25 & 0.749 & 0.0058 & 4.5 & 0.395 & 3.604 & 5.56 & 0.748 & 0.066 \\
\\ \tableline \\
2.248 / 2.123 & 0.10 & 0.11& 0.05 & 0.24 & 0.10 & 0.06 & 0.02 & 0.03 & 0.03 & 0.15 \\ 
2.034 / 2.222 & 1.58 & 1.52 & 1.56 & 2.09 & 1.63 & 1.49 & 1.18 & 1.29 & 1.08 & 1.03 \\
2.248 / 2.073 & 1.07 & 1.09 & 1.09 & 0.73 & 0.61 & \nodata & \nodata & \nodata & \nodata & \nodata \\
2.123 / 2.407 & 1.25 & 1.38 & 1.30 & 1.59 & 0.88 & 1.23 & 1.05 & 1.12 & 0.93 & 0.40 \\
\enddata
\end{deluxetable}


\clearpage

\begin{thebibliography}


\bibitem[Appenzeller et 
al.(2005)]{Appenzeller05} Appenzeller, I., Bertout, C., \& Stahl, O.\ 2005, \aap, 434, 1005 

\bibitem[{Armitage} et~al.(2001){Armitage}, {Livio}, \& {Pringle}]{armitage01}
{Armitage}, P.~J., {Livio}, M., \& {Pringle}, J.~E., 2001,
  \mnras\/, 324, 705--711.


\bibitem[Aspin et al.(2009)]{Aspin09} Aspin, C., Reipurth, B., 
Beck, T.~L., et al.\ 2009, \apjl, 692, L67 

\bibitem[Aspin(2011)]{Aspin11} Aspin, C.\ 2011, \aj, 141, 196 

\bibitem[Banzatti et al.(2012)]{Banzatti12} Banzatti, A., Meyer, 
M.~R., Bruderer, S., et al.\ 2012, \apj, 745, 90 

\bibitem[Bautista 
\& Pradhan(1998)]{BP98} Bautista, M.~A., \& Pradhan, A.~K.\ 1998, \apj, 492, 650 


\bibitem[Bautista et al.(1996)]{Bautista96} Bautista, M.~A., Peng, 
J., \& Pradhan, A.~K.\ 1996, \apj, 460, 372 


\bibitem[Beck et al.(2008)]{Beck2008} Beck, T.~L., McGregor,
P.~J., Takami, M., \& Pyo, T.-S.\ 2008, \apj, 676, 472

\bibitem[Beckwith et al.(1990)]{Beckwith1990} Beckwith, S.~V.~W., 
Sargent, A.~I., Chini, R.~S., \& Guesten, R.\ 1990, \aj, 99, 924 

\bibitem[{Bell} \& {Lin}(1994)]{bell94}
{Bell}, K.~R., \& {Lin}, D.~N.~C., 1994 \apj\/, 427, 987--1004.


\bibitem[Bertin 
\& Arnouts(1996)]{Bertin96} Bertin, E., \& Arnouts, S.\ 1996, \aaps, 117, 393 


\bibitem[Blake et al.(2008)]{Blake08} Blake, C.~H., Bloom, 
J.~S., Latham, D.~W., et al.\ 2008, \pasp, 120, 860 

\bibitem[{{Bloom} {et~al.}(2006){Bloom06}, {Starr}, {Blake}, {Skrutskie}, \&
  {Falco}}]{Bloom06}
{Bloom}, J.~S., {Starr}, D.~L., {Blake}, C.~H., {Skrutskie}, M.~F., \& {Falco},
  E.~E. 2006, in Astronomical Society of the Pacific Conference Series, Vol.
  351, Astronomical Data Analysis Software and Systems XV, ed. {C.~Gabriel,
  C.~Arviset, D.~Ponz, \& S.~Enrique}, 751


\bibitem[Bloom et al.(2009)]{Bloom09} Bloom, J.~S., Perley, 
D.~A., Li, W., et al.\ 2009, \apj, 691, 723 


\bibitem[Bloom et al.(2011)]{Bloom11} Bloom, J.~S., Richards, 
J.~W., Nugent, P.~E., et al.\ 2011, arXiv:1106.5491 

\bibitem[{Boley} et~al.(2006){Boley}, {Mej{\'{\i}}a}, {Durisen}, {Cai},
  {Pickett}, \& {D'Alessio}]{boley06}
  {Boley}, A.~C., {Mej{\'{\i}}a}, A.~C., {Durisen}, R.~H., {Cai}, K., {Pickett},
    M.~K., \& {D'Alessio}, P., 2006, \apj\/, 651, 517--534.


\bibitem[{Bonnell} \& {Bastien}(1992)]{Bonnell92}
{Bonnell}, I., \& {Bastien}, P., 1992, \apjl\/, 401, L31--L34.


\bibitem[Brugel et al.(1981)]{BBM81} Brugel, E.~W., Boehm, 
K.~H., \& Mannery, E.\ 1981, \apjs, 47, 117 


\bibitem[Carpenter(2001)]{Carpenter01} Carpenter, J.~M.\ 2001, \aj, 121, 2851 

\bibitem[Chen et al.(2012)]{Chen12} Chen, W.~P., Hu, S.~C.-L., 
Errmann, R., et al.\ 2012, \apj, 751, 118 

\bibitem[{Clarke} \& {Syer}(1996)]{clarke96}
{Clarke}, C.~J., \& {Syer}, D., 1996 \mnras\/,  278, L23--L27.

\bibitem[Clayton(2012)]{Clayton12} Clayton, G.~C.\ 2012, arXiv:1206.3448 

\bibitem[Cody 
\& Hillenbrand(2010)]{CodyHillenbrand10} Cody, A.~M., \& Hillenbrand, L.~A.\ 2010, \apjs, 191, 389 


\bibitem[Cody 
\& Hillenbrand(2011)]{CodyHillenbrand11} Cody, A.~M., \& Hillenbrand, L.~A.\ 2011, \apj, 741, 9 


\bibitem[Connelley
\& Greene(2010)]{Connelley10} Connelley, M.~S., \& Greene, T.~P.\
2010, \aj, 140, 1214


\bibitem[Covey et al.(2011)]{Covey11} Covey, K.~R., 
Hillenbrand, L.~A., Miller, A.~A., et al.\ 2011, \aj, 141, 40 


\bibitem[D'Angelo 
\& Spruit(2010)]{DS10} D'Angelo, C.~R., \& Spruit, H.~C.\ 2010, \mnras, 406, 1208 

\bibitem[D'Angelo 
\& Spruit(2012)]{DS12} D'Angelo, C.~R., \& Spruit, H.~C.\ 2012, \mnras, 420, 416 
\bibitem[Doppmann et al.(2005)]{Doppmann2005} Doppmann, G.~W.,
Greene, T.~P., Covey, K.~R., \& Lada, C.~J.\ 2005, \aj, 130, 1145


\bibitem[Dougados et al.(2010)]{Dougados10} Dougados, C., 
Bacciotti, F., Cabrit, S., 
\& Nisini, B.\ 2010, Lecture Notes in Physics, Berlin Springer Verlag, 793, 213 


\bibitem[Edwards et al.(2003)]{Edwards03} Edwards, S., Fischer, 
W., Kwan, J., Hillenbrand, L., \& Dupree, A.~K.\ 2003, \apjl, 599, L41 


\bibitem[Edwards et al.(2006)]{Edwards06} Edwards, S., Fischer, 
W., Hillenbrand, L., \& Kwan, J.\ 2006, \apj, 646, 319 

\bibitem[Espaillat et al.(2011)]{2011ApJ...728...49E} Espaillat, C., 
Furlan, E., D'Alessio, P., et al.\ 2011, \apj, 728, 49 


\bibitem[Favata et
al.(2006)]{Favata2006} Favata, F., Bonito, R., Micela, G., et al.\
2006, \aap, 450, L17

\bibitem[Fitzpatrick(1999)]{Fitzpatrick99} Fitzpatrick, E.~L.\ 1999, \pasp, 111, 63


\bibitem[Flaherty et al.(2012)]{2012ApJ...748...71F} Flaherty, K.~M., 
Muzerolle, J., Rieke, G., et al.\ 2012, \apj, 748, 71 

\bibitem[Giannini et
al.(2006)]{Giannini2006} Giannini, T., McCoey, C., Nisini, B., et al.\
2006, \aap, 459, 821


\bibitem[Girardi et 
al.(2002)]{Girardi02} Girardi, L., Bertelli, G., Bressan, A., et al.\ 2002, \aap, 391, 195 


\bibitem[Goodson 
\& Winglee(1999)]{GW99} Goodson, A.~P., \& Winglee, R.~M.\ 1999, \apj, 524, 159 

\bibitem[Greene et al.(2010)]{Greene10} Greene, T.~P., Barsony,
M., \& Weintraub, D.~A.\ 2010, \apj, 725, 1100


\bibitem[Hamann 
\& Persson(1989)]{HP89} Hamann, F., \& Persson, S.~E.\ 1989, \apj, 339, 1078 

\bibitem[Hamilton et al.(2012)]{Hamilton12} Hamilton, C.~M., 
Johns-Krull, C.~M., Mundt, R., Herbst, W., 
\& Winn, J.~N.\ 2012, \apj, 751, 147 

\bibitem[Hartigan et al.(2000)]{Hartigan00} Hartigan, P., Bally, 
J., Reipurth, B., \& Morse, J.~A.\ 2000, Protostars and Planets IV, 841 

\bibitem[Hartmann 
\& Kenyon(1996)]{Hartmann96} Hartmann, L., \& Kenyon, S.~J.\ 1996, \araa, 34, 207 


\bibitem[Herter et al.(2008)]{Herter08} Herter, T.~L., 
Henderson, C.~P., Wilson, J.~C., et al.\ 2008, \procspie, 7014,  


\bibitem[Hillenbrand(1997)]{Hillenbrand97} Hillenbrand, L.~A.\ 1997, 
\aj, 113, 1733 

\bibitem[Hillenbrand et al.(2012)]{Hillenbrand12} Hillenbrand, L.~A., 
Knapp, G.~R., Padgett, D.~L., Rebull, L.~M., 
\& McGehee, P.~M.\ 2012, \aj, 143, 37 

\bibitem[Hinkley et al.(2012)]{Hinkley12} Hinkley, S., Rice, E., Hillenbrand, L., et al., 2012, \apj, submitted.

\bibitem[Isella et al.(2012)]{2012ApJ...747..136I} Isella, A., P{\'e}rez, 
L.~M., \& Carpenter, J.~M.\ 2012, \apj, 747, 136 

\bibitem[Koresko et al.(1991)]{Koresko91} Koresko, C.~D., 
Beckwith, S.~V.~W., Ghez, A.~M., Matthews, K., 
\& Neugebauer, G.\ 1991, \aj, 102, 2073 

\bibitem[Kurosawa 
\& Romanova(2012)]{KR12} Kurosawa, R., \& Romanova, M.~M.\ 2012, arXiv:1203.5154 
\bibitem[Kwan 
\& Fischer(2011)]{KF11} Kwan, J., \& Fischer, W.\ 2011, \mnras, 411, 2383 


\bibitem[Indebetouw et al.(2005)]{Indebetouw05} Indebetouw, R., 
Mathis, J.~S., Babler, B.~L., et al.\ 2005, \apj, 619, 931 

\bibitem[J{\o}rgensen et 
al.(2004)]{Jorgensen2004} J{\o}rgensen, J.~K., Hogerheijde, M.~R., van Dishoeck, E.~F., Blake, G.~A., \& Sch{\"o}ier, F.~L.\ 2004, \aap, 413, 993 

\bibitem[Juh{\'a}sz et al.(2012)]{Juhasz12} Juh{\'a}sz, A., 
Dullemond, C.~P., van Boekel, R., et al.\ 2012, \apj, 744, 118 

\bibitem[{Kley} \& {Lin}(1999)]{kley99}
{Kley}, W., \& {Lin}, D.~N.~C., 1999; \apj\/, 518, 833--847.

\bibitem[K{\'o}sp{\'a}l et 
al.(2011)]{Kospal11} K{\'o}sp{\'a}l, {\'A}., {\'A}brah{\'a}m, P., Acosta-Pulido, J.~A., et al.\ 2011, \aap, 527, A133 


\bibitem[Lada(1987)]{Lada87} Lada, C.~J.\ 1987, Star Forming Regions, 115, 1 

\bibitem[{{Law} {et~al.}(2009){Law}, {Kulkarni}, {Dekany}, {Ofek}, {Quimby},
  {Nugent}, {Surace}, {Grillmair}, {Bloom}, {Kasliwal}, {Bildsten}, {Brown},
  {Cenko}, {Ciardi}, {Croner}, {Djorgovski}, {van Eyken}, {Filippenko}, {Fox},
  {Gal-Yam}, {Hale}, {Hamam}, {Helou}, {Henning}, {Howell}, {Jacobsen},
  {Laher}, {Mattingly}, {McKenna}, {Pickles}, {Poznanski}, {Rahmer}, {Rau},
  {Rosing}, {Shara}, {Smith}, {Starr}, {Sullivan}, {Velur}, {Walters}, \&
  {Zolkower}}]{Law09}
{Law}, N.~M. {et~al.} 2009, \pasp, 121, 1395

\bibitem[Li 
\& McCray(1992)]{LiMcCray92} Li, H., \& McCray, R.\ 1992, \apj, 387, 309 


\bibitem[Lomb(1976)]{Lomb76} Lomb, N.~R.\ 1976, \apss, 39, 447 

\bibitem[Lorenzetti et al.(2012)]{Lorenzetti12} Lorenzetti, D., 
Antoniucci, S., Giannini, T., et al.\ 2012, arXiv:1202.4136 

\bibitem[{{McCarthy} {et~al.}(1998){McCarthy}, {Cohen}, {Butcher}, {Cromer},
  {Croner}, {Douglas}, {Goeden}, {Grewal}, {Lu}, {Petrie}, {Weng}, {Weber},
  {Koch}, \& {Rodgers}}]{McCarthy1998}
{McCarthy}, J.~K. {et~al.} 1998, in Society of Photo-Optical Instrumentation
  Engineers (SPIE) Conference Series, Vol. 3355, ed. {S.~D'Odorico}, 81


\bibitem[Meyer et al.(1997)]{Meyer97} Meyer, M.~R., Calvet, N., 
\& Hillenbrand, L.~A.\ 1997, \aj, 114, 288 


\bibitem[McLean et al.(1998)]{McLean98} McLean, I.~S., Becklin, 
E.~E., Bendiksen, O., et al.\ 1998, \procspie, 3354, 566 


\bibitem[{{Miller} \& {Stone}(1993)}]{Miller93}
{Miller}, J.~S., \& {Stone}, R.~P.~S. 1993, {Lick Obs. Tech. Rep. 66} (Santa
  Cruz: Lick Obs.)


\bibitem[Miller et al.(2011)]{Miller11} Miller, A.~A., 
Hillenbrand, L.~A., Covey, K.~R., et al.\ 2011, \apj, 730, 80 

\bibitem[Morales-Calder{\'o}n et al.(2011)]{Morales11} 
Morales-Calder{\'o}n, M., Stauffer, J.~R., Hillenbrand, L.~A., et al.\ 
2011, \apj, 733, 50 

\bibitem[Muzerolle et al.(1998)]{Muzerolle1998} Muzerolle, J., 
Hartmann, L., \& Calvet, N.\ 1998, \aj, 116, 2965 

\bibitem[Muzerolle et al.(2009)]{2009ApJ...704L..15M} Muzerolle, J., 
Flaherty, K., Balog, Z., et al.\ 2009, \apjl, 704, L15 

\bibitem[Najita et al.(1996)]{Najita1996} Najita, J., Carr, J.~S.,
Glassgold, A.~E., Shu, F.~H., \& Tokunaga, A.~T.\ 1996, \apj, 462, 919

\bibitem[Nomura et al.(2007)]{Nomura2007} Nomura, H., Aikawa, Y.,
Tsujimoto, M., Nakagawa, Y., \& Millar, T.~J.\ 2007, \apj, 661, 334


\bibitem[Ofek et al.(2012)]{Ofek12} Ofek, E.~O., Laher, R., 
Law, N., et al.\ 2012, \pasp, 124, 62 


\bibitem[{{Oke} {et~al.}(1995){Oke}, {Cohen}, {Carr}, {Cromer}, {Dingizian},
  {Harris}, {Labrecque}, {Lucinio}, {Schaal}, {Epps}, \& {Miller}}]{Oke1995}
{Oke}, J.~B. {et~al.} 1995, \pasp, 107, 375

\bibitem[{{Oke} \& {Gunn}(1982)}]{Oke1982}
{Oke}, J.~B., \& {Gunn}, J.~E. 1982, \pasp, 94, 586


\bibitem[Perley et al.(2010)]{Perley10} Perley, D.~A., Bloom, 
J.~S., Klein, C.~R., et al.\ 2010, \mnras, 406, 2473 

\bibitem[Plavchan et al.(2008)]{Plavchan08} Plavchan, P., Gee, 
A.~H., Stapelfeldt, K., \& Becker, A.\ 2008, \apjl, 684, L37 


\bibitem[Plavchan et al.(2010)]{Plavchan10} Plavchan, P., 
Laohakunakorn, N., Seifahrt, A., Staplefeldt, K., 
\& Gee, A.~H.\ 2010, Bulletin of the American Astronomical Society, 42, \#429.06 

\bibitem[Ralchenko et al.(2011)]{NIST}
Ralchenko, Yu., Kramida, A.E., Reader, J., and NIST ASD Team (2011). NIST Atomic Spectra Database (ver. 4.1.0), [Online]. Available: http://physics.nist.gov/asd [2012, June 21]. National Institute of Standards and Technology, Gaithersburg, MD.

\bibitem[{{Rau} {et~al.}(2009){Rau}, {Kulkarni}, {Law}, {Bloom}, {Ciardi},
  {Djorgovski}, {Fox}, {Gal-Yam}, {Grillmair}, {Kasliwal}, {Nugent}, {Ofek},
  {Quimby}, {Reach}, {Shara}, {Bildsten}, {Cenko}, {Drake}, {Filippenko},
  {Helfand}, {Helou}, {Howell}, {Poznanski}, \& {Sullivan}}]{Rau09}
{Rau}, A. {et~al.} 2009, \pasp, 121, 1334

\bibitem[Pueyo et al.(2012)]{Pueyo12} Pueyo, L., Hillenbrand, L., Gautam, V., Oppenheimer, B.R.,Monnier, J., Hinkley, S., Crepp, J., Roberts, L.C., Brenner, D., Zimmerman, N., Parry, I., Beichman, C., Dekany, R., 2, Shao, M., Burruss, R., Baranec, C., Cady, E., Roberts, J., Soummer, R.\ 2012, \apj, in press 

\bibitem[Rayner et al.(2003)]{Rayner03} Rayner, J.~T., Toomey, 
D.~W., Onaka, P.~M., et al.\ 2003, \pasp, 115, 362 

\bibitem[Reipurth 
\& Aspin(2010)]{Reipurth10} Reipurth, B., \& Aspin, C.\ 2010, Evolution of Cosmic Objects through their Physical Activity, 19 

\bibitem[Richards et al.(2011)]{Richards11} Richards, J.~W., 
Starr, D.~L., Butler, N.~R., et al.\ 2011, \apj, 733, 10 

\bibitem[Romanova et al.(2005)]{Romanova05} Romanova, M.~M., 
Ustyugova, G.~V., Koldoba, A.~V., 
\& Lovelace, R.~V.~E.\ 2005, \apjl, 635, L165 


\bibitem[Romanova et al.(2008)]{Romanova08} Romanova, M.~M., 
Kulkarni, A.~K., \& Lovelace, R.~V.~E.\ 2008, \apjl, 673, L171 

\bibitem[Romanova et al.(2009)]{Romanova09} Romanova, M.~M., 
Ustyugova, G.~V., Koldoba, A.~V., 
\& Lovelace, R.~V.~E.\ 2009, \mnras, 399, 1802 


\bibitem[Scargle(1982)]{Scargle82} Scargle, J.~D.\ 1982, \apj, 263, 835 

\bibitem[Semkov et 
al.(2010)]{Semkov10} Semkov, E.~H., Peneva, S.~P., Munari, U., Milani, A., \& Valisa, P.\ 2010, \aap, 523, L3 

\bibitem[Semkov \& Peneva(2012)]{SemkovPeneva12} Semkov, E.~H., \& Peneva, S.~P.\ 2012, \apss, 338, 95 


\bibitem[Sicilia-Aguilar et al.(2008)]{SiciliaAguilar08} 
Sicilia-Aguilar, A., Mer{\'{\i}}n, B., Hormuth, F., et al.\ 2008, \apj, 
673, 382 


\bibitem[Smith(1995)]{Smith1995} Smith, M.~D.\ 1995, \aap, 296, 789


\bibitem[{{Steidel} {et~al.}(2004){Steidel}, {Shapley}, {Pettini},
  {Adelberger}, {Erb}, {Reddy}, \& {Hunt}}]{Steidel2004}
{Steidel}, C.~C., {Shapley}, A.~E., {Pettini}, M., {Adelberger}, K.~L., {Erb},
  D.~K., {Reddy}, N.~A., \& {Hunt}, M.~P. 2004, \apj, 604, 534

\bibitem[Sullivan et al.(2011)]{Sullivan2011} Sullivan, M., 
Kasliwal, M.~M., Nugent, P.~E., et al.\ 2011, \apj, 732, 118 



\bibitem[Vogt et al.(1994)]{Vogt94} Vogt, S.~S., Allen, S.~L., 
  Bigelow, B.~C., et al.\ 1994, \procspie, 2198, 362 

\bibitem[Wang et al.(2004)]{Wang04} Wang, H., Apai, D., Henning, T., \& Pascucci, I.\ 2004, \apj, 601, L83

\bibitem[White 
\& Hillenbrand(2004)]{WH04} White, R.~J., \& Hillenbrand, L.~A.\ 2004, \apj, 616, 998 

\bibitem[van Boekel et 
al.(2010)]{2010A&A...517A..16V} van Boekel, R., Juh{\'a}sz, A., Henning, T., et al.\ 2010, \aap, 517, A16 


\bibitem[{Vorobyov} \& {Basu}(2005)]{vorobyov05}
{Vorobyov}, E.~I., \& {Basu}, S., 2005, \apjl\/, 633, L137--L140.


\bibitem[Wilson et al.(2004)]{Wilson04} Wilson, J.~C., 
Henderson, C.~P., Herter, T.~L., et al.\ 2004, \procspie, 5492, 1295 

\bibitem[Xiao et al.(2010)]{Xiao10} Xiao, L., Kroll, P., 
\& Henden, A.~A.\ 2010, \aj, 139, 1527 


\bibitem[Zechmeister \& K{\"u}rster(2009)]{Zechmeister09} Zechmeister, M., \& K{\"u}rster, M.\ 2009, \aap, 496, 577 

\bibitem[{Zhu} et~al.(2009){Zhu}, {Hartmann}, {Gammie}, \& {McKinney}]{zhu09}
{Zhu}, Z., {Hartmann}, L., {Gammie}, C., \& {McKinney}, J.~C., 2009
  \apj\/, 701, 620--634.


\bibitem[Zhu et al.(2010)]{Zhu10} Zhu, Z., Hartmann, L., 
Gammie, C.~F., et al.\ 2010, \apj, 713, 1134 



\end{thebibliography}
\end{document}